\documentclass[11pt,letterpaper]{article}

\usepackage[margin=1in]{geometry}

\usepackage[dvipsnames,usenames]{xcolor}
\usepackage[colorlinks=true,pdfpagemode=UseNone,urlcolor=RoyalBlue,linkcolor=RoyalBlue,citecolor=OliveGreen,pdfstartview=FitH]{hyperref}
\definecolor{rotblue}{RGB}{30,103,151}

\usepackage{microtype}
\usepackage[utf8]{inputenc}
\usepackage[english]{babel}
\usepackage{csquotes}
\usepackage{graphicx}
\usepackage{amsmath}
\usepackage{amsthm}
\usepackage{amssymb}
\usepackage{amsfonts}
\usepackage{paralist}
\usepackage{mathtools}
\usepackage{thmtools, thm-restate}
\usepackage{enumitem}
\usepackage{subcaption}
\hypersetup{
colorlinks,
linkcolor={red!50!black},
citecolor={blue!50!black},
urlcolor={blue!80!black}
}
\usepackage[capitalize,noabbrev]{cleveref}

\usepackage{thm-restate}

\DeclareMathOperator*{\argmax}{arg\,max}

\declaretheorem{theorem}
\declaretheorem[sibling=theorem]{lemma}
\declaretheorem[sibling=theorem]{claim}

\declaretheorem[sibling=theorem]{corollary}
\declaretheorem[sibling=theorem]{fact}

\newcommand{\eps}{\varepsilon}

\newcommand{\sep}{\mathrm{Separator}}
\newcommand{\Span}{\mathrm{span}}
\newcommand{\diam}{\mathrm{diam}}
\newcommand{\dist}{\mathrm{dist}}

\newcommand{\opt}{\mathrm{OPT}}

\title{Stacking Stars: Minimum-Area Packing with and without Rotation}

\author{Mikkel Abrahamsen\thanks{University of Copenhagen, Denmark. \texttt{miab@di.ku.dk}.}
\and
Jacobus Conradi\thanks{University of Copenhagen, Denmark. \texttt{jaco@di.ku.dk}.}
\and
Arindam Khan\thanks{Department of Computer Science and Automation, Indian Institute of Science, Bengaluru, India. \texttt{arindamkhan@iisc.ac.in}.}
\and
Jack Stade\thanks{University of Copenhagen, Denmark. \texttt{jast@di.ku.dk}.}
\and
Sampson Wong\thanks{University of Copenhagen, Denmark. \texttt{sampson.wong123@gmail.com}.}
\and
Shuyi Yan\thanks{University of Copenhagen, Denmark. \texttt{shya@di.ku.dk}.}
\and
Hanwen Zhang\thanks{University of Copenhagen, Denmark. \texttt{hazh@di.ku.dk}.}
}

\date{}

\begin{document}

\maketitle
\thispagestyle{empty}

\begin{abstract}
We study the problem of packing a given set of polygons into a minimum-area convex container, that is, placing them without overlap so as to minimize the area of their convex hull. 
Every previously known polynomial-time algorithm with a nontrivial approximation guarantee for packing polygons with arbitrary edge orientations required the polygons to be convex.
In fact, an algorithm was not even known when every input polygon was a quadrilateral.
We obtain such algorithms for the entire class of star-shaped polygons with unrestricted combinatorial complexity. 
The stars may be just translated or also rotated. 
In either case, we give an $O(\log n)$-approximation algorithm, where $n$ is the number of stars.
Likewise, we get $O(\log n)$-approximation algorithms for bin and strip packing, assuming that the diameter of the stars is at most some (small) constant.

Previous approximation algorithms for polygon packing focused on convex objects, and the analysis always relied on a density argument, i.e., if the area of the packed objects is at least a fraction $\rho$ of the area of the container, then we have a $(1/\rho)$-approximation.
However, star-shaped polygons may need an arbitrarily large area to be packed, even when the total area of the stars approaches $0$ and each star has bounded diameter.
This phenomenon prevents the use of density arguments.

We instead establish a strong structural property about any packing:
Even though the stars can fit together in complicated ways, we can reorganize them into simple stacks, replace these stacks by 
rectangles, and consider a packing of these rectangles.
The rearrangement costs only a factor of $O(\log n)$, which stems from the number of size groups. 

In the translational setting, we use an approximation algorithm for asymmetric TSP to construct stacks and then apply a convex-polygon packing algorithm to their convex hulls.
In the rotational setting, we prove an additional structural property: restricting all radii in a stack to have exactly the same orientation costs only a constant factor.
The radius is a segment from a point in the kernel to a furthest vertex.
It follows that we can rotate all stars so that their radius are horizontal and then apply the translational algorithm.

We complement our algorithms by showing that both the translational and the rotational versions are APX-hard, even for constant complexity stars.
\end{abstract}

\newpage
\clearpage

\tableofcontents

\newpage
\clearpage
\pagenumbering{arabic} 

\section{Introduction}\label{sec:intro}

\begin{figure}
\centering
\includegraphics[width=\textwidth,page=16]{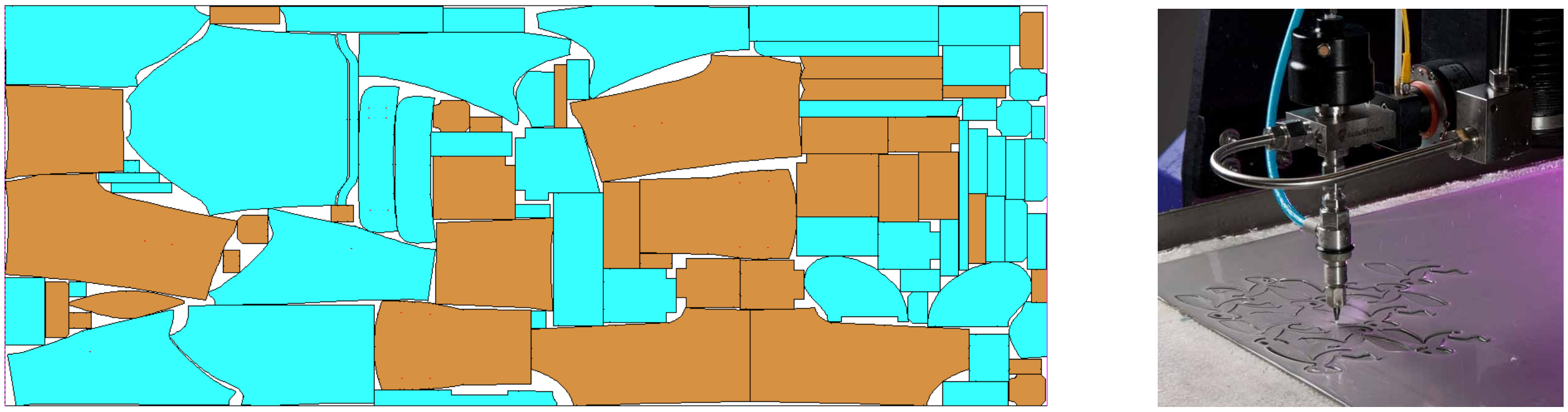}
\caption{Examples of applications involving irregular packing.
Left: Garment pieces arranged on fabric, kindly provided by \href{https://www.mirisys.com/}{MIRISYS} and produced by their automatic nesting software.
Right: Waterjet cutting. The image was cropped from a \href{https://commons.wikimedia.org/wiki/File:Retro_Systems_Waterjet.jpg}{photograph} by Steve Brown Photography, licensed under CC BY 3.0.}
\label{fig:intro}
\end{figure}

Packing problems are among the most fundamental and practically motivated problems in computational geometry and combinatorial optimization.
They arise in a wide range of real-world scenarios, for instance when cutting shaped pieces from fabric, arranging parts for milling, waterjet cutting and laser cutting, and placing objects for 3D printing (see \cref{fig:intro}).
In many of these applications, the objects are irregular and nonconvex, and their edges are not restricted to a fixed set of orientations.

The bulk of the theoretical work on geometric packing has focused on axis-parallel rectangles; for surveys with hundreds of references, see~\cite{CHRISTENSEN201763,DBLP:journals/sigact/Stee12,DBLP:journals/sigact/Stee15,Csirik1998}.
A smaller body of theoretical work considers packing convex polygons
with unrestricted edge orientations.
This includes algorithms for a constant number of pieces%
~\cite{DBLP:conf/jcdcg/AltH00,DBLP:journals/ijcga/TangWC06,
DBLP:journals/algorithmica/AhnC12,DBLP:journals/comgeo/ParkBAA16}
and approximation algorithms for arbitrary-sized input sets%
~\cite{altapproximating,alt_convexOffline_corr,
abrahamsen2023onlinesorting,kurpisz2023improved,DBLP:journals/corr/abs-2607-22179}.
Related results concern geometric knapsack%
~\cite{DBLP:conf/icalp/MerinoW20}
and the exact complexity of packing under rotations%
~\cite{DBLP:journals/theoretics/AbrahamsenMS24}.
For convex objects, classical techniques such as density arguments are effective.

Packing nonconvex polygons remains poorly understood from the viewpoint of
polynomial-time approximation algorithms.
Early geometric methods capable of handling arbitrary simple polygons were
developed by Milenkovic and by Daniels and Milenkovic%
~\cite{MILENKOVIC19993,DBLP:conf/stoc/Milenkovic96,
DBLP:journals/algorithmica/Milenkovic97,
milenkovich1999translational,DBLP:journals/algorithmica/DanielsM97}.
They gave exact algorithms for several containment and enclosure problems,
as well as algorithms allowing a controlled geometric error, but with running
times exponential in the number of pieces.
Further exact approaches to irregular nesting have been developed using
constraint programming, mixed-integer programming, branch-and-bound and
global optimization%
~\cite{CarravillaRibeiroOliveira2003,FischettiLuzzi2009,
AlvarezValdesMartinezTamarit2013,Jones2014}.
Some of these methods solve a discretized version of the geometric problem,
whereas methods addressing continuous placements generally apply only to
relatively small instances.
More recently, Gerlach, Hennies and Kleist%
~\cite{DBLP:journals/corr/abs-2603-22098}
initiated the study of online packing of nonconvex orthogonal polygons.

The scarcity of polynomial-time worst-case results stands in sharp contrast
to the extensive operations-research literature on irregular cutting and
packing. 
A series of surveys and bibliographies collectively covering hundreds of
works documents mathematical models, geometric techniques, exact approaches,
heuristics and metaheuristics%
~\cite{dyckhoff1992cutting,sweeney1992cutting,
DOWSLAND1995506,bennell2008geometry,bennell2009tutorial,
hopper2001review,leao2020irregular,guo2022twodimensional}.
Optimization-based approaches may certify optimality on manageable
instances, sometimes only after discretizing the possible placements,
whereas heuristic methods are designed to produce high-quality solutions
in practice but typically provide no worst-case approximation guarantees.
This research has led to multiple software systems used in practice.%
\footnote{Examples of prominent commercial systems include
\href{https://www.sigmanest.com/}{SigmaNEST},
\href{https://www.hypertherm.com/hypertherm/pronest/pronest-cadcam-nesting-software/}{Hypertherm ProNest},
\href{https://www.lantek.com/uk/Lantek-Expert-Nesting-Software}{Lantek Expert},
\href{https://www.autodesk.com/products/inventor-nesting/overview}{Autodesk Inventor Nesting}, and
\href{https://www.mirisys.com/}{MIRISYS}.
Representative open-source systems include
\href{https://deepnest.io/}{Deepnest} and
\href{https://svgnest.com/}{SVGnest}.}
The 2024 edition of the CG:SHOP Challenge (Computational Geometry: Solving Hard Optimization Problems) also concerned a packing problem involving irregular and nonconvex objects~\cite{DBLP:journals/corr/abs-2403-16203}.
Thus, the theoretical gap does not reflect a lack of interest in irregular packing.
Rather, there is a striking contrast between a mature experimental and operations-research literature and the very limited theory of polynomial-time approximation.

Every previously known polynomial-time algorithm with an approximation factor of $o(n)$, i.e., better than what is obtained by packing the convex hulls, for packing an arbitrary number $n$ of polygons with arbitrary edge orientations requires the polygons to be convex.
In fact, no such algorithm was known even when every input polygon was a quadrilateral, the simplest type of polygon that can be nonconvex.
We obtain such algorithms for the entire class of star-shaped polygons, without restricting their combinatorial complexity.
Recall that a polygon $P$ is said to be \emph{star-shaped} if there is a point $q\in P$ that sees all of $P$, i.e., for every point $p\in P$, we have $qp\subseteq P$.
The set of points for which this holds is called the \emph{kernel} of $P$.
The \emph{radius} of $P$ is the segment from a fixed point $q$ in the kernel to a furthest vertex.
In concurrent and independent work, de Oliveira Silva, Dell'Arriva, and Miyazawa~\cite{silva2026beyond} give asymptotic approximation schemes for translational bin packing of nonconvex polygons from classes that satisfy certain structural conditions.
Their main example of such a nonconvex class consists of the homothetic copies of a fixed finite set of star-shaped polygons, so only a fixed finite set of edge orientations occurs.

We focus primarily on the \emph{minimum-area convex-container problem} (MACC).
Given a set of \(n\) polygons, the goal is to place them without overlap so as to minimize the area of their convex hull%
~\cite{altapproximating,alt_convexOffline_corr}
(related online variants require the container to be an axis-parallel rectangle or square%
~\cite{DBLP:conf/compgeom/AbrahamsenB21,
DBLP:journals/algorithmica/FeketeH17}).
Equivalently, we seek a minimum-area convex container whose shape and the placement of the objects are both chosen by the algorithm.
We analyze two variants.
In the \emph{translational} variant, the objects may only be translated.
In the \emph{rotational} variant, they may be both translated and rotated.

\subsection{Our Contribution}

Our main result is the following.

\begin{theorem}\label{thm:main}
There are polynomial-time $O(\log n)$-approximation algorithms for both translational and rotational packing of $n$ star-shaped polygons into a minimum-area convex container.
\end{theorem}

The translational algorithm is built from simple steps.
We first pack the stars into convex \emph{mini-containers} of total area $O(\log n)\cdot\opt$, and then pack the mini-containers as convex objects using existing methods by Alt, de Berg, and Knauer~\cite{altapproximating,alt_convexOffline_corr}.
All nonconvex interactions are handled inside the mini-containers; at the global level only convex geometry remains.

To create the mini-containers, the stars are grouped by size and shape, and similar stars are nested into \emph{stacks}, much like a stack of bowls.
The order within a stack is found using an approximation algorithm for the asymmetric traveling salesperson problem~\cite{DBLP:journals/jacm/SvenssonTV20,TraubVygen2024}.
The convex hulls of the stacks serve as the mini-containers.

In the rotational case, we show that we lose only an extra constant factor by rotating the stars in each mini-container so they have horizontal radii.
Our algorithm thus rotates the stars to have horizontal radii and then applies the translational algorithm.

The constants in our analysis are large, mainly due to the very fine grouping of the stars and pessimistic choices of scaling factors.
However, the idea of building stacks of similar stars is natural, and we believe that it could also be the basis for practical heuristics, which would be interesting to evaluate experimentally.

The mini-container approach extends to other packing problems for which algorithms are known for convex polygons.
In \emph{bin packing}, the goal is to pack the objects into a minimum number of unit-square bins; in \emph{strip packing}, the objects are packed into a unit-width strip, and the goal is to minimize the used length of the strip.
Rearranging the packing produced in \Cref{thm:main} gives the following.
The details are in \Cref{app:cor:bin-strip}.

\begin{corollary}\label{cor:bin-strip}
There exists a constant $\delta>0$ such that, if every polygon has diameter at most $\delta$, there are polynomial-time $O(\log n)$-approximation algorithms for both translational and rotational bin packing and strip packing of $n$ star-shaped polygons.
\end{corollary}

For stars of comparable size, the algorithms in fact give $O(1)$-approximations, that is, if the convex hulls of all input stars have areas within a constant factor of each other (in particular, if all stars are identical).
The logarithmic factor is inherent to any algorithm that groups stars by size before packing them into convex mini-containers, as we show in \cref{sec:technical-overview}.
The grouping step itself does not use that the polygons are stars, and we formalize it as a reduction for general polygons, proved in \cref{app:thm4}. We believe it can be useful for designing algorithms for other classes of polygons.

\begin{restatable}{theorem}{binreduction}
\label{thm:general-macc-to-bin}
For some constants $\epsilon,\delta\in(0,1)$, if there exists a polynomial-time $\alpha$-approximation translational bin packing algorithm for polygons of diameter at most $\epsilon$ and convex hull area at least $\delta\epsilon^2$, then there exists a polynomial-time $O(\alpha\log n)$-approximation algorithm for translational packing of $n$ polygons into a minimum-area convex container. 
\end{restatable}

Finally, we complement the algorithmic results by proving APX-hardness, even for stars of comparable sizes and constant complexity.

\begin{theorem}\label{thm:hardness}
The translational and rotational variants of minimum-area convex-container packing are APX-hard.
In the instances produced by our reduction, each star has at most $17$ vertices and the convex hulls are fat and have diameters within a constant factor of each other.
\end{theorem}

The next section gives an overview of the techniques behind these results.

\section{Technical overview}
\label{sec:technical-overview}

Our overall approach is to partition the stars and to pack each set of the partition into convex \emph{mini-container}.
We will then use the techniques of Alt, de Berg, and Knauer~\cite{altapproximating,alt_convexOffline_corr} to pack the mini-containers.
In particular, we use the following result as a black box.

\begin{theorem}[Theorem 2.2 from \cite{altapproximating,alt_convexOffline_corr}]
\label{thm:convex-packing}
There exists an $O(N\log N)$-time algorithm for packing a given set of convex polygons $C_1,\dots,C_n$ of total complexity $N$ into an axis-aligned rectangle $R$ such that
\[
|R| \le c\cdot\left(\sum_{i=1}^n |C_i|+\ell_{\max}h_{\max}\right),
\]
where $|\cdot|$ denotes area, $c$ is a constant, and $\ell_{\max}$ and $h_{\max}$ are the maximum horizontal and vertical spans of $C_1,\dots,C_n$, respectively.
\end{theorem}

The goal is to obtain a packing whose convex hull is $O(\log n)\cdot \opt$, where $\opt$ is the area of a minimum convex container for all the stars.
To this end, we need two properties:
(1) the total area of the mini-containers is $O(\log n)\cdot\opt$, and (2) for each mini-container, its extend in every direction is proportional to that of each of the stars that it contains.
The latter requirement ensures the bound $\ell_{\max}h_{\max}=O(\opt)$ in \Cref{thm:convex-packing} in the coordinate system where the longest radius is horizontal.

\paragraph{Stacks.}
For now, consider a much simpler setting: we are given $n$ stars $P_1,\ldots,P_n$ that are all copies of the same star $P$. One might ask: \emph{For which $P$ does packing the convex hulls of the $P_i$, instead of the stars themselves, give a good approximation?}

\begin{figure}[t]
\centering
\includegraphics[page=17,width=\textwidth]{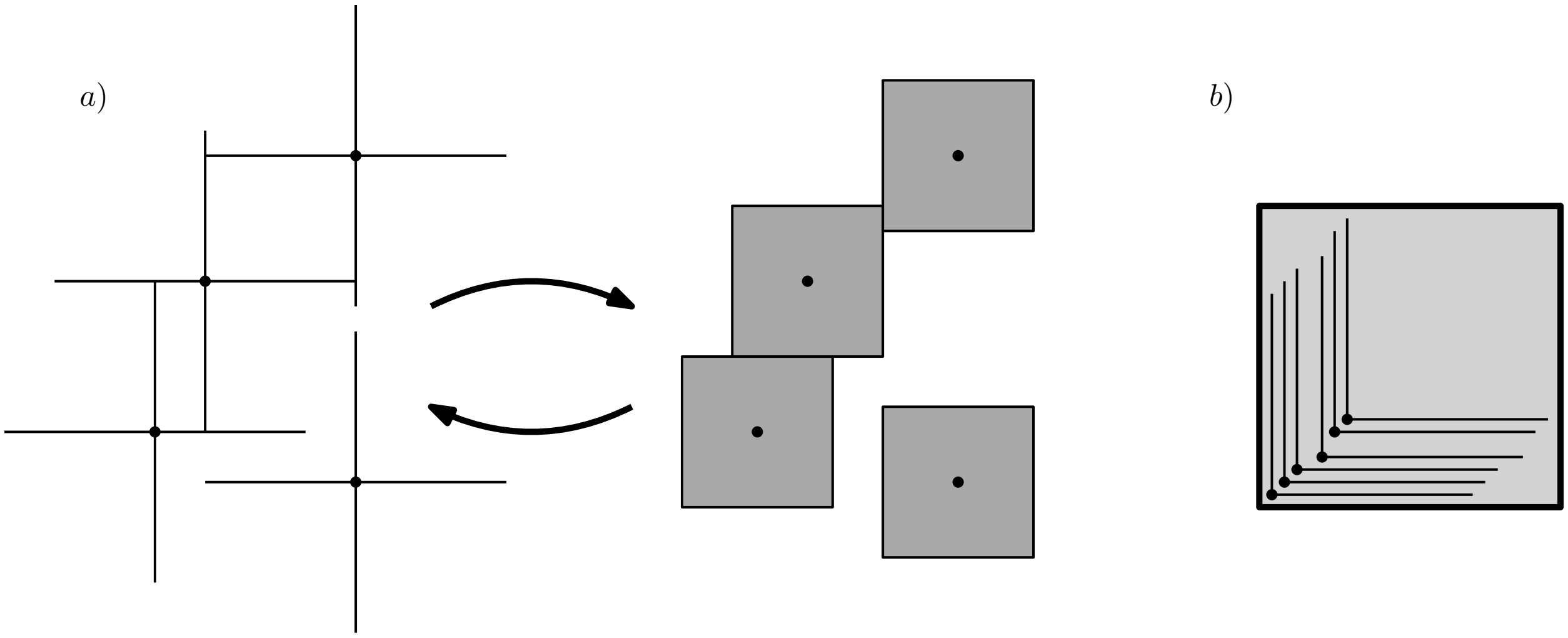}
\caption{a) Duality between packing copies of a plus-shaped polygon and axis-aligned squares. b) a packing of many L-shapes as a stack in a constant-sized container.}
\label{fig:duality}
\end{figure}

The answer is surprisingly nuanced. If $P$ is convex, the answer is of course ``always''. But there are also polygons of zero area for which the answer is yes. Consider the polygon $P$ given by the union of the two segments $\{0\}\times[-1,1]$ and $[-1,1]\times\{0\}$, forming a plus. For two copies of $P$, no matter how they are arranged, their origins never have $\ell_\infty$-distance less than $1$. One can, in fact, replace $P$ by the square $Q=[-0.5,0.5]\times[-0.5,0.5]$: any packing of $n$ copies of $P$ is also a packing of $n$ copies of $Q$, and vice versa; see \cref{fig:duality} a). The convex hull of $P$ has twice the area of $Q$, so packing the convex hulls of copies of $P$ is, up to constants, as good as packing the copies of $P$.

If we instead take $P$ to be the union of the two segments $\{0\}\times[0,1]$ and $[0,1]\times\{0\}$, forming an L, the answer is very different. For any $\varepsilon>0$, one can pack $n$ copies of $P$ into the convex container $[0,1+\varepsilon]\times[0,1+\varepsilon]$ by forming a \emph{stack} of nested copies of $P$ (see \cref{fig:duality} b)), whereas packing the convex hulls requires a container of area $\Theta(n)$.

Our main structural insight is that this stacking behavior is \emph{the only} structure that prevents us from replacing stars by their convex hulls, at a loss of $O(\log n)$.

\begin{figure}[!h]
\centering
\includegraphics[page=12]{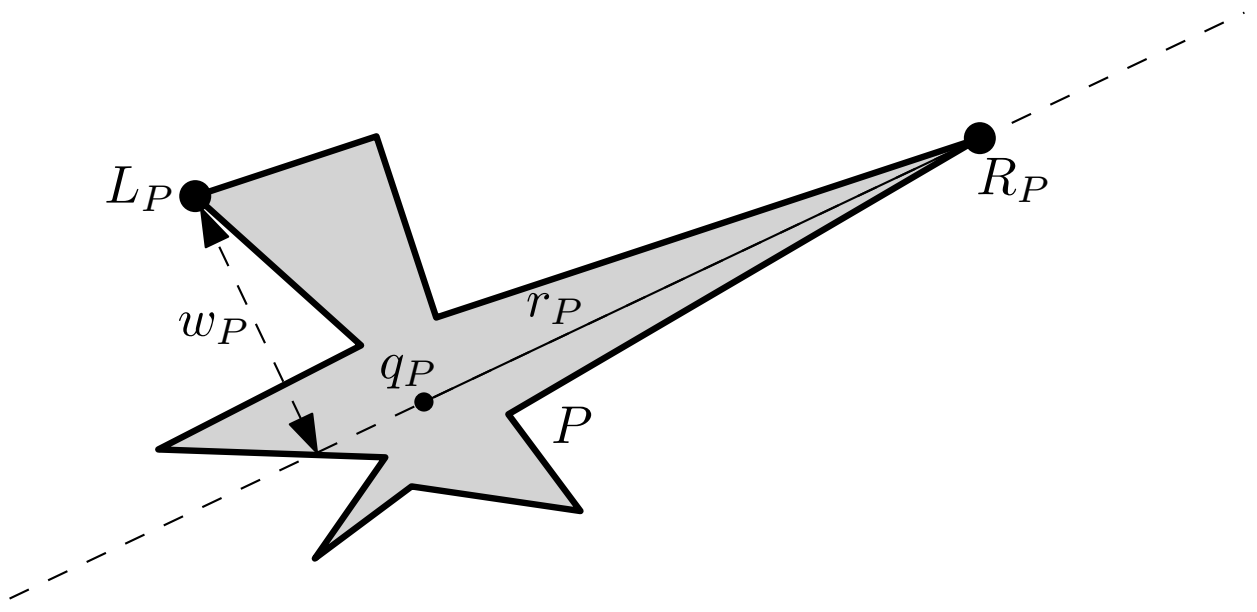}
\caption{A star $P$ with reference point $q_P$, radius $q_PR_P$ of length $r_P$, and width $w_P$ attained at $L_P$.}
\label{fig:radiuswidth}
\end{figure}

\paragraph{Size groups and fine types.} For arbitrary instances we do not have the luxury of congruent polygons, but we still want to treat the input polygons as somewhat congruent objects. For this, fix for every star $P$ a point $q_P$ in its kernel, its \emph{reference point}. The \emph{radius} of $P$ is the segment from $q_P$ to a farthest point $R_P$ of $P$ (ties are broken arbitrarily), and $r_P$ is its length. The \emph{width} $w_P$ of $P$ is the largest distance of a point of $P$ from the line through the radius, and $L_P$ is a point attaining it; see \cref{fig:radiuswidth}. Let $r_{\max}$ and $w_{\max}$ be the maximum radius length and width of any input star.

We partition the input stars into size groups $\mathcal{G}_{a,b}$, which contains all stars with radius and width of one \emph{dyadic scale}, i.e., stars $P$ such that $r_P\in(r_{\max}/2^{a+1},r_{\max}/2^{a}]$ and $w_P\in(w_{\max}/2^{b+1},w_{\max}/2^{b}]$.
Polygons with $a>\log n$ or $b>\log n$ are packed via their convex hulls: each of these hulls has area $O(r_{\max}w_{\max}/n)$, so together they need area $O(r_{\max}w_{\max})=O(\opt)$. We are left with polygons in $O(\log^2 n)$ \emph{size groups}, where the size group of $a,b$ has \emph{scales} $r=r_{\max}/2^a$ and $w=w_{\max}/2^b$.

We partition the size groups further into \emph{fine types}. The \emph{orientation} of a star is the direction of its radius, directed away from $q_P$. We split $[0,2\pi)$ into intervals of length $\Theta(w/r)$, and the fine type of a star is the interval containing its orientation, together with the side of the radius on which $L_P$ lies; the precise parameters are fixed in \cref{sec:star-types}. Polygons of one fine type are similar enough that the stacks observed above make sense. %

We would now like to pack every fine type independently. The problem is that there might be up to $n$ nonempty fine types, and bounding the area of each of them independently against $\opt$ could result in a $\Theta(n)$-approximation. Luckily, this is not the case.

\paragraph{A structural theorem for fine types.} Consider one fixed size group. Our first main technical contribution is the following theorem.

\begin{theorem}
\label{thm:star-structural-decomposition}
Let $\mathcal P$ be a set of star-shaped polygons in one size group with scales $r,w$, and let $\Pi$ be a packing of $\mathcal P$ in a convex container $C$.
Then $\mathcal P$ can be partitioned into a disjoint union of \underline{parts}
\[
\mathcal P
=
\mathcal P_1\mathbin{\dot\cup}\cdots
\mathbin{\dot\cup}\mathcal P_t
\]
such that in every part $\mathcal P_j$, all the stars are of the same fine type, and, without changing the placement inherited from $\Pi$, are contained in a (not necessarily axis-aligned) rectangle $B_j$ of dimensions $O(r)\times O(w)$.
Moreover,
\[
\sum_{j=1}^t |B_j| = O(|C|),
\]
where $|C|$ is the area of $C$.
\end{theorem}

\begin{figure}[!h]
\centering
\includegraphics[page=3]{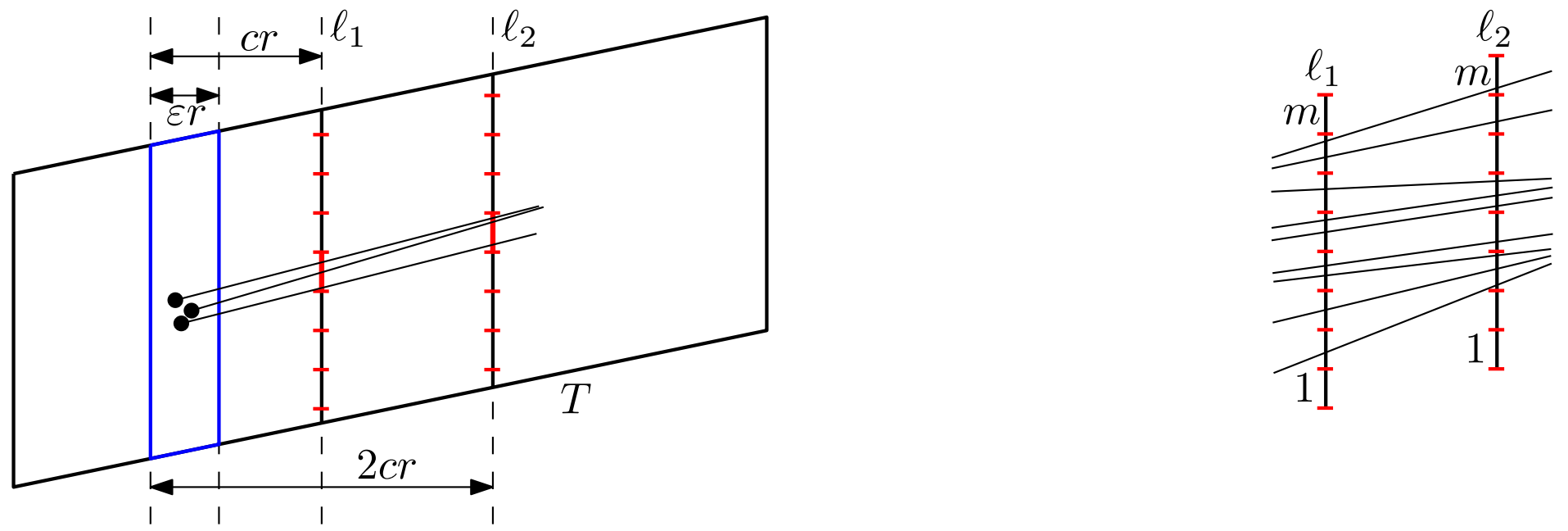}
\caption{Left: The enclosing parallelogram is cut into vertical strips of width $\varepsilon r$; one strip is shown. Two test lines $\ell_1,\ell_2$ are placed to its right. The reference points and radii assigned to one pair $(a,b)$ of intervals are shown.
Right: The occupied interval pairs form a monotone sequence in $[m]^2$.}
\label{fig:stabbing-segment}
\end{figure}

We sketch the proof here; \cref{sec:star-local-decomposition} proves the slightly stronger \cref{thm:star-local-decomposition-strong}.
First, split the polygons of the size group by orientation into $(\pi/100)$-radian classes; there are constantly many of these. Within one class, all radii can be considered to be nearly horizontal, and we may take $C$ to be a parallelogram with vertical sides of length $H$ and horizontal span $W$.
Cut $C$ into vertical strips of width $\varepsilon r$, and consider the stars with reference point in a fixed strip. Each of their radii has length more than $r/2$ and is nearly horizontal, so it crosses two vertical \emph{test lines} $\ell_1,\ell_2$ placed $\Theta(r)$ to the right of the strip; see \cref{fig:stabbing-segment}. Partition each test line into $m=O(H/w)$ intervals of length $\varepsilon w$, and assign each star the pair $(a,b)$ of intervals crossed by its radius. Two radii ordered one way on $\ell_1$ and the other way on $\ell_2$ would cross in between, which is impossible since every radius lies inside its star. Hence the occupied pairs are monotone in both coordinates, and there are at most $2m=O(H/w)$ of them rather than $m^2$.

The stars assigned the same pair $(a,b)$ and of the same fine type form one part $\mathcal P_j$. 
For stars assigned to the same pair $(a,b)$, their reference points lie in a strip of width $\varepsilon r$, and their radii pass through the same two intervals of length $\varepsilon w$ at distance $\Theta(r)$, so their orientations differ by $O(w/r)$ and they belong to constantly many fine types. They also fit, in place, into a rectangle of dimensions $O(r)\times O(w)$. There are $O(W/r)$ strips, each with $O(H/w)$ occupied pairs, each contributing area $O(rw)$, so the rectangles over all parts have total area $O(WH)=O(|C|)$. %

\begin{figure}[!h]
\centering
\includegraphics[page=8]{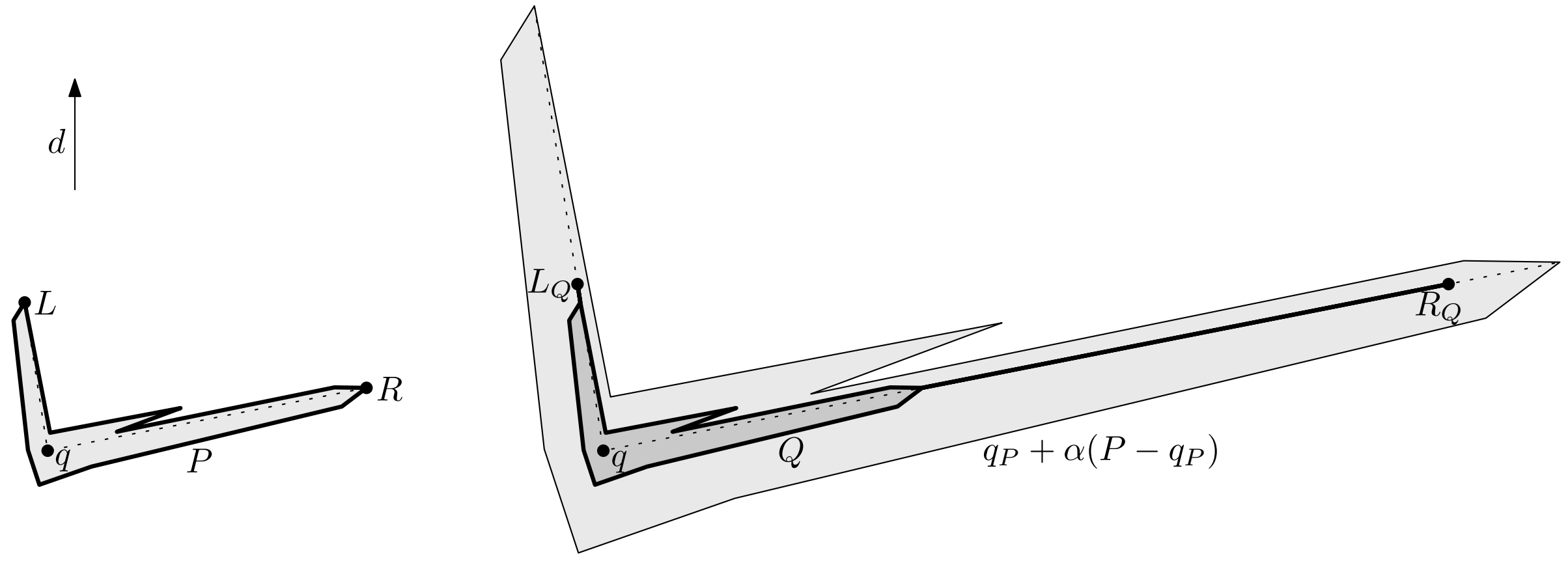}
\caption{A star $P$ (left) and its bowl $Q(P)$ (right). The arms of the bowl are the extensions of $q_PR_P$ and $q_PL_P$ to height $h$ in direction $d$. The bowl sits inside a copy of $P$ scaled by the constant $\alpha$.
}
\label{fig:star-to-bowl}
\end{figure}

\paragraph{Stacks of bowls.} By \cref{thm:star-structural-decomposition} it suffices to pack each fine type separately. As alluded to above, polygons of one fine type are ``congruent enough'' that we expect them to admit a nice stack. To find such a stack, we first align the stars of one fine type with each other, as follows.
Fix a fine type of a size group with scales $r$ and $w$, and let $d$ be a direction, depending only on the fine type, that is nearly perpendicular to all its radii. For a star $P$, extend the two segments $q_PR_P$ and $q_PL_P$ beyond $R_P$ and $L_P$ until both endpoints are at height exactly $h=\Theta(w)$ above $q_P$ in direction $d$.
The value $h$ is common for all stars of the same fine type.
We call these two extended segments the \emph{arms} of $P$, and the polygon $Q(P)$ obtained by adding the arms to $P$ the \emph{bowl} of $P$; see \cref{fig:star-to-bowl}. A bowl (i) extends at most a distance $h$ below $q_P$ in direction $d$, (ii) extends exactly a distance $h$ above $q_P$, and (iii) satisfies
\begin{equation}
\label{eqn:star-bowl-bigstar}
    P\subset Q(P)\subset q_P+\alpha(P-q_P)
\end{equation}
for a universal constant $\alpha$; see \cref{lem:star-to-bowl} for the detailed proof. By \cref{eqn:star-bowl-bigstar}, packing the bowls instead of the stars costs at most a factor $\alpha^2$, and any packing of bowls contains a valid packing of the stars.
This is where star-shapedness of our pieces is crucial:
star-shapedness ensures that scaling with $q_P$ fixed results in a superset of the original polygon $P$.

Stacks of bowls are (informally) defined as follows: a sequence of bowls $Q_1,\ldots,Q_j$ forms a stack if the reference point of each $Q_i$ lies in the triangle spanned by the arms of $Q_{i-1}$. \cref{sec:star-bowls} gives the formal definition, phrased via \emph{separators}.

\paragraph{Stacking bowls is (almost) optimal.} It turns out that packing bowls arbitrarily and packing bowls into stacks are essentially the same.

\begin{theorem}
\label{thm:stacking-overview}
Let $\mathcal P$ be a set of star-shaped polygons of one fine type, and let $\Pi$ be a packing of $\mathcal P$ in a convex container $C$. The bowls of the stars in $\mathcal{P}$ can be packed into strict stacks in mini-containers with total area $O(|C|)$.
\end{theorem}

\begin{figure}[!h]
\centering
\includegraphics[page=1]{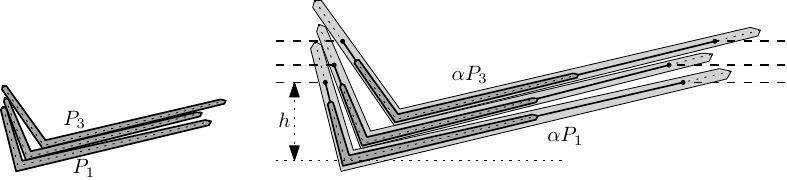}
\caption{Left: A part $\mathcal P_j$ with the points $q_i,R_i,L_i$ marked. Right: The scaled stars and the packing of the corresponding bowls, which form a stack.}
\label{fig:induced-bowl-packing}
\end{figure}

We sketch the proof. 
The details are in \cref{sec:local-bowl-stacks}; see in particular \cref{lem:localized-stars-to-bowl-stacks}.
We first replace the stars of a part by their bowls:
Scale the packing of $\mathcal P_j$ by the factor $\alpha$ and place each bowl $Q(P)$ at the scaled copy of $q_P$; see \cref{fig:induced-bowl-packing}.
Each bowl will lie inside its scaled star by \cref{eqn:star-bowl-bigstar}, so the bowls do not overlap and fit into the scaled copy of $B_j$.
Now take two bowls $Q_1,Q_2$ of the part with $q_2$ higher than $q_1$ in direction $d$. We will show that, if $q_2$ were outside the triangle spanned by the arms of $Q_1$, an arm of $Q_2$ would cross an arm of $Q_1$, which is impossible. Hence the bowls of $\mathcal P_j$, ordered along $d$, form a strict stack.
The scaled rectangles $\alpha B_j$, of dimensions $O(r)\times O(w)$, are therefore mini-containers holding strict stacks, and by \cref{thm:star-structural-decomposition} their total area is $\alpha^2\sum_j|B_j|=O(|C|)$.

\begin{figure}[!h]
\centering
\includegraphics[page=9]{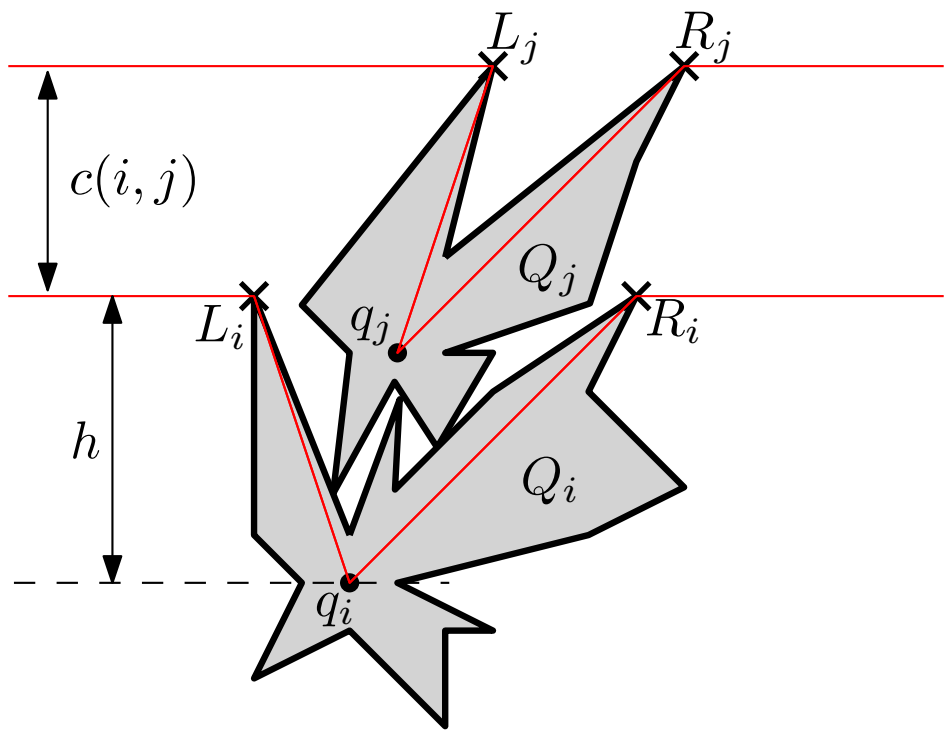}
\caption{Two stacked bowls and the height $c(i,j)$ needed to place $Q_j$ above $Q_i$. The cost $c$ itself is not a metric, but $c'(i,j)=\min(c(i,j),h)$ is.}
\label{fig:bowl-metric}
\end{figure}

\paragraph{Computing approximately optimal stacks.} Stacking bowls is, algorithmically, very easy; see \cref{sec:algorithm} for the details. Fix a fine type, and for two bowls $Q_i,Q_j$ let $c(i,j)$ be the smallest amount by which $Q_j$ has to be placed higher than $Q_i$, in direction $d$, so that $Q_i,Q_j$ form a stack; see \cref{fig:bowl-metric}. Translations orthogonal to $d$ are free, so $c(i,j)$ is a single number, and $c(i,j)\le 2h$. The height of a stack with a fixed order is the sum of $c(i,j)$ over consecutive bowls, and we need to choose the order minimizing it. This is a asymmetric metric TSP instance for which a constant-factor approximation~\cite{DBLP:journals/jacm/SvenssonTV20,TraubVygen2024} gives a stack of height $H$ within a constant factor of the minimum; see \cref{lem:bowl-TSP}.

Finally, the tall stack can be split into short stacks.
We therefore cut the stack every $\Theta(h)$ along $d$ and enclose each piece in a rectangle of dimensions $O(r)\times O(w)$; see \cref{lem:chunk-path}. These rectangles are the mini-containers. Their total area will be $O(rH)$. Moreover, each mini-container has spans within a constant factor of the spans of a star inside it.

\paragraph{Algorithm summary.}
All the ingredients are now in place for the $O(\log^2 n)$-approximation.
The algorithm sets aside the small stars, splits the remaining polygons into their fine types, turns them into bowls, stacks the bowls, and cuts the stacks into mini-containers.
Then \cref{thm:convex-packing} packs the mini-containers together with the convex hulls of the small stars into a rectangle. See \cref{sec:translational-final} for the details.

In the case of strip and bin packing, we get \cref{cor:bin-strip} by slicing this rectangle along vertical and horizontal lines.
We then rearrange the slices, each with its mini-containers, in the strip or the bins. See \cref{app:cor:bin-strip} for the details.

\paragraph{\boldmath Shaving a $\log$-factor.}
The factor $O(\log^2 n)$ comes from packing the $O(\log^2 n)$ size groups independently of each other. Some loss here is unavoidable, as we will show at the end of this paragraph, but we can show that the total loss is actually only $O(\log n)$. Recall that $\mathcal G_{a,b}$ is the size group with scales $r=r_{\max}/2^a$ and $w=w_{\max}/2^b$.
For a fixed $k$, the stars in the groups with $a+b=k$ have comparable convex hull areas, $\Theta(r_{\max}w_{\max}/2^k)$, but very different shapes, ranging from long and thin ($a$ small) to short and fat ($a$ large).
A charging argument exploits both facts to show that for every $k$,
\[
\sum_{a+b=k}\opt(\mathcal G_{a,b})=O(\opt).
\]
Summing over the $O(\log n)$ values of $k$ gives $\sum_{a,b}\opt(\mathcal G_{a,b})=O(\log n)\cdot\opt$, which is exactly the bound we need on the total area of the mini-containers.

\begin{figure}[!h]
\centering
\includegraphics[page=5]{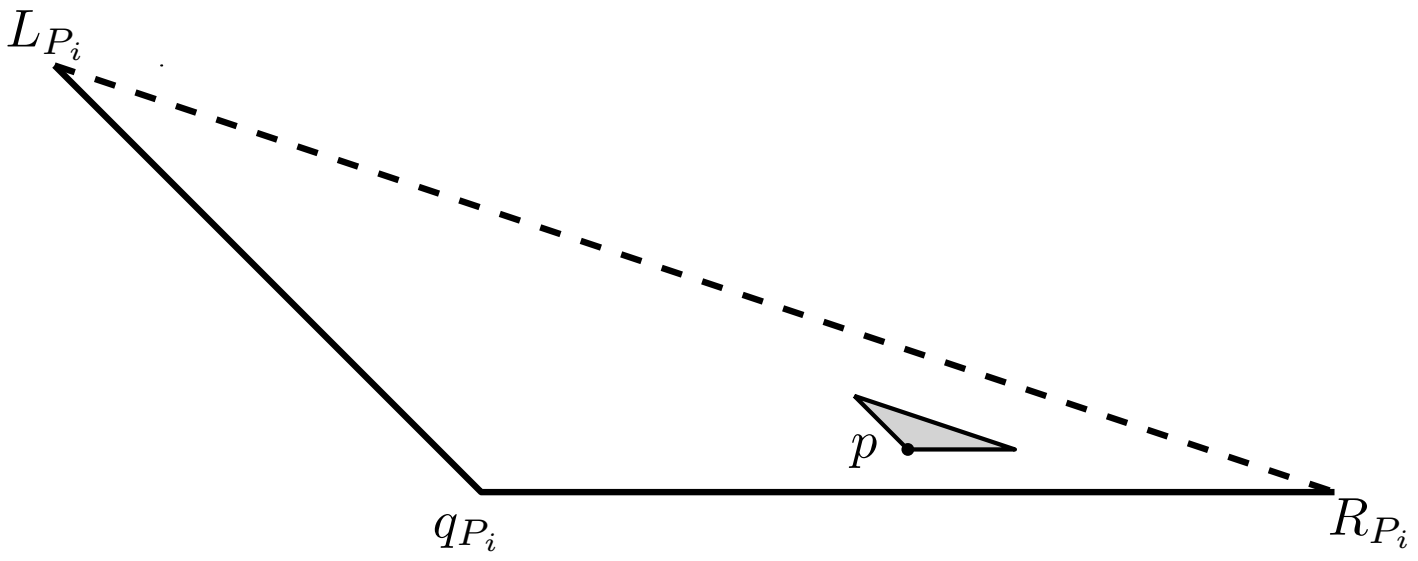}
\caption{The core of a star $P_i$ is the gray triangle.}
\label{fig:disjoint-cores}
\end{figure}

The charging argument works as follows.
Give every star $P$ a \emph{core}, which is a small copy of the triangle $q_PR_PL_P$ sitting above the middle of the radius; see \cref{fig:disjoint-cores}. A core has area $\Theta(r_Pw_P)$, so for $a+b=k$ all cores have the same area, and we can charge the area of a packing to the cores it contains. Two things make this work.
Cores of stars from different groups with $a+b=k$ cannot overlap: a short and fat star whose core covers a point of a long and thin star's core would have to cross the thin star's radius; see \cref{lem:core-disjoint} for the details.
And within one group, all stars whose cores overlap the core of $P$ fit into one rectangle of dimensions $O(r)\times O(w)$, because their radii cannot cross that of $P$; see \cref{lem:core-rectangle} for the details.
So, a packing of all groups with $a+b=k$ contains as many disjoint cores as the groups do separately, and each group can be packed into a constant times as many rectangles as it has disjoint cores.
The details are included in \cref{sec:star-local-grouping}.

\begin{figure}[!h]
\centering
\includegraphics[scale=0.5,page=6]{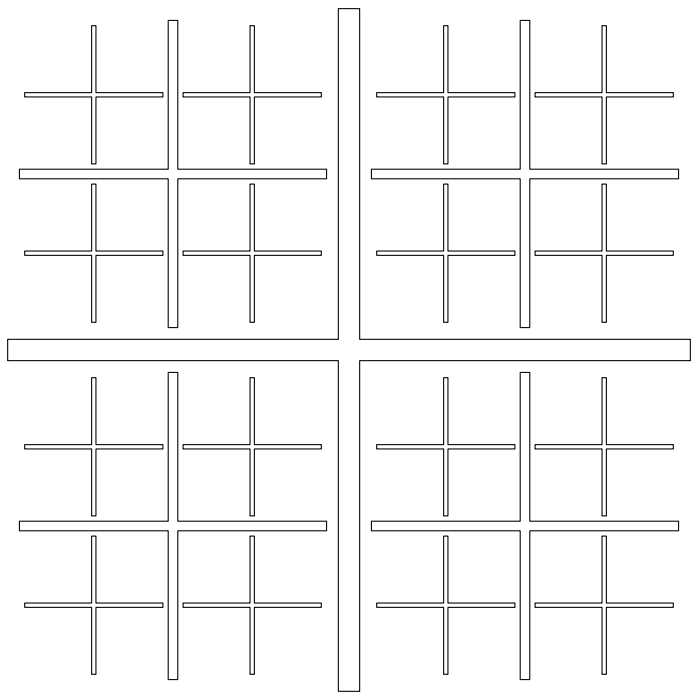}
\caption{Pluses of $\Theta(\log n)$ different scales nested into a region of constant area. Packing each scale into its own convex mini-containers requires area $\Omega(1)$ per scale.}
\label{fig:lowerbound}
\end{figure}

The $\log n$ in the approximation factor is inherent to our approach. Consider, for $k=0,\ldots,K$, $4^k$ copies of a plus, scaled by $2^{-k}$; see \cref{fig:lowerbound}.
For a single scale we saw earlier that pluses behave like squares, and packing their convex hulls is fine. Across scales, however, the pluses of scale $2^{-k}$ nest into the gaps between the pluses of scale $2^{-k+1}$, so all $\Theta(4^K)$ pluses fit into a container of area $O(1)$, whereas each scale on its own needs area $\Omega(1)$. Any algorithm that packs size groups into separate convex mini-containers thus loses $\Omega(\log n)$ on this instance, and any mini-container based $o(\log n)$-approximation has to mix size groups within a mini-container.

\begin{figure}[t]
\centering
\includegraphics[page=10]{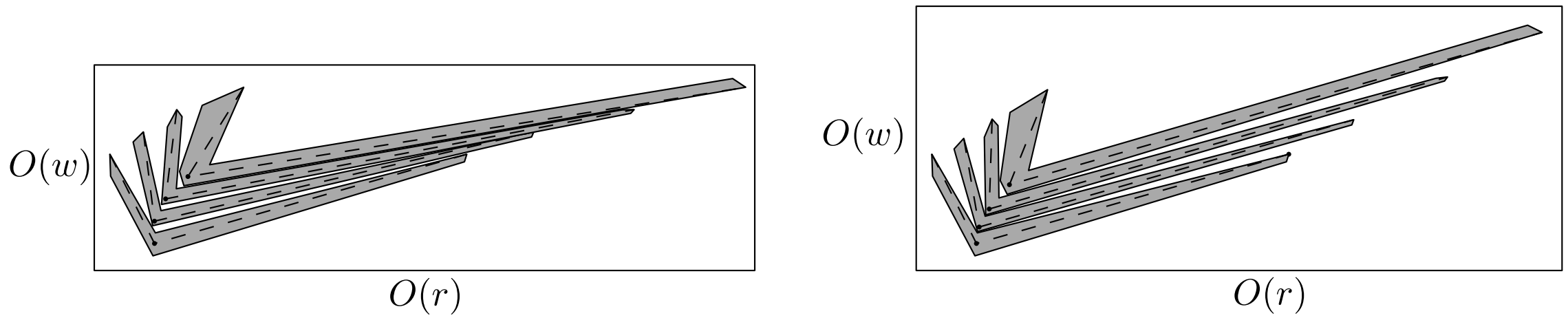}
\caption{Left: A stack of one fine type in its rectangle. Right: Rotating all radii to the same orientation only scales the rectangle by a constant factor.}
\label{fig:rotation}
\end{figure}

\paragraph{Rotations.} In the rotational variant the algorithm chooses the orientation of each star, so fine types are no longer a property of the input. We get around this with a preprocessing step, after which we simply apply our translation-only algorithm. For a size group $\mathcal{G}$, let $\widehat{\mathcal{G}}$ be obtained by rotating every star so that its radius is horizontal. We show in \cref{cor:rotation-algorithm-interface} that
\[
\opt_{\mathrm{trans}}(\widehat{\mathcal{G}})=\Theta(\opt_{\mathrm{rot}}(\mathcal{G})),
\]
where $\opt_{\mathrm{trans}}$ and $\opt_{\mathrm{rot}}$ are the optimum values under translation, and under translation \emph{and} rotation respectively.
This allows us to simply run the translational algorithm on $\widehat{\mathcal{G}}$; see \cref{fig:rotation}. The lower bound is trivial, since a translational packing of $\widehat{\mathcal{G}}$ is a rotational packing of $\mathcal{G}$. For the upper bound, take an optimal rotational packing of $\mathcal{G}$, freeze the chosen rotations, and apply \cref{thm:star-structural-decomposition,thm:stacking-overview} to the rotated stars (\cref{lem:rot-initial-containers}). This gives stacks of bowls in rectangles of dimensions $O(r)\times O(w)$ with total area $O(\opt_{\mathrm{rot}}(\mathcal{G}))$, in which the radii are already within an angle of $O(w/r)$ of each other. It remains to make them exactly parallel without blowing up the rectangles. We show that, with $\theta_i$ the direction of the $i$th radius in stack order,
\[
\sum_i|\theta_{i+1}-\theta_i|=O(w/r),
\]
because the arms of consecutive bowls cannot cross, which lets us charge each change of direction to the height of the stack, which is $O(w)$; see \cref{lem:rot-variation}. We then straighten the stack bowl by bowl, each time rotating the whole suffix above the current bowl so that the next radius becomes parallel, and shifting it to restore the stack (\cref{lem:rot-sweep}). The total displacement is $O(r\sum_i|\theta_{i+1}-\theta_i|)=O(w)$ across the stack and $O(r)$ along it, so the rectangles only grow by a constant factor; see \cref{lem:rotation-straightening}.

\paragraph{APX-hardness.} We reduce from MAX-VERTEX-DISJOINT-PATH-COVER. 
Here we are given a directed graph $G=(V,E)$ on $V=\{1,\ldots,n\}$ with in- and out-degree at most $2$, and the goal is to find a set of vertex-disjoint paths covering all vertices and using as many edges as possible. Equivalently, we want an ordering $u_1,\ldots,u_n$ of the vertices maximizing the number of consecutive pairs $(u_i,u_{i+1})$ that are edges. Bosboom, Demaine, Demaine, Hesterberg, Manurangsi and Yodpinyanee~\cite{DBLP:journals/jip/BosboomDDHMY17} showed that it is NP-hard to distinguish graphs with a Hamiltonian path, that is, a path cover with $|V|-1$ edges, from graphs where every path cover has fewer than $(1-\eps)(|V|-1)$ edges, for $\eps<\frac1{1396640}$ (restated as \cref{lem:apx:disjpathcover}).

\begin{figure}[t]
\centering
\includegraphics[page=7]{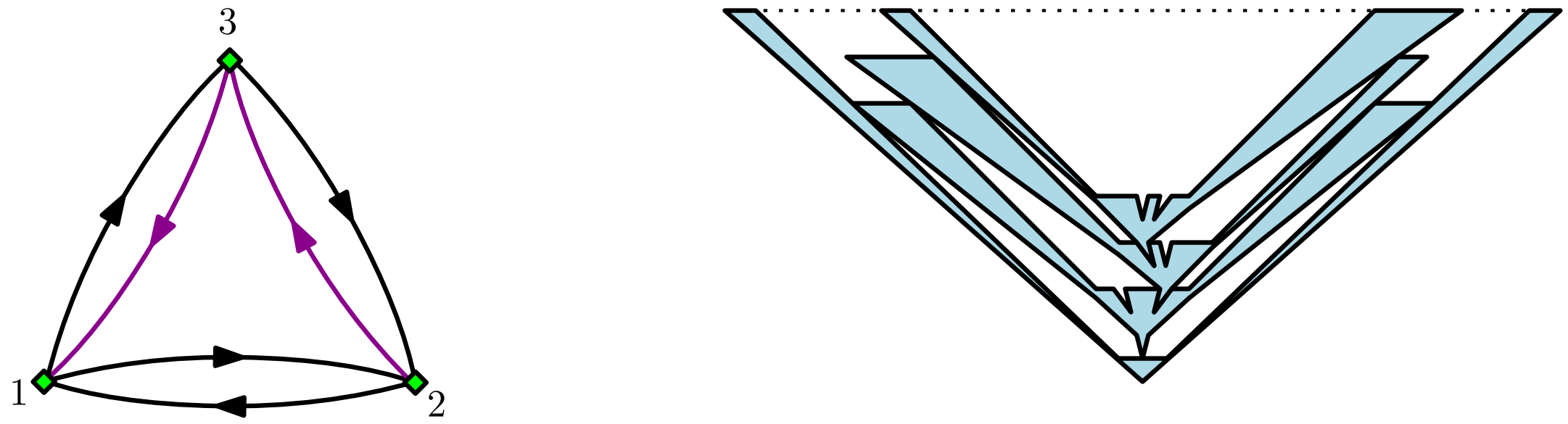}
\caption{Left: A graph $G$ with in- and out-degree at most $2$ and a Hamiltonian path. Right: The corresponding stack of the pieces $S_i$ inside the convex hull of the container piece $C$.}
\label{fig:apx:goodstack:overview}
\end{figure}

We create a star $S_i$ for every vertex $i$, designed so that any near-optimal packing is a stack; see \cref{fig:apx:goodstack:overview}. Two long arms give $S_i$ the overall shape of a wedge, which makes the stars stackable. Below the arms is a spike whose angle encodes $i$, and between the arms are up to two slits whose angles encode the out-neighbours of $i$. The spike of $S_j$ fits into a slit of $S_i$ if and only if $(i,j)\in E$. If it does not fit, stacking $S_j$ on top of $S_i$ costs about the height of the spike extra (\cref{lem:apx:soundness}). A single \emph{container piece} $C$ with even longer arms forces all $S_i$ to form a stack between its arms in any near-optimal packing (\cref{lem:apx:similarangles,lem:apx:sinsideC,lem:apx:formsstack}). If this stack is a Hamiltonian path, that is, every spike sits in a slit of the star directly below, then the stack fits inside the convex hull of $C$ (\cref{lem:apx:completeness}). Otherwise it sticks out above $C$ by an amount proportional to the number of unmatched pairs. We show, for some constant $\delta>0$, that it is NP-hard to distinguish instances where everything fits inside the convex hull of $C$ from instances that need a $(1+\delta)$ factor more area; see \cref{thm:apx:main}.

\section{Algorithm for translational packing}
\label{sec:star-packing}

In this section, we present the details of our $O(\log n)$ approximation for packing stars translationally into a minimum-area convex container.
\begin{itemize}
\item In \cref{sec:star-types}, we define some parameters of stars and show how to partition them into size groups and fine types.
\item In \cref{sec:star-local-decomposition}, we prove a structural theorem for the stars in one size group, which splits a size group into fine types without increasing the cost by more than a constant factor.
\item In \cref{sec:star-bowls}, we show how to transform the original stars into bowls.
\item In \cref{sec:local-bowl-stacks}, we prove that packing bowls of each fine type into a stack would only increase the optimal cost by at most a constant factor.
\item In \cref{sec:algorithm}, we show how to approximately compute the optimal stacks.
\item In \cref{sec:star-local-grouping}, we show that packing each size group separately would only introduce an $O(\log n)$ factor.
\item In \cref{sec:translational-final}, we put everything together to prove our main theorem for translational packing.
\end{itemize}

Throughout this section, all placements use translations only. Changes of coordinates only serve for simplicity of analysis and do not really rotate the stars.
We write $\opt(\mathcal P)$ for the infimum area of a convex container admitting a packing of $\mathcal P$, and $\opt'(\mathcal P)$ for the infimum total area of one or more convex containers admitting a packing of $\mathcal P$. By definition, $\opt'(\mathcal P)\le\opt(\mathcal P)$.

\subsection{Parameters, size groups, and fine types}
\label{sec:star-types}

\begin{figure}[!h]
\centering
\includegraphics[page=2]{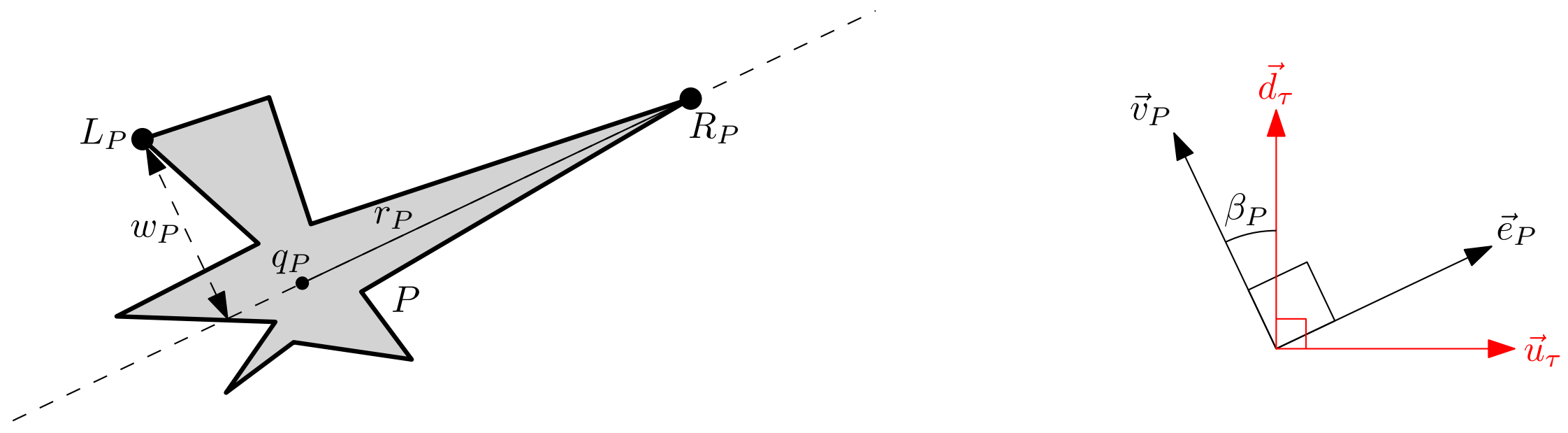}
\caption{Left: The reference point, radius, and transverse width of a star $P$.
Right: The black vectors are the direction of the radius and the perpendicular for the star $P$ from the left.
The red vectors are the vectors defining the fine type of $P$.}
\label{fig:star-shape-parameters}
\end{figure}

For a star-shaped polygon $P$, recall that the \emph{kernel} of $P$ is the set of points in $P$ from which all points in $P$ are visible.
It is well-known that the kernel is convex.
We choose an arbitrary reference point $q_P$ deterministically in the kernel, for instance, the lexicographically maximal kernel vertex. 
Let $R_P$ be an arbitrary farthest point in $P$ from $q_P$, and let $r_P$ be the \emph{radius length}, i.e., the distance between $q_P$ and $R_P$:
\[
R_P\in\argmax_{z\in P}\|zq_P\|, \qquad r_P=\|R_Pq_P\|.
\]
Let $\Vec{e}_P = (R_P - q_P) / r_P$ be the unit vector along the \emph{orientation} of $P$, i.e., the direction of the radius.
Let $L_P$ be an arbitrary point in $P$ that maximizes the distance from the radius line $\overline{q_{P}R_{P}}$, and let \[w_P=\dist(L_P, \overline{q_{P}R_{P}})\] be the \emph{width} of $P$.
Let $\Vec{v}_P$ be the unit vector perpendicular to $\Vec{e}_P$ pointing toward $L_P$.
See \cref{fig:star-shape-parameters} for an illustration.

Note that $w_P \le r_P$ and $\diam(P) \le 2r_P$.
Without loss of generality, we assume that $L_P$ lies to the left of the directed radius $\overrightarrow{q_P R_P}$.
Let $T_P$ denote the triangle $\triangle q_PL_PR_P$.

\paragraph{Size groups.}
Let $r_{\max}=\underset{P}{\max}~r_P$ and $w_{\max}=\underset{P}{\max}~w_P$.
We partition the stars into size groups $\mathcal G_{a,b}$ for integers $a,b \ge 0$ according to their radius lengths and widths.
Each $\mathcal G_{a,b}$ consists of stars with $r_P\in(r/2,r]$ and $w_P\in(w/2,w]$, where $r=r_{\max}/2^a$ and $w=w_{\max}/2^b$.
We will pack each group separately.

\begin{lemma}
\label{lem:rmax-wmax}
For a set of stars $\mathcal P$, $\opt(\mathcal P) = \Omega(r_{\max} w_{\max})$.
\end{lemma}

\begin{proof}
The convex hull of a star of width $w_P$ contains a disk of radius $\Omega(w_P)$. So in any packing, the container must have spans of $\Omega(w_{\max})$ in all directions and a span of $r_{\max}$ in one direction.
\end{proof}

Without loss of generality, we only consider the groups where $a,b\le \log_2 n$. All other stars can be packed individually, since the total area of them is $O(r_{\max} w_{\max})$.

\paragraph{Fine types.}
From now on, we only consider one size group $\mathcal G_{a,b}$ with scales $r=r_{\max}/2^a$ and $w=w_{\max}/2^b$.
We are going to further partition the stars into fine types according to their orientations.

Define $N=100\lceil r/w\rceil$ as the number of fine types in the size group and 
\begin{equation}
\label{eq:delta-bounds}
\delta = \frac{2\pi}{N}
\le \frac{\pi}{50}
\end{equation}
as the angle range of each fine type.
Let $D=\{(\cos k\delta,\sin k\delta):0\le k<N\}$ be the set of $N$ equally distributed unit vectors.

We create a fine type $\tau$ for each vector $\Vec{d}_\tau\in D$, and let $\Vec{u}_\tau$ be the unit vector obtained by rotating $\Vec{d}_\tau$ by $90$ degrees clockwise.
For each star $P$, we associate it with the unique fine type $\tau$ such that the clockwise angular displacement $\beta_P$ from $\Vec{v}_P$ toward $\Vec{d}_\tau$ is in $[\delta,2\delta)$. 
In other words,
\begin{equation*}
\Vec{d}_\tau = \sin\beta_P \Vec{e}_P + \cos\beta_P \Vec{v}_P,
\qquad
\Vec{u}_\tau = \cos\beta_P \Vec{e}_P - \sin\beta_P \Vec{v}_P,
\qquad
\beta_P \in [\delta,2\delta).
\end{equation*}
See \cref{fig:star-shape-parameters} (right) for an illustration.

The orientations of the stars of each fine type form an angular interval of length $\delta=\Theta(w/r)$.
Consequently, an interval shorter than $\delta$ intersects with at most two fine types.

\subsection{A structural theorem for fine types}
\label{sec:star-local-decomposition}

We start with an elementary enclosure lemma.

\begin{lemma}
\label{lem:convex-to-vertical-parallelogram}
Every compact convex set $C$ of positive area is contained in a vertical
parallelogram (that is, a parallelogram with two vertical sides) $T$ with $|T|\le2|C|$.
\end{lemma}

\begin{proof}
Choose a leftmost point $p$ and a rightmost point $q$ of $C$.
Apply a vertical shear making $pq$ horizontal, and let $T'$ be the
axis-aligned bounding rectangle of the sheared set. If its width and
height are $W$ and $H$ respectively, the triangles with common base $pq$ and apices at a
highest and a lowest point have disjoint interiors and total area
$WH/2$. They lie in the sheared convex set. The inverse shear maps
$T'$ to the required parallelogram and preserves area. 
\end{proof}

The next theorem captures the main structural insight for the original stars.
This is a stronger version of \cref{thm:star-structural-decomposition} in the sense that it records more local information about the reference points.
This is needed later when we transform the stars into bowls.

\begin{theorem}
\label{thm:star-local-decomposition-strong}
Let $\mathcal P$ be a set of stars in one size group with scales $r,w$, and let $\Pi$ be a packing of $\mathcal P$ in a convex container $C$.
For any $0 < c < 0.1$ and $0<\varepsilon< c/100$, there exists a partition
\[
\mathcal P = \mathcal P_1\mathbin{\dot\cup} \cdots \mathbin{\dot\cup}\mathcal P_t, 
\]
such that each part $\mathcal P_j$ only contains stars of one fine type $\tau_j$ and we have the following properties (under the placement inherited from $\Pi$):
\begin{enumerate}[label=(\roman*),nosep]
\item In each $\mathcal{P}_j$, the reference points span at most $10\varepsilon r$ along $\Vec{u}_{\tau_j}$ and $10\varepsilon w$ along $\Vec{d}_{\tau_j}$.

\item All stars in $\mathcal P_j$ lie in a rectangle $B_j$ of dimensions $O(r)\times O(w)$, whose sides are parallel to $\Vec{u}_{\tau_j}$ and $\Vec{d}_{\tau_j}$ respectively.
Moreover,
\begin{equation*}
\sum_{j=1}^t |B_j| = O_{\varepsilon}(|C|).
\end{equation*}
\end{enumerate}
\end{theorem}

\begin{proof}
We first divide the stars into $O(1)$ groups such that in each group, the orientations of the stars span at most $\pi/100$.
Consider one such group. Choose coordinates such that the clockwise angular displacement from each orientation to the positive $x$-axis is in $[0,\pi/100)$.
Let $T$ be the enclosing vertical parallelogram of $C$ from \cref{lem:convex-to-vertical-parallelogram}.
Let $R$ be the horizontal span of $T$ and $W$ be the vertical side length of $T$. Then $RW=|T|\le 2|C|$.
Moreover, we have $R=\Omega(r)$ and $W=\Omega(w)$.

Then, we partition $T$ into $O(R/(\eps r))$ vertical strips of width $\eps r$, and consider each strip separately.
For all stars whose reference points lie in a strip $x\in[X_k,X_k+\varepsilon r)$, since $0<c<0.1$ and $0<\eps<c/100$, their radii must cross the two vertical lines
\[
\ell_{k,1}:x=x_k+cr,\qquad \ell_{k,2}:x=x_k+2cr.
\]
Now we partition each of $\ell_{k,1}\cap T$ and $\ell_{k,2}\cap T$ into $m=O(W/(\eps w))$ intervals of lengths at most $\varepsilon w$.
We say a pair $(a, b)$ is occupied if there exists a star whose radius crosses the $a$-th interval on $\ell_{k,1}$ and the $b$-th interval on $\ell_{k,2}$, ordered from bottom to top.
Notice that if $a<a'$ and $b>b'$, then $(a,b)$ and $(a',b')$ cannot be both occupied, since otherwise the radii must intersect. So, there are at most $2m=O(W/(\eps w))$ occupied pairs in each vertical strip.
See \Cref{fig:stabbing-segment2} for an illustration.

\begin{figure}
\centering
\includegraphics[page=3]{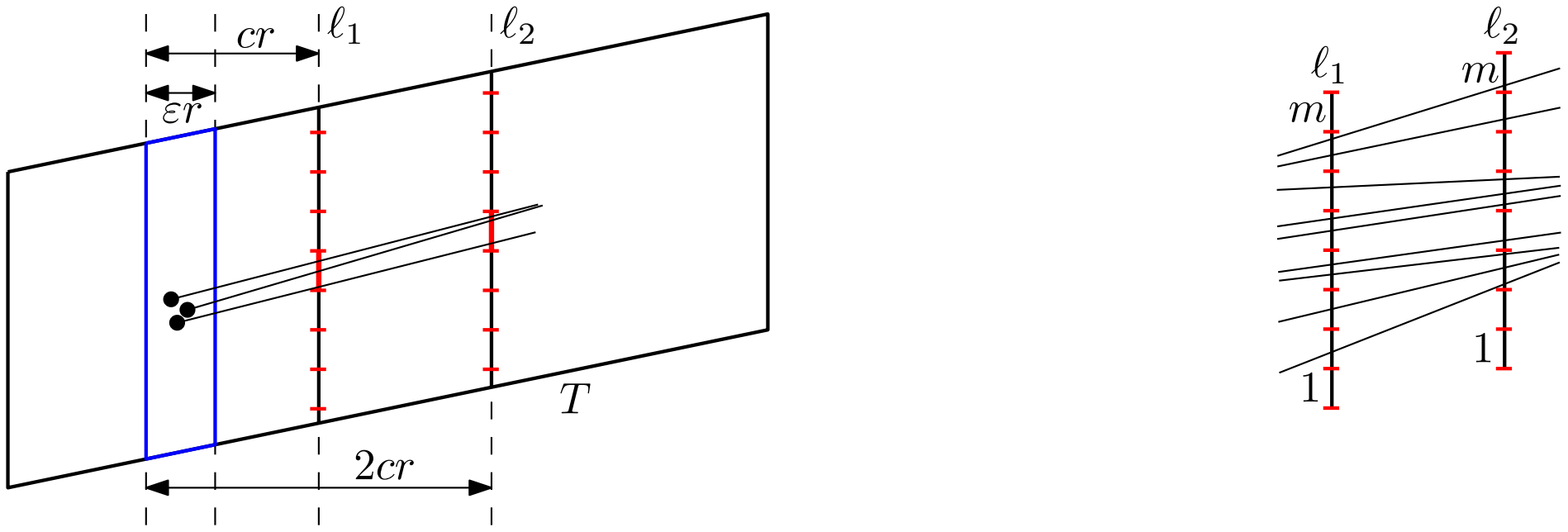}
\caption{Left: The enclosing parallelogram is cut into vertical strips of width $\varepsilon r$; one strip is shown. Two test lines $\ell_1,\ell_2$ are placed to its right. The reference points and radii assigned to one pair $(a,b)$ of intervals are shown.
Right: The occupied interval pairs form a monotone sequence in $[m]^2$.
}
\label{fig:stabbing-segment2}
\end{figure}

Consider each occupied pair $(a,b)$. Note that the slopes of the radii of the stars corresponding to $(a,b)$ differ by at most $2\eps w/(cr)$. Since $\eps<c/100$, these stars belong to at most $2$ fine types.
Therefore, we can partition all stars in $\mathcal{P}$ into $t=O(RW/(\eps^2 rw))$ subsets $\mathcal{P}_j$ such that the stars in each $\mathcal{P}_j$ correspond to the same occupied pair in the same vertical strip and share the same fine type $\tau_j$.

Consider each $\mathcal P_j$.
Consider any two stars $P_1,P_2\in\mathcal P_j$. Write $q_{P_1}=(x_1,y_1)$ and $q_{P_2}=(x_2,y_2)$. Since $P_1$ and $P_2$ are in the same vertical strip,
\begin{equation}
\label{eqn:direction-x}
|x_1-x_2|\le \eps r.
\end{equation}
Let $p=(x_p,y_p)$ be the point on the line $\overline{q_{P_1}R_{P_1}}$ such that $y_p=y_2$. Then
\begin{equation}
\label{eqn:direction-vP1}
|(q_{P_2}-q_{P_1})\cdot\vec{v}_{P_1}| = \dist(q_{P_2},\overline{q_{P_1}R_{P_1}}) \le \|pq_{P_2}\| \le 3\eps w
\end{equation}
since $P_1$ and $P_2$ correspond to the same occupied pair and $p$ and $q_{P_2}$ are at most $cr$ far from $\ell_{\cdot,1}$.
Recall that the angle between $\vec{e}_{P_1}$ and the positive $x$-axis is at most $\pi/100$. So, by \cref{eqn:direction-x,eqn:direction-vP1}, we have
\begin{equation}
\label{eqn:direction-eP1}
|(q_{P_2}-q_{P_1})\cdot\vec{e}_{P_1}| \le 2\eps r.
\end{equation}
Recall that the angle from $\vec{e}_{P_1}$ to $\vec{u}_{\tau_j}$ is $\beta_{P_1} < 2\delta < (\pi/25)w/r$. So, by \cref{eqn:direction-vP1,eqn:direction-eP1}, we have
\[
|(q_{P_2}-q_{P_1})\cdot\vec{u}_{\tau_j}| \le 3\eps r, \qquad
|(q_{P_2}-q_{P_1})\cdot\vec{d}_{\tau_j}| \le 4\eps w,
\]
which proves (i).
Combining (i) with that all stars in $\mathcal P_j$ share the same fine type $\tau_j$, it is straightforward to see that the dimensions of the bounding rectangle with sides parallel to $\vec{u}_{\tau_j}$ and $\vec{d}_{\tau_j}$ are $O(r)\times O(w)$.
The total area is $O(trw)=O(RW/\eps^2)=O_\eps(|C|)$.
\end{proof}

\subsection{Stacks of bowls}
\label{sec:star-bowls}

We are now going to pack each fine type separately. We focus on one fine type $\tau$ with orthogonal unit vectors $\Vec{d}_\tau$ and $\Vec{u}_\tau$.
The first step is to transform each star into a \emph{bowl} which has useful properties.

\paragraph{Bowls.} Let $\Vec{d},\Vec{u}$ be orthogonal unit vectors.
We say that a (possibly weakly simple) star-shaped polygon $Q$ is a \emph{bowl} aligned along $(\Vec{d},\Vec{u})$ with height $h>0$ if it has a specified kernel point $q_Q$ and specified points $L_Q,R_Q$ such that
\[
\Vec{d} \cdot (L_Q - q_Q) = \Vec{d} \cdot(R_Q - q_Q) = h,
\qquad \Vec{u} \cdot L_Q < \Vec{u} \cdot R_Q,
\qquad \forall z \in Q, -h \le \Vec{d} \cdot(z - q_Q)\le h.
\]
We call the two segments $q_Q L_Q$ and $q_Q R_Q$ the two finite arms of $Q$.
Note that a bowl is also a star; since $q_Q,L_Q,R_Q$ also follow their original definition, e.g., $R_Q$ is one of the farthest points from $q_Q$.

\begin{figure}[!h]
\centering
\includegraphics[page=4]{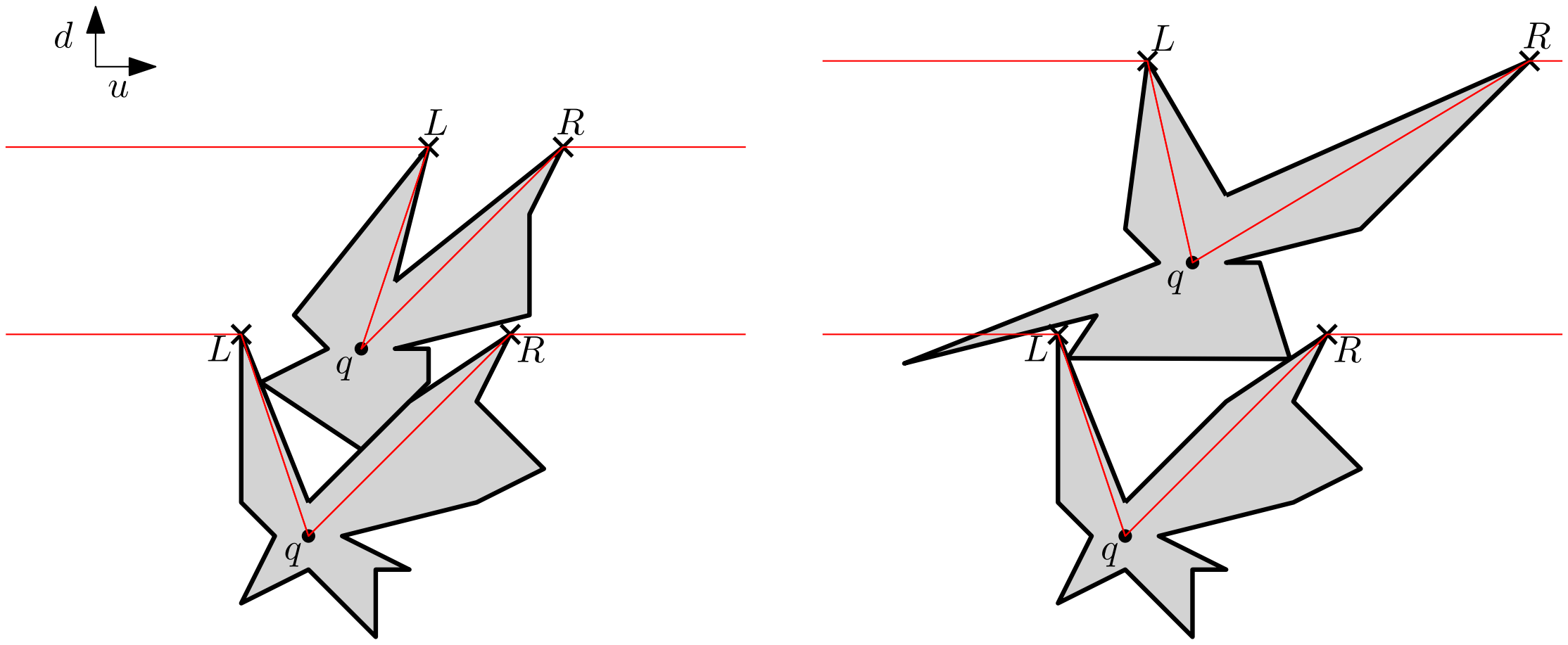}
\caption{Bowls and their separators. The finite arms belong to the
bowls; the two tails extend indefinitely perpendicular to the alignment
direction. Disjoint separators define the stack order. The left shows a \emph{strict stack}, and the right shows a non-strict stack. 
}
\label{fig:separator}
\end{figure}

\paragraph{Separators and stacks.} Given such a bowl $Q$, we define its \emph{separator} to be the union of these two arms and the two outward rays perpendicular to $\Vec{d}$:
\[
\sep(Q)=\{L_Q - t\Vec{u}: t \ge 0\} \cup L_Q q_Q \cup q_Q R_Q \cup \{R_Q + t\Vec{u} : t \ge 0\}.
\]
We say that a packing of bowls with the same directions $\Vec{d},\Vec{u}$ and height $h$ forms a \emph{stack} if their separators are pairwise disjoint.
We say that the packing forms a \emph{strict stack} if, in addition, each separator does not intersect with any other bowl.
See \cref{fig:separator} for an illustration.

Without loss of generality, we assume that $\Vec{d}$ is upward and $\Vec{u}$ is right. 
Then a separator is a simple curve with two horizontal ray tails at the same height, which divides the plane into an upper region and a lower region. So, in a stack, we can order these disjoint separators (and hence the bowls) vertically. 

\begin{lemma}
\label{lem:star-to-bowl}
For every star $P$ of fine type $\tau$, there is a bowl $Q(P)=Q_{\tau}(P)$ with directions $\Vec{d}_\tau,\Vec{u}_\tau$ and height $h = 100w$, such that
\begin{equation}
\label{eq:star-bowl-containment}
P\subseteq Q_{\tau}(P)\subseteq q_P+\alpha(P-q_P)
\end{equation}
for some universal constant $\alpha>1$.
Moreover, 
\begin{itemize}
\item $q_{Q_{\tau}(P)} = q_P$;
\item $R_{Q_{\tau}(P)}$ lies on the ray $\overrightarrow{q_P R_P}$, $L_{Q_{\tau}(P)}$ lies on the ray $\overrightarrow{q_P L_P}$; 
\item In coordinates with axes $\{\Vec{u}_\tau,\Vec{d}_\tau\}$ and origin $q_Q$, we have
\begin{equation}
\label{eq:bowl-arm-bounds}
R_Q=(h a_Q,h),\qquad L_Q=(h b_Q,h),
\qquad
7\frac rw\le a_Q\le32\frac rw,
\quad -4\frac rw\le b_Q\le3\frac rw.
\end{equation}
\end{itemize}
\end{lemma}

\begin{proof}
Consider a star $P$ of fine type $\tau$.
Note that $L_P-q_P=x\Vec{e}_P + w_P\Vec{v}_P$ for some $|x|\le r$.
By \cref{eq:delta-bounds} and $\beta_P\in[\delta, 2\delta)$, $w_P\in(w/2,w]$, we have
\begin{equation}
\label{eqn:bowl-1}
\Vec{d}_\tau\cdot(R_P-q_P) = r_P\sin\beta_P \in (0.01w, 0.13w)
\end{equation}
and
\begin{equation}
\label{eqn:bowl-2}
\Vec{d}_\tau\cdot(L_P-q_P) = x\sin\beta_P + w_P\cos\beta_P \in (0.49w, 1.13w). 
\end{equation}
For every $p\in P$, write $p - q_P = s\Vec{e}_P + t\Vec{v}_P$, where $|s|\le r$ and $|t|\le w$. Similarly, $|\Vec{d}_\tau \cdot (p-q_P)| \le 1.13w$. 

Let $h=100w$.
We extend the rays $q_PR_P$ and $q_PL_P$ to points $R_Q$ and $L_Q$ such that
\[
\Vec{d}_\tau \cdot (R_Q - q_P) = \Vec{d}_\tau \cdot (L_Q - q_P) = h ,
\]
and let the bowl $Q = Q_{\tau}(P)$ be $P \cup qR_Q \cup qL_Q$.
See Figure~\ref{fig:bowl} for an illustration.
By \cref{eqn:bowl-1,eqn:bowl-2}, both extensions are bounded by a universal constant factor.
It is straightforward to check that $R_Q$ and $L_Q$ match their definitions.

\begin{figure}[!h]
    \centering
    \includegraphics[page=11]{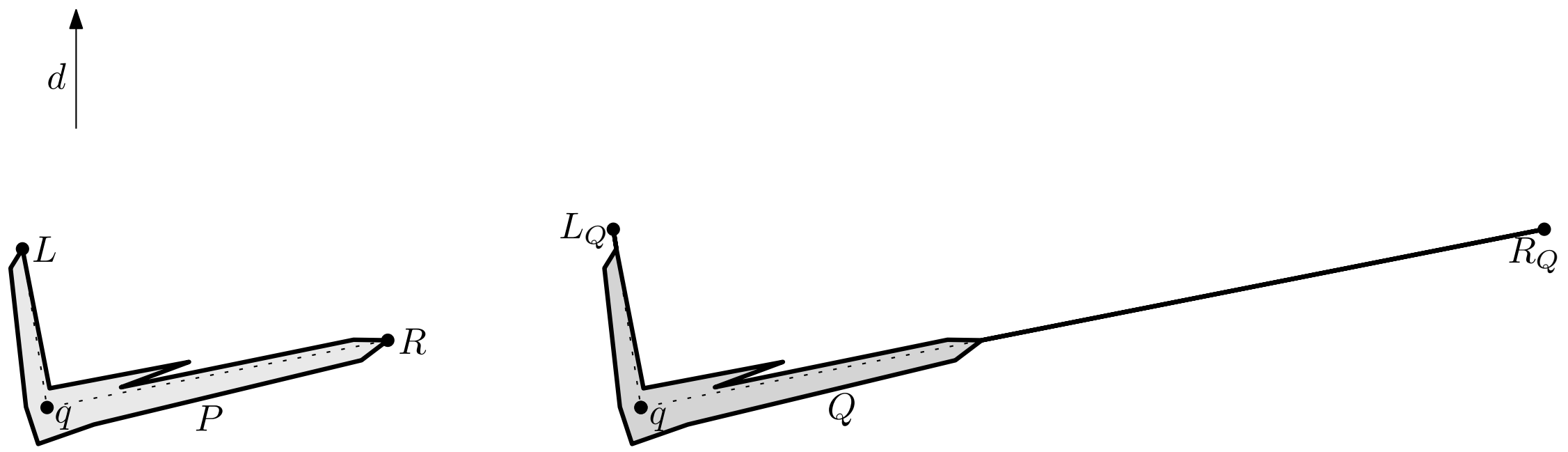}
    \caption{The star $P$ and the bowl $Q(P)$.}
    \label{fig:bowl}
\end{figure}

In coordinates with axes $\{\Vec{u}_\tau,\Vec{d}_\tau\}$, let $R_Q=(ha_Q,h)$ and $L_Q=(hbQ,h)$. Then we have
\[
a_Q=\cot\beta_P,
\qquad
b_Q=\frac{x\cos\beta_P-w_P\sin\beta_P}{x\sin\beta_P+w_P\cos\beta_P}.
\]
\cref{eq:bowl-arm-bounds} follows from \cref{eq:delta-bounds} and $\beta_P\in[\delta, 2\delta)$, $|x|\le r$, $w_P\in(w/2,w]$.
\end{proof}

\cref{eq:star-bowl-containment} shows that replacing the stars with bowls will only increase the optimum by at most a constant factor $\alpha^2$. 
To see this, we can take the optimal packing of $\alpha$-times-larger input stars, which has area $\alpha^2 \opt(\mathcal{P})$. 
Then a feasible packing of the bowls follows by matching the reference points with the corresponding scaled stars.
Therefore, it suffices to compute a constant approximate packing for the bowls. Then, to obtain a packing of the original stars, we place each $P_i$ at the location of $Q(P_i)$, sharing the same reference point. 

\subsection{Stacking bowls is (almost) optimal}
\label{sec:local-bowl-stacks}

We now show that the structural packing from \cref{thm:star-local-decomposition-strong} can be used to obtain a stack packing of bowls.
Let $\alpha$ be the constant from \cref{eq:star-bowl-containment}, and apply \cref{thm:star-local-decomposition-strong} to a packing of the scaled stars $\mathcal{P}'=\alpha\mathcal{P}$.
For each part $\mathcal{P}'_j$ with bounding rectangle $B_j$, replace every scaled star $\alpha P\in\mathcal{P}'_j$ by its corresponding bowl $Q(P)$, keeping the reference point fixed.
Since $Q(P)\subseteq\alpha P$, this gives a feasible packing of $\mathcal{Q}_j=\{Q(P):\alpha P\in\mathcal{P}'_j\}$ inside $B_j$.
See \cref{fig:induced-bowl-packing} for an illustration.

It remains to show that the bowls in each rectangle $B_j$ form a stack.
Fix a part $\mathcal{P}'_j$, and let $\tau$ be its common fine type, with scales $\alpha r$ and $\alpha w$.
All bowls in $\mathcal{Q}_j$ therefore share the directions $\Vec{d}_{\tau}$ and $\Vec{u}_{\tau}$.
Moreover, \cref{thm:star-local-decomposition-strong} guarantees that their reference points are close together.
We use this property in the next lemma to show that their separators are pairwise disjoint, and hence that the bowls form a stack.

\begin{lemma}
\label{lem:localized-stars-to-bowl-stacks}
Let $\alpha$ be the constant in \cref{lem:star-to-bowl}.
Apply \cref{thm:star-local-decomposition-strong} to a packing of $\mathcal{P}'=\alpha \mathcal{P}$ in a convex container $C$.
For every $j$, the bowls
\[
\mathcal Q_j=\{Q(P):\alpha P\in\mathcal P'_j\}
\]
admit a stack packing in a rectangle $\widehat B_j$ of dimensions $O(r)\times O(w)$, with sides parallel to their common directions $\Vec{u}_{\tau_j},\Vec{d}_{\tau_j}$, such that
\[
\sum_j|\widehat B_j|=O(|C|).
\]
\end{lemma}

\begin{proof}
Fix any $j$.
Let $\mathcal P'_j,B_j$ be defined in \cref{thm:star-local-decomposition-strong}.
For each $P'=\alpha P\in \mathcal{P}'_j$, place the bowl $Q(P)$ such that $q_{Q(P)}=q_{P'}$.
By \cref{eq:star-bowl-containment}, we have $Q(P)\subseteq P'$, so the bowls are feasible and lie in $\widehat B_j=B_j$. The area bound follows from \cref{thm:star-local-decomposition-strong}(ii).
The only remaining thing is to show that the induced packing for $\mathcal{Q}_j$ forms a stack.

Now we use the coordinate system along $\vec{u}_{\tau_j}$ and $\vec{d}_{\tau_j}$. For any point $p$, we use $(x(p),y(p))$ to denote its coordinates.
Consider any pair of bowls $Q_1$ and $Q_2$ and translate $q_{Q_1}$ to the origin.
Write $q_{Q_2}=(X,Y)$ where $Y\ge 0$ without loss of generality.
By \cref{thm:star-local-decomposition-strong}(i), $|X|\le10\eps r$ and $Y\le10\eps w$.
Recall that $y(L_{Q_1})=y(R_{Q_1})=h >10\eps w$ by \cref{lem:star-to-bowl}, so $q_{Q_2}$ is below the line $\overline{L_{Q_1}R_{Q_1}}$.

\begin{claim}
\label{clm:qQ2-triangle}
$q_{Q_2}$ is in the triangle $T_{Q_1}$.
\end{claim}

\begin{proof}
We will show that if $q_{Q_2}$ is to the left of $q_{Q_1}L_{Q_1}$, then $q_{Q_2}R_{Q_2}$ intersects with $q_{Q_1}L_{Q_1}$, and similarly, if $q_{Q_2}$ is to the right of $q_{Q_1}R_{Q_1}$, then $q_{Q_2}L_{Q_2}$ intersects with $q_{Q_1}R_{Q_1}$.

Recall that $a_{Q_1},b_{Q_1},a_{Q_2},b_{Q_2}$ are defined in \cref{eq:bowl-arm-bounds}.
If $q_{Q_2}$ is to the left of $q_{Q_1}L_{Q_1}$, then $X<b_{Q_1}Y$. The two rays $\overrightarrow{q_{Q_2}R_{Q_2}}$ and $\overrightarrow{q_{Q_1}L_{Q_1}}$ intersect at $(\cdot,Y+t)$ where
\[ t=\frac{b_{Q_1}Y-X}{a_{Q_2}-b_{Q_1}} \le \frac{30\eps r - (-10\eps r)}{7r/w-3r/w} = 10\eps w < h-Y, \]
so this is an intersection of the two segments.

If $q_{Q_2}$ is to the right of $q_{Q_1}R_{Q_1}$, then $X>a_{Q_1}Y$. The two rays $\overrightarrow{q_{Q_2}L_{Q_2}}$ and $\overrightarrow{q_{Q_1}R_{Q_1}}$ intersect at $(\cdot,Y+t)$ where
\[ t=\frac{X-a_{Q_1}Y}{a_{Q_1}-b_{Q_2}} \le \frac{10\eps r - 7\eps r}{7r/w-3r/w} = \frac{3\eps w}{4} < h-Y, \]
so this is an intersection of the two segments.
\end{proof}

By \cref{clm:qQ2-triangle}, if $Y>0$, then $q_{Q_2}$ is above $\sep(Q_1)$. Furthermore, the two arms of $Q_2$ can only leave the triangle through $L_{Q_1}R_{Q_1}$ to avoid intersections. So the two separators are disjoint.
If $Y=0$, then $q_{Q_1}=q_{Q_2}$ and $y(L_{Q_1})=y(R_{Q_1})=y(L_{Q_2})=y(R_{Q_2})=h$. By the feasibility, the two separators must also be (weakly) disjoint.
Since any pair of separators are disjoint, the bowls form a stack.
\end{proof}

In summary, we show that there exists a packing of the bowls as stacks into small rectangles with a total area comparable to the optimal packing of the original stars into multiple convex containers.

\begin{corollary}
\label{thm:box-star-packing}
Let $\mathcal P$ be a set of stars in the same size group and let $\mathcal Q=\{Q(P):P\in\mathcal P\}$.
Then there exists a packing of $\mathcal Q$ into rectangles of total area $O(\opt'(\mathcal P))$, each containing a stack of bowls and having dimensions $O(r)\times O(w)$ with sides parallel to $\Vec{u}_\tau,\Vec{d}_\tau$, where $\tau$ is the common fine type of the stars in it.
\end{corollary}
\begin{proof}
Consider the optimal packing of the original stars into multiple convex containers.
For each container $C$, we apply \cref{thm:star-local-decomposition-strong,lem:localized-stars-to-bowl-stacks} to $C$ and the stars in it.
\end{proof}

Let $\opt_{\mathrm{stack}}(\mathcal Q_\tau)$ denote the infimum area of packing all bowls of type $\tau$ into a convex container, under the constraint that they must form a stack.
We have the following lemma.

\begin{lemma}
\label{cor:bowl-stack-packing-lowerbound}
For a set of stars $\mathcal P$ in the same size group,
\begin{equation*}
\sum_\tau\opt_{\mathrm{stack}}(\mathcal Q_\tau)
=O(\opt'(\mathcal P)).
\end{equation*}
\end{lemma}
\begin{proof}
Consider the packing of $\mathcal Q$ in \cref{thm:box-star-packing}.
For each fine type $\tau$, all the rectangles that contain bowls of this type have dimensions $O(r)\times O(w)$ along the same directions $\Vec{u}_\tau,\Vec{d}_\tau$.
We put them one above another along $\vec{d}_\tau$. If there are $t_{\tau}$ rectangles, then they fit in an $O(r)\times O(t_{\tau}w)$ large rectangle.
It is straightforward to see that the bowls in this large rectangle still form a stack, and the area of it is at most a constant times the total area of the initial rectangles.
Summing over the types, the total area is still $O(\opt'(\mathcal P))$.
\end{proof}

\subsection{Computing approximately optimal stacks}
\label{sec:algorithm}

In this subsection, we consider bowls from one fine type $\tau$.
Note that the separators of the bowls in a stack are disjoint, so they have a natural order along $\Vec{d}_\tau$.
Therefore, to pack a set of bowls into a stack, the key part is to compute the order.
We model this as an asymmetric TSP problem.

Recall that the bowls have the same heights $h=100w$. Without loss of generality, we assume $\Vec{d}_\tau$ is upward. We start with a simple lower bound of the area of a stack.

\begin{lemma}
\label{lem:stack-min-area}
If a stack has vertical span $H$, then its minimum convex container $C$ has area $\Omega(rH)$.
\end{lemma}

\begin{proof}
Let $P$ be a star in the stack. The span of $P$ along $\vec{e}_P$ is at least $r/2$, and the angle between $\vec{e}_P$ and $\vec{u}_\tau$ is at most $2\delta \le \pi/25$ by \cref{eq:delta-bounds}.
So the horizontal span of the stack is $\Omega(r)$.
\end{proof}

The next lemma shows that a tall stack can be split into short strict stacks with bounded total area.

\begin{lemma}
\label{lem:chunk-path}
In polynomial time, a stack of vertical span $H$ can be partitioned into $O(H/h)$ strict stacks, each fitting in an $O(r)\times O(h)$ rectangle with sides parallel to $\Vec{u}_\tau$ and $\Vec{d}_\tau$.
\end{lemma}

\begin{proof}
Partition the reference points into $O(H/h)$ horizontal strips of height $h/2$.
For each strip, consider the bowls whose reference points are in this strip. The vertical span of them is $O(h)$.
Since they form a stack, their separators are disjoint. Then these reference points must all be in the triangle $T_Q$ where $Q$ is the lowest bowl, which means the horizontal span of these bowls is $O(r)$. So they fit in an $O(r)\times O(h)$ rectangle.

Now we prove they form a strict stack. Let $Q_1$ and $Q_2$ be two bowls in the same strip such that $\sep(Q_1)$ is (weakly) below $\sep(Q_2)$. Clearly, $Q_1$ cannot intersect with $\sep(Q_2)$ since the bowls have the same heights.
On the other hand, $q_{Q_2}$ is in the triangle $T_{Q_1}$. Since $q_{Q_2}$ is the reference point of $Q_2$, for any point $p\in Q_2$, $pq_{Q_2}$ cannot intersect with both $q_{Q_1}L_{Q_1}$ and $q_{Q_1}R_{Q_1}$. So $Q_2$ does not intersect with $\sep(Q_1)$.
\end{proof}

We remark that the total area of the rectangles in \cref{lem:chunk-path} is $O(rH)$, although the area of the convex hull of the original stack may be much larger. This will be important for our algorithm.
By \cref{lem:chunk-path,lem:stack-min-area}, when considering stack packing, we only need to care about the height of the stack.

The next corollary shows that we can restrict our attention to strict stacks at only a constant-factor increase in height.

\begin{corollary}
\label{lem:strict-stack}
If some bowls admit a stack of vertical span $H$, then they admit a strict stack of vertical span $O(H)$.
\end{corollary}
\begin{proof}
We simply put the rectangles in \cref{lem:chunk-path} one above another along $\vec{d}_\tau$.
\end{proof}

For two bowls $Q_i,Q_j$, we write $Q_i\prec Q_j$ if we place them in a strict stack with $Q_i$ below $Q_j$. Note that if $Q_i\prec Q_j\prec Q_k$, then the $\sep(Q_j)$ separates $Q_i$ from $Q_k$. So, when we construct a strict stack, we only need to make sure that each adjacent pair of bowls do not intersect.

\begin{lemma}
\label{lem:bowl-TSP}
In polynomial time, a bowl class $\mathcal Q_\tau$ can be packed into
$O(r)\times O(h)$ rectangles with sides parallel to $\Vec{u}_\tau$ and $\Vec{d}_\tau$ and total area $O(\opt_{\mathrm{stack}}(\mathcal Q_\tau))$.
\end{lemma}
\begin{proof}
Suppose the optimal stack packing of $\mathcal Q_\tau$ has vertical span $H$.
By \cref{lem:strict-stack}, $\mathcal Q_\tau$ also admits a strict stack of vertical span $H'=O(H)$.
We only need to approximate the minimum height of the strict stack.

Let $\mathcal Q_\tau=\{Q_1,Q_2,\dots,Q_m\}$.
For $i,j\in[1,m]$, let $c(i,j)\le 2h$ be the minimum increase of the height (i.e. the minimum value of $(q_{Q_2}-q_{Q_1})\cdot\vec{d}_\tau$) when placing $Q_i\prec Q_j$.
In particular, let $c(i,i)=0$.
Notice that for any strict stack of three bowls $Q_i\prec Q_j\prec Q_k$, removing $Q_j$ gives us a strict stack of the other two bowls $Q_i\prec Q_k$. So $c(i,k)\le c(i,j)+c(j,k)$. In other words, $c$ is a directed metric.

Since the bowls admit a strict stack of vertical span $H'$, by the definition of $c$, they admit a Hamiltonian path of length at most $H'$. So they admit a Hamiltonian cycle of length at most $H'+2h=O(H)$.
We apply the constant-factor approximation for metric asymmetric TSP in \cite{DBLP:journals/jacm/SvenssonTV20}, then delete the most expensive edge. This gives us a Hamiltonian path of length $O(H)$ in polynomial time.

From the Hamiltonian path, it is straightforward to get a strict stack of $\mathcal Q_\tau$ of vertical span $O(H)$.
Then we partition them into $O(r)\times O(h)$ rectangles by \cref{lem:chunk-path}.
The total area of the rectangles is $O(rH)$.
On the other hand, $\opt_{\mathrm{stack}}(\mathcal Q_\tau)=\Omega(rH)$ by \cref{lem:stack-min-area}.
\end{proof}

We summarize our conclusion for packing one size group in the following lemma.

\begin{lemma}
\label{cor:star-size-class-algorithm}
For a set of stars $\mathcal P$ in the same size group, one can compute in polynomial time a packing into rectangular mini-containers of total area
$O(\opt'(\mathcal P))$. Each mini-container has dimensions $O(r)\times O(w)$ with sides parallel to $\Vec{u}_\tau,\Vec{d}_\tau$, where $\tau$ is the common fine type of all stars in this mini-container.
\end{lemma}
\begin{proof}
We first partition the stars into fine types and compute the corresponding bowls by \cref{lem:star-to-bowl}.
Then we compute the mini-containers for each bowl class separately by \cref{lem:bowl-TSP}. The total area of the mini-containers is $O(\opt'(\mathcal P))$ by \cref{cor:bowl-stack-packing-lowerbound}.
Finally, we replace each bowl with the original star at the same reference point.
\end{proof}

For a direction $z$, let $\Span_z(S)$ denote the span of $S$ in that direction. The following corollary shows that the span of a mini-container in any direction is comparable to any star in it.

\begin{corollary}
\label{lem:star-mini-container-spans}
Every mini-container $B$ in \cref{cor:star-size-class-algorithm} satisfies
\[
\Span_z(B)=O(\Span_z(P))
\]
for every direction $z$ and every star $P$ in $B$.
\end{corollary}
\begin{proof}
The mini-container has dimensions $O(r)\times O(w)$ along $\Vec{u}_\tau,\Vec{d}_\tau$.
The star has dimensions $\Omega(r)\times \Omega(w)$ along $\vec{e}_P,\vec{v}_P$.
The angle between $\vec{u}_\tau$ and $\vec{e}_P$ ($\vec{d}_\tau$ and $\vec{v}_P$) is less than $2\delta=O(w/r)$.
\end{proof}

\subsection{Shaving a log-factor}\label{sec:star-local-grouping}

Recall that there are $O(\log^2 n)$ size groups, and we pack the stars of each size group separately. This would naturally introduce an approximation ratio of $O(\log^2 n)$.

In this subsection, we show that packing the size groups separately increases the total area by only an $O(\log n)$ factor. The key observation concerns stars with much larger radii but much smaller widths than others. We associate a triangular core with each star and show that the cores of such stars must be disjoint in any feasible packing. These disjoint cores give an area lower bound that allows us to divide the size groups into $O(\log n)$ classes. This improves the final approximation ratio to $O(\log n)$.

\begin{figure}[!h]
    \centering
    \includegraphics[page=13]{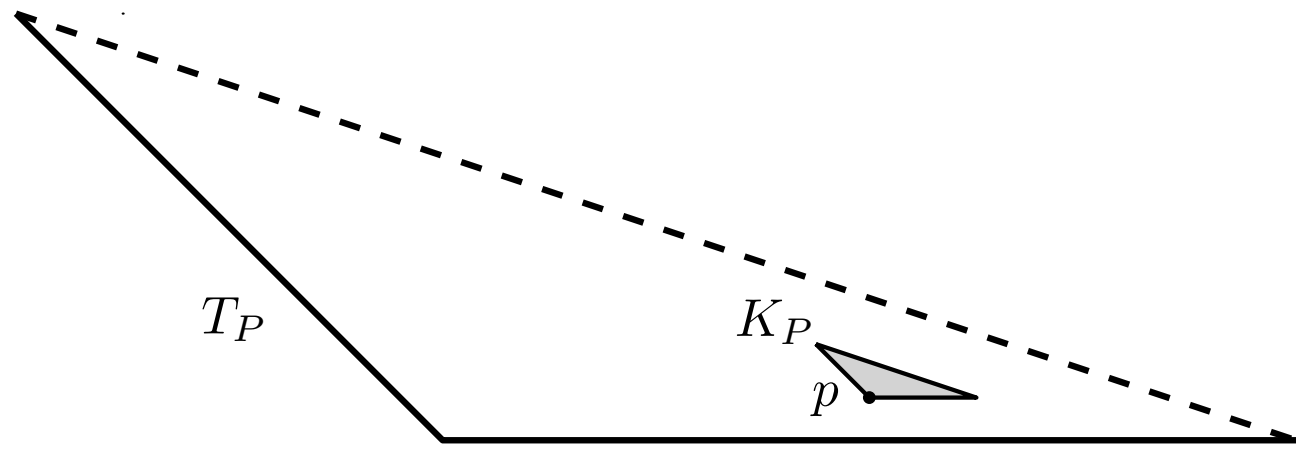}
    \caption{The core of a star $P$ is the gray triangle $K_P$.}
    \label{fig:core2}
\end{figure}

\paragraph{Cores.}
Let $q'_P=(q_P+R_P)/2+w_P\vec{v}_P/10$, $R'_P=q'_P+(R_P-q_P)/10$ and $L'_P=q'_P+(L_P-q_P)/10$. We define the \emph{core} of $P$, denoted by $K_P$, as the triangle $\triangle q'_PR'_PL'_P$.
In other words, it is a $0.1$-scale copy of $T_P$ located ``above'' the middle point of the radius. See Figure~\ref{fig:core2} for an illustration.

By definition, we have
\[ K_P\subseteq T_P \]
and
\begin{equation}
\label{eqn:core-area}
|K_P|=\Theta(r_Pw_P).
\end{equation}
It is straightforward to check that, for any point $p\in K_P$,
\begin{equation}
\label{eqn:core-circle}
\dist(p,q_PR_P),\dist(p,q_PL_P),\dist(p,L_PR_P)\ge w_P/20.
\end{equation}
Consider the coordinates along $\vec{e}_P$ and $\vec{v}_P$ centered at $q_P$. We further have 
\begin{equation}
\label{eqn:core-bound}
K_P\subseteq[0.4r_P,0.6r_P]\times[0.1w_P,0.2w_P].
\end{equation}

The next lemma shows that the cores of stars with very different shapes must be disjoint.

\begin{lemma}
\label{lem:core-disjoint}
Let $P_1,P_2$ be two stars such that $r_{P_1}>10r_{P_2}$ and $w_{P_2}>10w_{P_1}$. In any packing, $K_{P_1}\cap K_{P_2}=\emptyset$.
\end{lemma}

\begin{proof}
Suppose for contradiction that $K_{P_1}\cap K_{P_2}\neq\emptyset$.
Consider the coordinates along $\vec{e}_{P_1}$ and $\vec{v}_{P_1}$ centered at $q_{P_1}$.
Let $p=(x_p,y_p)$ be an arbitrary point in $K_{P_1}\cap K_{P_2}$.
By \cref{eqn:core-bound}, $x_p\in[0.4r_{P_1},0.6r_{P_1}]$ and $y_p\in[0.1w_{P_1},0.2w_{P_1}]$. Since $p\in K_{P_2}\subseteq T_{P_2}$, any point in $T_{P_2}$ is at a distance at most $2r_{P_2}<0.2r_{P_1}$ from $p$.
So, $T_{P_2}$ is in the strip $0.2r_{P_1}<x<0.8r_{P_1}$.
Since $q_{P_1}R_{P_1}$ cut this strip at $y=0$ and $y_p>0$, $T_{P_2}$ lies in $y\ge 0$ due to connectivity of $P_2$. 

On the other hand, consider the disk $D$ centered at $p$ with radius $w_{P_2}/20>0.5w_{P_1}$. Since $y_p\le 0.2w_{P_1}$, $D$ intersects $y<0$. By \cref{eqn:core-circle}, $D\subseteq T_{P_2}$. So $T_{P_2}$ intersects $y<0$, which is a contradiction.
\end{proof}

The next lemma proves an upper bound of packing stars in the same size group with intersecting cores. Note that the lower bound of packing each single core is $\Omega(r)\times\Omega(w)$.

\begin{lemma}
\label{lem:core-rectangle}
Let $P$ be a star in the size group with scales $r$ and $w$. There is an $O(r)\times O(w)$ rectangle that contains all stars $P'$ in the same group such that $K_{P'}\cap K_P\neq\emptyset$.
\end{lemma}

\begin{proof}
Consider the coordinates along $\vec{e}_{P}$ and $\vec{v}_{P}$ centered at $q_{P}$.
Consider one such $P'$. Let $p=(x_p,y_p)$ be an arbitrary point in $K_{P'}\cap K_P$. 
Let $\theta$ be the angle between the two radius lines $\overline{q_PR_P}$ and $\overline{q_{P'}R_{P'}}$.
For each radius line, by \cref{eqn:core-bound}, the distance from $p$ is at most $0.2w$, and the perpendicular projection of $p$ lies at distance at least $0.2r$ from both $q_P$ ($q_{P'}$) and $R_{P}$ ($R_{P'}$). See Figure~\ref{fig:theta}.

\begin{figure}[!h]
    \centering
    \includegraphics[page=18]{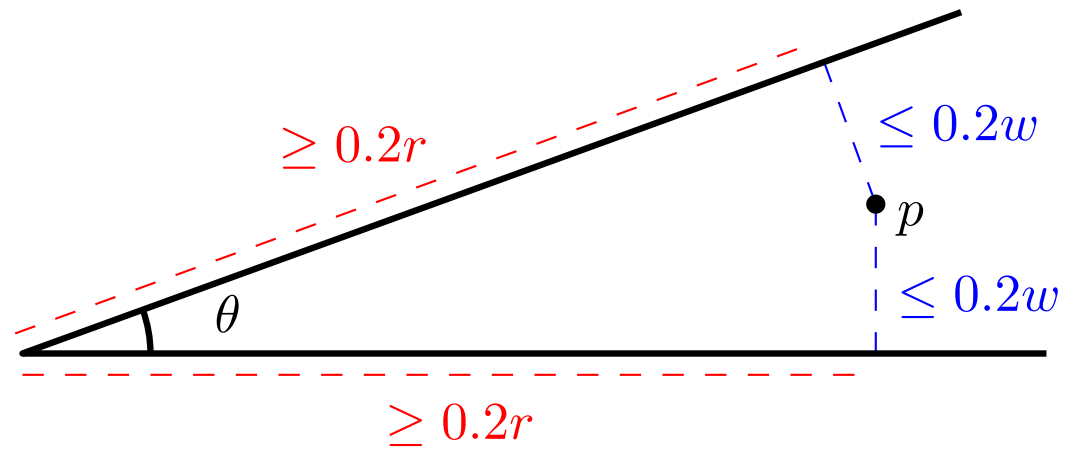}
    \caption{The point $p$ and the two radius lines.}
    \label{fig:theta}
\end{figure}

The two radii $q_PR_P$ and $q_{P'}R_{P'}$ are part of the stars, so they cannot intersect. Therefore, we have $0.4w/\sin\theta \ge 0.2r$, so $\sin\theta \le 2w/r$.
Consider any point $z=(x_z,y_z)\in P'$. Since $p\in T_{P'}$, $|x_z-x_p|\le 2r\cos\theta + 2w\sin\theta = O(r)$ and $|y_z-y_p|\le 2r\sin\theta + 2w\cos\theta = O(w)$. Combining these with $p\in K_p$ and \cref{eqn:core-bound} proves the lemma.
\end{proof}

Combining the above lemmas, we can prove that, for all the stars with similar sizes, packing each group separately will only increase the cost by a constant factor.

\begin{lemma}
\label{lem:similar-size}
Let $\mathcal{P}$ be a set of stars such that the sizes $r_P\cdot w_P$ for all $P\in\mathcal{P}$ differ by at most a constant factor. Then,
\[ \sum_{\mathcal G:\text{ size group}}\opt'(\mathcal G) = O(\opt'(\mathcal P)). \]
\end{lemma}

\begin{proof}
Recall that each size group $\mathcal{G}_{a,b}$ has $r=r_{\max}/2^a$ and $w=w_{\max}/2^b$. We divide all size groups into $O(\log n)$ classes, such that each class $\mathcal{C}_{i,j}$ ($i\in[0,O(\log n)], j\in[0,4]$) consists of the groups $\mathcal{G}_{a,b}$ where $a+b=i$ and $a \text{ mod } 5 = j$.
Since all stars have similar sizes $r_P\cdot w_P$, there are only a constant number of classes that have non-empty size groups. So it suffices to look at one such class $\mathcal{C}_{i,j}$.

Consider the optimal packing of $\mathcal{C}_{i,j}$. Let $I(\mathcal{P'})$ denote the size of the maximum independent set of the cores of $\mathcal{P'}$ (that is, the maximum number of cores that are disjoint).
For each class $\mathcal{C}_{i,j}$, by \cref{lem:core-disjoint}, the cores of stars from different size groups are disjoint. So
\begin{equation}
\label{eqn:IG-IC}
\sum_{\mathcal{G}\in\mathcal{C}_{i,j}}I(\mathcal{G}) = I(\mathcal{C}_{i,j}).
\end{equation}
Since each core has an area of $\Omega(r_{\max}w_{\max}/2^i)$ by \cref{eqn:core-area} and lies in the convex hull of the star, we have
\begin{equation}
\label{eqn:IC}
\opt(\mathcal{C}_{i,j}) = \Omega(I(\mathcal{C}_{i,j})r_{\max}w_{\max}/2^i).
\end{equation}
On the other hand, by \cref{lem:core-rectangle}, we have
\begin{equation}
\label{eqn:IG}
\opt(\mathcal{G}) = O(I(\mathcal{G})r_{\max}w_{\max}/2^i).
\end{equation}
Combining \cref{eqn:IG-IC,eqn:IC,eqn:IG} proves the lemma.
\end{proof}

We conclude this subsection with the following corollary.

\begin{corollary}
\label{lem:logn}
For a set of stars $\mathcal{P}$,
\[ \sum_{\mathcal G:\text{ size group}}\opt'(\mathcal G) = O(\log n)\cdot\opt(\mathcal P). \]
\end{corollary}

\begin{proof}
Recall that stars with sizes smaller than $r_{\max}w_{\max}/n$ can be packed individually at a cost of $\opt(\mathcal P)$. Then it simply follows from \cref{lem:similar-size}.
\end{proof}

\subsection{The complete translational algorithm}
\label{sec:translational-final}

Finally, we put everything together to get an $O(\log n)$-approximate algorithm.
We first give a lemma that follows easily from our previous results.

\begin{lemma}
\label{lem:total-minicontainer-area}
For a set of stars $\mathcal{P}$, in polynomial time, we can compute a packing of $\mathcal{P}$ into a set of convex mini-containers of total area $O(\log n)\cdot\opt(\mathcal P)$.
Every mini-container $B$ satisfies
\begin{equation}
\label{eqn:span-mini-container}
    \Span_z(B)=O(\Span_z(P))
\end{equation}
for every direction $z$ and every star $P$ that $B$ contains.
\end{lemma}

\begin{proof}
We pack all stars with $r_P<r_{\max}/n$ or $w_P<w_{\max}/n$ individually into its convex hull. The total area of them is $O(r_{\max}w_{\max})=O(\opt(\mathcal P))$ by \cref{lem:rmax-wmax}.
For these mini-containers, \cref{eqn:span-mini-container} trivially holds.

The remaining stars are partitioned into $O(\log^2 n)$ size groups.
We pack each size group separately by \cref{cor:star-size-class-algorithm}. The total area is
\[ \sum_{\mathcal G:\text{ size group}}O(\opt'(\mathcal G)). \]
By \cref{lem:logn}, this equals $O(\log n)\cdot\opt(\mathcal P)$.
For these mini-containers, \cref{eqn:span-mini-container} follows from \cref{lem:star-mini-container-spans}.
\end{proof}

We can now bound the approximation factor of our algorithm.

\begin{theorem}
\label{thm:translational-star-packing}
For a set of stars $\mathcal{P}$, in polynomial time, we can compute a packing of $\mathcal{P}$ into a rectangle of area $O(\log n)\cdot\opt(\mathcal P)$.
\end{theorem}

\begin{proof}
We use the packing of \Cref{lem:total-minicontainer-area} into convex mini-containers of total area $O(\log n)\cdot\opt(\mathcal P)$.
Then we apply \cref{thm:convex-packing} to these mini-containers in the coordinate system where the longest radius (whose length is $r_{\max}$) is horizontal.

Let $\ell_{\max}$ and $h_{\max}$ be the maximum horizontal and vertical spans of these mini-containers respectively.
It only remains to show that $\opt(\mathcal P) = \Omega(\ell_{\max}h_{\max})$.
By \cref{eqn:span-mini-container}, $r_{\max}=\Omega(\ell_{\max})$ and there exists a star $P$ whose vertical span is $h_P=\Omega(h_{\max})$.
Since the minimum convex container of $\mathcal P$ must contain the longest radius, which is a horizontal segment of length $r_{\max}$, and the star $P$, we have $\opt(\mathcal P)=\Omega(r_{\max}h_P)=\Omega(\ell_{\max}h_{\max})$.
\end{proof}

\section{Algorithm for rotational packing}
\label{sec:structural-rotation}

The preceding section gives an algorithm for packing stars by
translations: it replaces stars by bowls, arranges bowls in stacks, and
packs the resulting rectangular mini-containers. Let us now consider the setting, in which we are also allowed both translations and rotations in order to pack stars.
We will show that, within each size group, requiring all radius vectors
to point in the same direction increases the minimum container area by
only a constant factor. We can then rotate the input stars to this
direction and apply the translational algorithm of
\cref{sec:star-packing}, including the stack-packing procedure of
\cref{lem:bowl-TSP}.

Fix a size group $\mathcal G=\mathcal G_{a,b}$ from the preceding
section. As before, $r_{\max}$ and $w_{\max}$ denote the maximum input
radius and width. The scales of the group are
$r=r_{\max}/2^{a}$ and $w=w_{\max}/2^{b}$.
The normalized bowls have height $h=100w$.
For a star $P_i$, let $q_i$ denote its reference point and $R(P_i)$
its distinguished radius endpoint, as in the preceding section.
Its directed radius vector is $R(P_i)-q_i$. We choose these
reference points and endpoints once and carry them along under all
transformations. Constants below
are independent of the number of pieces and the ratio $r/h$.

We first apply the previous construction to a packing whose chosen
rotations are held fixed. It gives small rectangles containing stacks
whose radius directions are already close. The remaining task is to
make these directions equal. After a preliminary division into locally separated
blocks, we take every third bowl and enclose it between its two neighboring
separators. We process the resulting enclosures from bottom to top,
rotating all remaining enclosures about the current reference point
at each step. The sum of the successive angle changes bounds the total
upward translation, and hence the increase in the rectangle's height.

\begin{lemma}%
\label{lem:rot-initial-containers}
Fix a constant $\varepsilon$ with $0<\varepsilon\le10^{-6}$.
Given a rotational packing of one size group into a convex container
$C$, the stars can be replaced by containing bowls and packed, using
translations of their fixed orientations, into rectangular mini-containers
of dimensions $O(r)\times O(h)$ and total area $O(|C|)$.
Within each mini-container, let $d$ denote the common bowl-alignment
direction and let $u$ denote the perpendicular direction pointing towards
the distinguished radius. The mini-container may be replaced by a rectangle
whose sides are parallel to $u$ and $d$, with respective lengths
$O(r)$ and $O(h)$. These rectangles can all be enlarged to common
dimensions $\kappa r\times\kappa h$, for a fixed constant $\kappa$, while retaining
the total-area bound. We have:
\begin{enumerate}
\item the bowls have common height $h$, alignment direction $d$,
and pairwise disjoint separators;
\item their radii are $\Theta(r)$ and their widths are $\Theta(h)$;
\item their directed radius angles lie in an interval of length at most
$\varepsilon h/r$;
\item each bowl contains its corresponding star, with the same reference
point and distinguished radius ray.
\end{enumerate}
Moreover, no point of one bowl lies in the interior of another bowl
in the same mini-container. This includes the two added radial arms,
which may have zero width.
All hidden constants are universal after fixing $\varepsilon$.
\end{lemma}
\begin{proof}
Fix the orientations of the stars in the given packing and regard it as
a translational instance, keeping the chosen reference points and
radius witnesses; fine types are computed from these orientations.
Apply \cref{thm:box-star-packing}. It returns the bowls $Q_i=Q(P_i)$ of
\cref{lem:star-to-bowl}, packed into rectangles of dimensions
$O(r)\times O(w)=O(r)\times O(h)$, each with sides parallel to the
directions $u,d$ of the common fine type of its bowls, and of total
area $O(|C|)$. Each rectangle holds a stack of bowls of one fine type;
the stack property gives height $h$ and pairwise disjoint separators.
Each $Q_i$ contains $P_i$ together with its reference point and
distinguished radius ray, and has radius $\Theta(r)$ and width
$\Theta(h)$, by \cref{lem:star-to-bowl}. This gives items~1, 2
and~4, but with the angular span of item~3 replaced by that of a fine
type, which we now reduce.

By \cref{eq:delta-bounds}, the directed radius angles of the stars of
one fine type lie in an interval of length
$\delta=2\pi/N\le2\pi w/(100r)=(2\pi/10^{4})\,h/r$.
Split this interval into
$m=\lceil2\pi/(10^{4}\varepsilon)\rceil$ consecutive subintervals of
length at most $\varepsilon h/r$, and split each stack into the at most
$m$ subsequences of bowls whose angles fall into the same subinterval,
each in its inherited order. A subsequence of a stack is again a stack:
its separators are among the original ones and hence remain pairwise
disjoint, its bowls still have height $h$ and alignment direction $d$,
and it lies in the rectangle of the original stack. We place each
subsequence, by a translation, in its own copy of that rectangle and
call the result a mini-container. This multiplies the number of
rectangles, and hence their total area, by at most $m=O(1/\varepsilon)$,
so the total area is still $O(|C|)$. Within each mini-container the
directed radius angles now lie in an interval of length at most
$\varepsilon h/r$, which is item~3.

In the proof of \cref{lem:localized-stars-to-bowl-stacks} each $Q_i$ is
placed inside the dilated star $F(P_i)$, and these are polygons with
pairwise disjoint interiors; hence no point of $Q_i$ lies in the
interior of $Q_j$ for $i\ne j$, and this persists under the translations
into the mini-containers, including the passage to subsequences.

Finally, \cref{thm:box-star-packing} allows all rectangles to be enlarged
to common dimensions $\kappa r\times\kappa h$ while keeping the area
estimate; this applies verbatim to the copies above. The constants
depend on $\varepsilon$ but not on $r/h$ or on the number of stars.
\end{proof}

It therefore suffices to straighten the containing bowls.

\paragraph{Coordinates and separators.}
Fix one mini-container with bowls $Q_1,\ldots,Q_k$. Choose coordinates
with horizontal axis along the common type direction $u$ and vertical
axis along the alignment direction $d$, so that $u=(1,0)$ and
$d=(0,1)$. Signed rotations are measured in these coordinates, which
may reverse the orientation of the ambient plane. Write
$q_i=(x_i,t_i)$, $L_i=L(Q_i)$, and $R_i=R(Q_i)$.
Let $\sigma_i=\sep(Q_i)$ be the separator defined in the preceding
section: the chain $L_iq_iR_i$, together with the horizontal rays
extending outward from $L_i$ and $R_i$. We call these rays its tails
and the segments $q_iL_i,q_iR_i$ its finite arms.
The separator divides the plane into an upper and a lower component.
Write $\operatorname{Above}(\sigma_i)$ and
$\operatorname{Below}(\sigma_i)$ for their closures, with the upper
component containing points of sufficiently large vertical coordinate.
Order the bowls by their separators from bottom to top. Their tails have
heights $t_i+h$, so their reference-point heights $t_i$ follow this
order as well. Below the tails, the upper component is precisely the
triangle $q_iL_iR_i$.
Define the normalized horizontal coordinates $A_i$ and $B_i$ of
the two arm endpoints by $R_i-q_i=(rA_i,h)$ and $L_i-q_i=(rB_i,h)$.
Using the estimates in \cref{eq:bowl-arm-bounds} and $h=100w$, we obtain
\begin{equation}\label{eq:rot-arm-bounds}
700\le A_i\le3200
\qquad\qquad
-400\le B_i\le300.
\end{equation}
In particular, $A_i-B_i\ge 400$, a margin we will use repeatedly.
Throughout, bowls and separators may touch along their boundaries, and
the stack order between separators is understood weakly at such
contacts, as in the preceding section; what makes contacts harmless is
the nonpenetration property of \cref{lem:rot-initial-containers}, which
keeps the zero-width arms of one bowl out of the interior of every
other. Note that two separators can coincide
only if their bowls share the reference point and both arm directions,
and then, since the two bowls have disjoint interiors and are both
star-shaped about the common point, at most one of them has interior
points outside the common closed triangle $qLR$; we place that bowl
below the other in the stack. Finally, let $\theta_i$ be the angle that
$R_i-q_i$ makes with the positive horizontal axis. All these angles lie
in a short common interval, and we take real representatives there, so
that differences of angles never wrap around.
\begin{figure}[t]
\centering
\includegraphics[page=1]{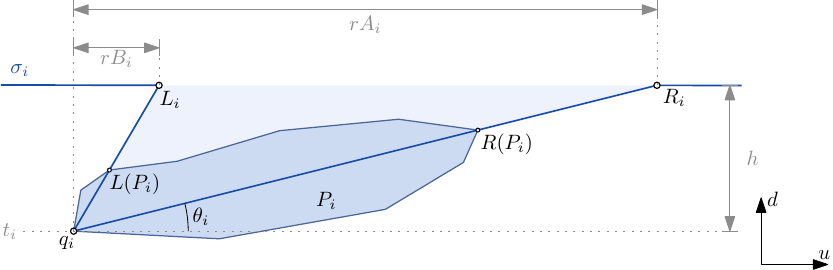}
\caption{A bowl $Q_i=P_i\cup q_iL_i\cup q_iR_i$ in the coordinates $(u,d)$ of its mini-container, with its separator $\sigma_i$ (two finite arms and two horizontal tails at height $t_i+h$), the triangle $q_iL_iR_i$, and the normalized arm coordinates $A_i,B_i$ of \cref{eq:rot-arm-bounds}.}
\label{fig:rot-anatomy}
\end{figure}
The next lemma bounds the sum of successive angle changes. This will
bound the total displacement in the rotation procedure.

\begin{lemma}%
\label{lem:rot-variation}
For the bowls in stack order,
\[
\sum_{i=1}^{k-1}|\theta_{i+1}-\theta_i|=O(h/r).
\]
The same bound holds for every subsequence in the inherited order.
\end{lemma}
\begin{figure}[t]
\centering
\includegraphics[page=2]{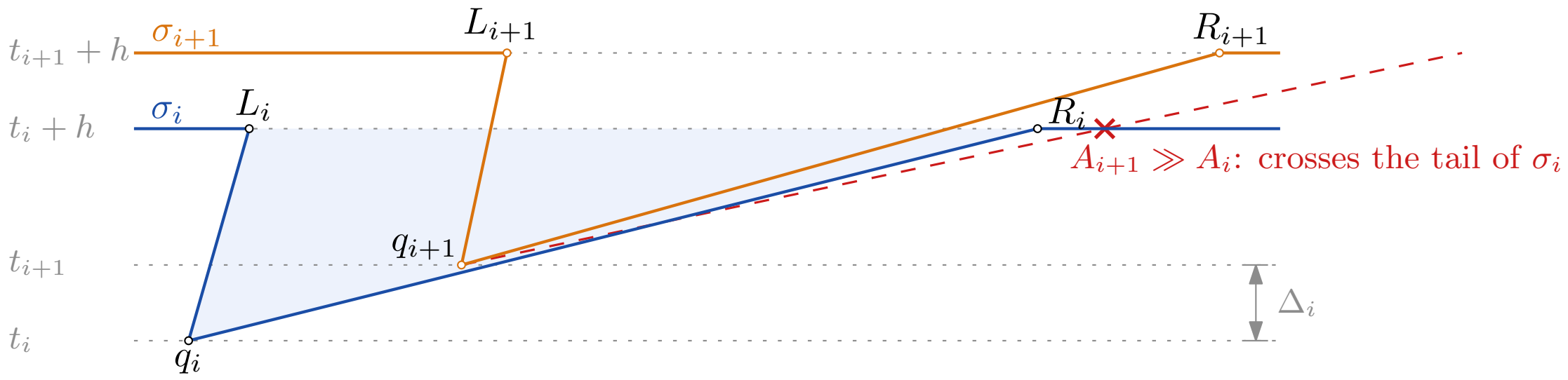}
\caption{Two consecutive separators of a stack. Since $q_{i+1}$ lies in the triangle $q_iL_iR_i$, the right arm of $\sigma_{i+1}$ must exit through the top chord at or left of $R_i$; a steeper arm (dashed) would cross the right tail of $\sigma_i$. This bounds $A_{i+1}-A_i$ by $O(\Delta_i/h)$.}
\label{fig:rot-variation}
\end{figure}
\begin{proof}
For each $i<k$, put $\Delta_i=t_{i+1}-t_i$. If $\Delta_i<h$, the point $q_{i+1}$ lies in the triangle
$q_iL_iR_i$, because it is above $\sigma_i$ and below its tails.
At height $t_i+h$, the right arm of $\sigma_{i+1}$ must lie no farther
right than $R_i$; otherwise it would cross the right tail of
$\sigma_i$. Therefore
$x_{i+1}+rA_{i+1}(1-\Delta_i/h)\le x_i+rA_i$.
Since $q_{i+1}$ lies in the triangle, we also have
$x_{i+1}\ge x_i+rB_i\Delta_i/h$. It follows that
$A_{i+1}-A_i\le (A_{i+1}-B_i)\Delta_i/h\le2M\Delta_i/h$.
For $\Delta_i\ge h$ the last bound on the positive part is automatic
from \cref{eq:rot-arm-bounds}. Thus, in both cases,
$\max\{A_{i+1}-A_i,0\}\le2M\Delta_i/h$.

The reference points lie in a rectangle of height $O(h)$, so
$\sum_i\Delta_i=O(h)$ and the sum of the positive changes of $A_i$
is $O(1)$. The negative changes have the same total up to
$A_1-A_k=O(1)$. Consequently,
$\sum_i|A_{i+1}-A_i|=O(1)$.

Since $A_i\ge700$, the radius angle is
$\theta_i=\arctan\bigl(h/(rA_i)\bigr)$. For every $A\ge700$,
\[
\left|\frac{d}{dA}\arctan\!\left(\frac{h}{rA}\right)\right| =\frac{h/r}{A^2+(h/r)^2}\le\frac{h}{700^2r}.
\]
The mean value theorem and the bound on the total variation of $A_i$
give
\[
\sum_i|\theta_{i+1}-\theta_i| \le\frac{h}{700^2r}\sum_i|A_{i+1}-A_i| =O(h/r).
\]
A subsequence has no greater total variation: each of its jumps is
bounded by the sum of the original consecutive jumps it skips, and
these skipped ranges are disjoint.
\end{proof}

Although the separators of a stack are disjoint,
a separator may still pass through the interior
of another bowl. This is inconvenient for
rotations: moving a separator does not
automatically preserve the position of a bowl
lying across it.
We therefore split the stack into a constant
number of consecutive blocks in which every
separator avoids the interior of every other
bowl. This lets us enclose individual bowls
between neighboring separators and move
these enclosures as rigid objects.

\begin{lemma}%
\label{lem:rot-strict}
A stack in a mini-container can be partitioned into $O(1)$ consecutive
blocks such that, within each block, no separator meets the interior
of another bowl. Boundary contacts are allowed.
\end{lemma}
\begin{proof}
Partition the reference points by the height intervals
$[\nu h/2,(\nu+1)h/2)$; this gives $O(1)$ consecutive blocks. Fix $i<j$
in one block, so $0\le t_j-t_i<h/2$. In the $(u,d)$-frame of the fine
type, every point of $P_k$ has $d$-coordinate within $1.13w$ of $t_k$
(a point at signed distances $x',\omega'$ along and across the radius
line has $d$-coordinate $x'\sin\beta+\omega'\cos\beta$ with
$|x'|\le r$, $|\omega'|\le w$, $\sin\beta\le\frac{\pi}{25}\frac wr$).
Hence the interiors of $Q_i$ and $Q_j$ lie below height
$t_i+h/2+1.13w<\min(t_i,t_j)+h$, where both pairs of tails are, and
the finite arms are excluded by
\cref{lem:rot-initial-containers}.
\end{proof}

\paragraph{Enclosing every third bowl.}
We now enclose each retained bowl between its neighboring
separators. These bounded enclosures can be moved as rigid objects,
carrying the bowls inside them.

\begin{lemma}%
\label{lem:rot-enclosures}
Apart from at most two end bowls, a locally separated block can be partitioned into
three subsequences. In each subsequence, every bowl lies in a bounded
polygonal region, called its enclosure. Consecutive enclosures are
separated by two of the original separators, and all enclosures lie
in an $O(r)\times O(h)$ rectangle.
\end{lemma}
\begin{proof}
Index the block by $1,\ldots,\ell$ and set aside its first and last
bowls, which are packed individually at cost $O(rh)$. Let $S$ be a
vertical strip of width $O(r)$ containing all bowls and all finite arms,
and for $2\le i\le\ell-1$ put
\[
E_i=S\cap\operatorname{Above}(\sigma_{i-1})\cap\operatorname{Below}(\sigma_{i+1}).
\]
Inside $S$ each separator is a simple arc joining the two walls, so $E_i$
is a bounded polygon of horizontal span $O(r)$ and vertical span $O(h)$.

We claim $Q_i\subseteq E_i$. Its arms lie on $\sigma_i$, which lies
weakly between $\sigma_{i-1}$ and $\sigma_{i+1}$. The rest of $Q_i$ is
the star $P_i$, whose interior is connected and by \cref{lem:rot-strict}
meets neither $\sigma_{i-1}$ nor $\sigma_{i+1}$, so it lies on one side
of each; since $q_i$ is between the two separators, that side is the
right one unless $q_i$ itself lies on $\sigma_{i-1}$ or $\sigma_{i+1}$.
In that degenerate case, $P_i$ contains the two segments
$q_iR_{P_i}$ and $q_iL_{P_i}$ in the directions $(rA_i,h)$ and
$(rB_i,h)$; if $P_i$ were on the wrong side of, say, $\sigma_{i-1}$,
both segments would lie on $\sigma_{i-1}\cap\sigma_i$, forcing
$\sigma_i=\sigma_{i-1}$, and then $Q_i$ would be the bowl with interior
points outside the common triangle, which the tie-break convention
places below $Q_{i-1}$, not above. The same argument applies to
$\sigma_{i+1}$.

Finally, group the indices by residue modulo $3$. Between $E_i$ and
$E_{i+3}$ lie $\sigma_{i+1}$ and $\sigma_{i+2}$, so the enclosures of a
class are pairwise disjoint, separated by two original separators, and
in stack order.
\end{proof}

Each enclosure is now moved as a rigid object, carrying its bowl with
it. Thus feasibility reduces to keeping the enclosures disjoint. At a
step of the procedure, the \emph{prefix} consists of the enclosures
already aligned, and the \emph{suffix} consists of all remaining
enclosures in their original order.

\begin{figure}[t]
\centering
\includegraphics[page=3]{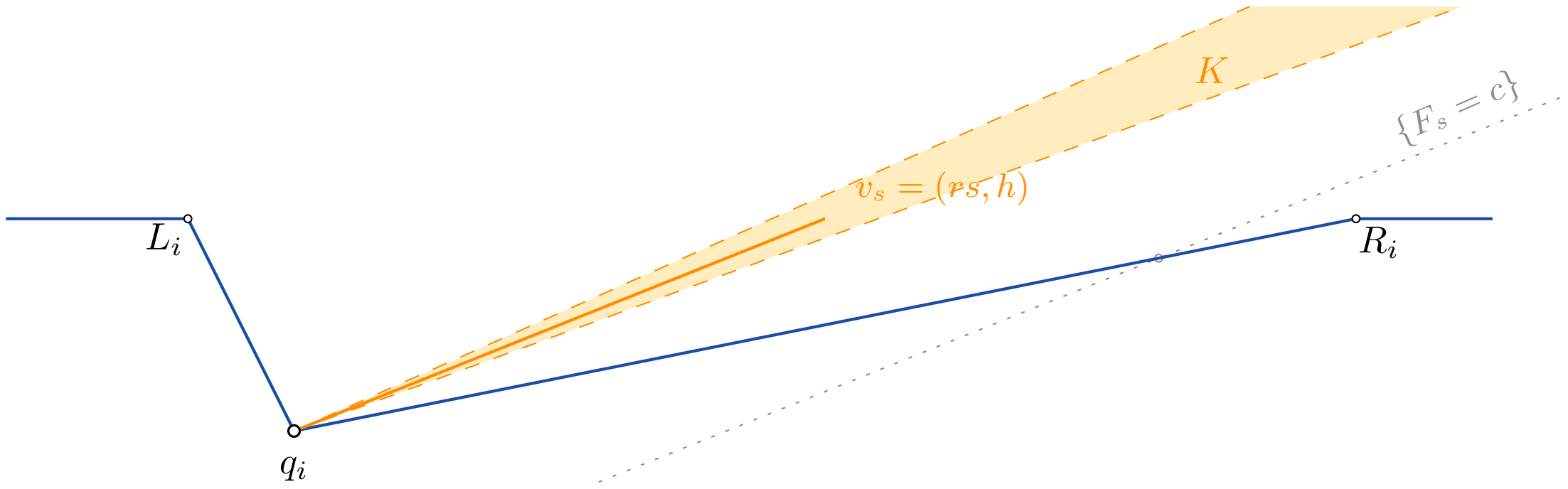}
\caption{The cone $K$ of admissible translation directions for a
separator. Along $\sigma_i$, from the left tail to the right tail, the
functional $F_s$ is strictly increasing, with increments of at least
$100r$ on the two arms; hence every line parallel to $v_s$ meets
$\sigma_i$ exactly once.}
\label{fig:rot-admissible-cone}
\end{figure}

\begin{figure}[t]
\centering
\includegraphics[page=4]{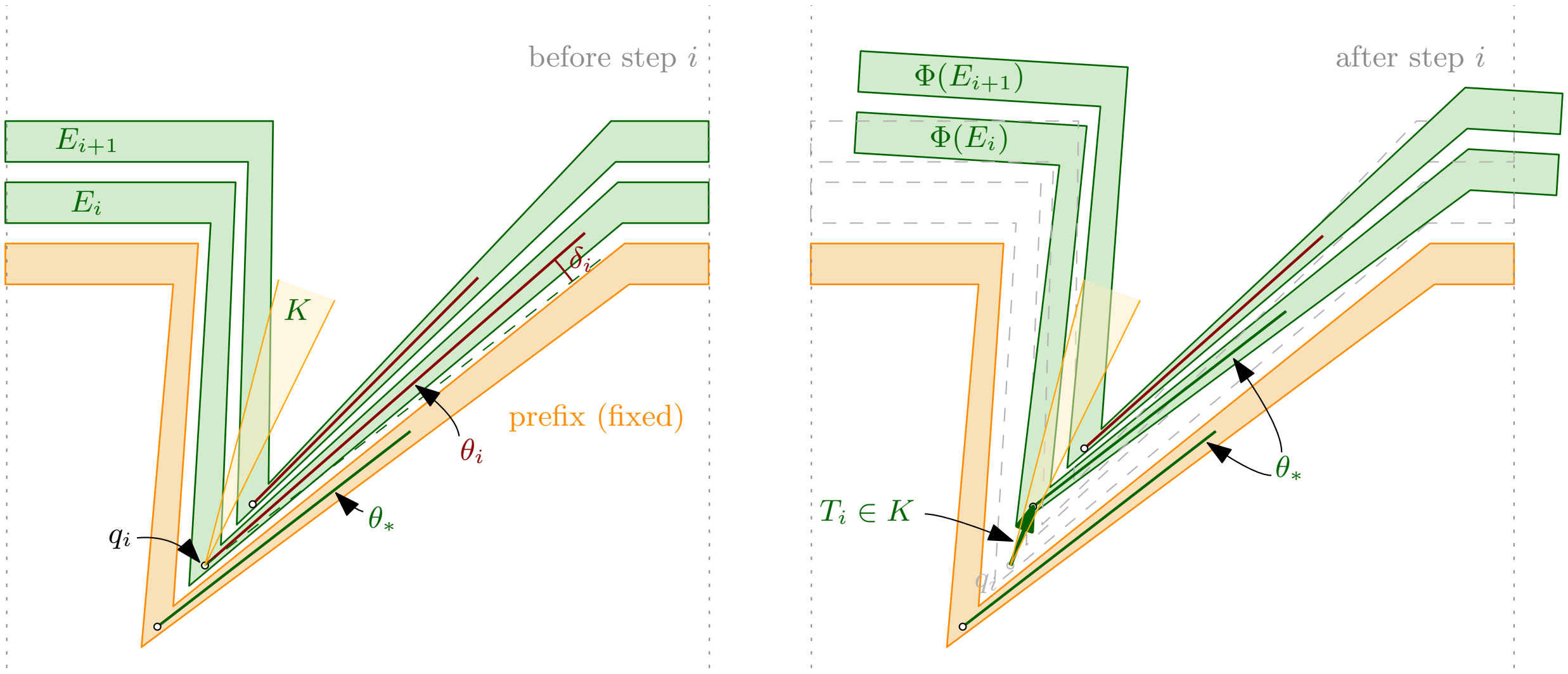}
\caption{One step of the sweep, for three enclosures of a residue class: the prefix (orange, already aligned, radius at $\theta_*$) and the suffix $E_i,E_{i+1}$ (green). $\delta_i=\theta_*-\theta_i$ is the angle between the arm of bowl~$i$ and the common direction. Light green is $E_i+K$ and $E_{i+1}+K$. The step rotates the suffix about $q_i$ through $\delta_i$ and translates it by $T_i\in K$, which moves every suffix point by a vector of~$K$. Afterwards bowl~$i$ points at $\theta_*$ and the prefix is unchanged.}
\label{fig:rot-sweep}
\end{figure}

\begin{lemma}%
\label{lem:rot-sweep}
Let $E_1,\ldots,E_m$ be one subsequence of enclosures from
\cref{lem:rot-enclosures}, relabeled in their inherited order.
For each $i$, let $q_i$ be the reference point of its bowl and let
$\theta_i$ be its original radius angle, as above. Let $I_0$ be a
common interval of length at most $\varepsilon h/r$ containing these
angles. For any prescribed $\theta_*\in I_0$, the enclosures can be
repacked in an $O(r)\times O(h)$ rectangle with disjoint interiors,
in their original order, and with all radius vectors pointing in
direction $\theta_*$.
\end{lemma}
\begin{proof}
Write $H=h/r$; then $0<H<200$. By \cref{eq:rot-arm-bounds},
$700\le A_i\le3200$ and $-400\le B_i\le300$. Fix $I=[400,600]$ with
midpoint $s_0=500$, and $J=(450,550)$. For $s\in I$ let $v_s=(rs,h)$ and
$F_s(x,y)=x-(s/H)y$, so that the lines parallel to $v_s$ are the level
sets of $F_s$. Along a separator, from the left tail to the right tail,
$F_s$ increases by $r(s-B_i)\ge100r$ on the left arm, by
$r(A_i-s)\ge100r$ on the right arm, and strictly along both tails
(\cref{fig:rot-admissible-cone}). Hence every line parallel to $v_s$
meets each separator exactly once, and it meets the separators in stack
order. A common rotation through an angle $|\gamma|\le\varepsilon H$
changes $F_s(z)$ by at most $\sqrt{1+(s/H)^2}\,|\gamma|\,\|z\|
\le\varepsilon(H+s)\|z\|$, which is below $100r$ for an arm
($\|z\|\le3400r$) and below $1$ for a unit tail direction once
$\varepsilon\le10^{-6}$; so the same conclusions hold after any such
rotation. Let
\[
K=\{\lambda(rs,h):\lambda>0,\ s\in J\},
\]
an open cone with $K+K\subseteq K$, since the parameter of a sum is a
weighted average.

\emph{The invariant.} For $i<j$, $p\in E_i$ and $z\in E_j$ we maintain $p-z\notin K.$
It holds for the original family, and after any common translation and
common rotation by at most $\varepsilon H$: a line in a direction of $K$
meets the upper bounding separator of $E_i$ no later than the lower
bounding separator of $E_j$, so it cannot pass through $E_j$ before
$E_i$, as $p-z\in K$ would require. The invariant forbids a common
interior point, since an open disk in $E_i\cap E_j$ contains two points
whose difference is a small vector of $K$.

\emph{The step.} Put $\theta_0=\theta_*$ and
$\delta_i=\theta_{i-1}-\theta_i$ for $i=1,\ldots,m$. At step $i$ the
prefix $E_1,\ldots,E_{i-1}$ is already aligned and fixed; rotate the
suffix $E_i,\ldots,E_m$ about its current reference point $q_i$ through
$\delta_i$, so that bowl~$i$ now points at $\theta_*$, and then
translate the suffix by $T_i=c\tau_i(rs_0,h)$ with
$\tau_i=|\delta_i|/H$ and a universal constant $c$ fixed below
(\cref{fig:rot-sweep}). Since the suffix has only undergone common
rigid motions, its diameter is still that of the original
$O(r)\times O(h)$ rectangle, at most $Lr$, so the rotation displaces
each of its points by at most $Lr|\delta_i|=L\tau_i h$. Writing the
net displacement of a point as
$\tau_i\big(r(cs_0+e_x),\,h(c+e_y)\big)$ with $|e_x|\le LH\le200L$ and
$|e_y|\le L$, its direction parameter $s'=(cs_0+e_x)/(c+e_y)$ satisfies
$|s'-s_0|\le 200L(1+s_0)/(c-200L)$, which is below $50$ for $c$ large.
Thus every point of the suffix moves by a vector of $K$ (if
$\delta_i=0$ nothing moves).

The invariant survives the step. Within the prefix nothing changed.
Within the suffix the motion is common, and its total rotation since the
original configuration is $\theta_*-\theta_i$, of absolute value at most
$\varepsilon H$ as both angles lie in $I_0$. For $p$ in the prefix and
$z$ in the suffix with new position $z'$, we have $z'-z\in K$; if
$p-z'$ were in $K$ (or zero), then $p-z=(p-z')+(z'-z)\in K$, against the
invariant before the step. Hence the prefix and the moved suffix are
disjoint, and after all $m$ steps the enclosures have pairwise disjoint
interiors, in their original order, with every radius at $\theta_*$.

\emph{Size.} By \cref{lem:rot-variation} applied to the subsequence,
$\sum_{i=1}^m|\delta_i|\le|\theta_*-\theta_1|+\sum_{i\ge2}|\delta_i|
=O(h/r)$. The translations total
$\sum_i(T_i)_y=cr\sum_i|\delta_i|=O(h)$ vertically and
$\sum_i(T_i)_x=cs_0(r^2/h)\sum_i|\delta_i|=O(r)$ horizontally, and the
rotations displace any point by at most $Lr\sum_i|\delta_i|=O(h)=O(r)$
in total. So the final configuration lies in an $O(r)\times O(h)$
rectangle.
\end{proof}

\begin{theorem}%
\label{thm:rotation-straightening}
\label{lem:rotation-straightening} %
For a mini-container from \cref{lem:rot-initial-containers}, all its stars
can be rotationally repacked into a rectangle with side lengths $O(r)$
and $O(h)$. Their directed radii then point in one common direction,
and the side of length $O(r)$ is parallel to that direction.
\end{theorem}
\begin{proof}
\Cref{lem:rot-strict} splits the mini-container into $O(1)$ locally
separated blocks. Fix a target angle $\theta_*$ that occurs among the
bowls; it lies in the common angle interval of every block. In each
block apply \cref{lem:rot-enclosures} and then \cref{lem:rot-sweep} to
each of the three classes with this $\theta_*$, and rotate each of the
$O(1)$ end bowls to $\theta_*$ individually. This yields $O(1)$
rectangles of span $O(r)$ along $u$ and $O(h)$ along $d$, each holding
bowls whose radii all point in direction $\theta_*$; the rectangles are
placed independently.

Since $\theta_*$ is the radius angle of some bowl, $A_i\ge700$ in
\cref{eq:rot-arm-bounds} gives $|\sin\theta_*|\le h/(700r)$. A rectangle
with spans $W=O(r)$ along $u$ and $V=O(h)$ along $d$ therefore has span
$W|\cos\theta_*|+V|\sin\theta_*|=O(r)+O(h^2/r)=O(r)$ in direction
$\theta_*$ and span $W|\sin\theta_*|+V|\cos\theta_*|=O(h)$
perpendicular to it, using $h<200r$. Rebox each packing in a rectangle
aligned with $\theta_*$ of these dimensions, stack the $O(1)$ boxes
perpendicular to $\theta_*$, and replace each bowl by the star it
contains. The result is a rectangle of side lengths $O(r)$ along and
$O(h)$ across the common radius direction, as claimed.
\end{proof}

We conclude with the main theorem from this section, which enables our algorithm for the rotational case.

\begin{theorem}%
\label{thm:structural-with-rotation}
For one size group $\mathcal G$, every rotational packing into a
convex container $C$ yields, for some integer $m\ge1$, a rotational
packing into rectangles $C_1,\ldots,C_m$ with the following properties:
\begin{enumerate}
\item each rectangle has dimensions $O(r)\times O(w)$;
\item all directed distinguished radii point horizontally to the right
in every rectangle;
\item after enlarging the rectangles to common dimensions
$\kappa r\times\kappa w$ for a fixed universal constant $\kappa$, their total area
is $O(|C|)$.
\end{enumerate}
Consequently, the entire group can be packed into a single convex
container of area $O(|C|)$ with all directed radii horizontal.
\end{theorem}
\begin{proof}
\Cref{lem:rot-initial-containers} gives $N$ mini-containers of common
dimensions $\kappa r\times\kappa h$ with $N\kappa^2rh=O(|C|)$. By
\cref{thm:rotation-straightening}, the stars of each mini-container can
be repacked into a rectangle with side $C_1r$ along their common radius
direction and side $C_1h$ across it, for a universal constant $C_1$.
Rotate each such packing as a whole until its radii point to the right;
its rectangle is then axis-parallel, of width $C_1r$ and height $C_1h$.
Enlarge all $N$ rectangles to $C_1r\times C_1h$ and stack them
vertically. The result has width $C_1r$, height $NC_1h$ and area
$NC_1^2rh=O(|C|)$; as $h=100w$, this is the claim.
\end{proof}

\begin{corollary}%
\label{cor:rotation-algorithm-interface}
Let $\widehat{\mathcal G}$ be obtained by rotating each input star so
that its radius vector points in the positive horizontal direction.
Write $\opt_{\mathrm{trans}}$ and
$\opt_{\mathrm{rot}}$ for the minimum area of a convex
container when translations alone, or translations and rotations,
respectively, are allowed. Then
\[
\opt_{\mathrm{rot}}(\mathcal G) \le \opt_{\mathrm{trans}}(\widehat{\mathcal G}) \le \kappa'\opt_{\mathrm{rot}}(\mathcal G)
\]
for a universal constant $\kappa'$. The common per-group translational
algorithm from the preceding section consequently applies to
$\widehat{\mathcal G}$ with only constant additional loss. 
\end{corollary}
\begin{proof}
$\opt_{\mathrm{rot}}(\mathcal G) \le \opt_{\mathrm{trans}}(\widehat{\mathcal G})$ holds as any translational packing of $\widehat{\mathcal{G}}$ is a rotational packing of $\mathcal{G}$. 
Apply \cref{thm:structural-with-rotation} to rotational packings
approaching the infimum. In the resulting packings every star is a
translate of its copy in $\widehat{\mathcal G}$, which gives the
displayed bound. The algorithm then runs the translational pipeline of
\cref{lem:star-to-bowl,thm:box-star-packing,lem:bowl-TSP} on
$\widehat{\mathcal G}$; the enclosures of this section serve only the
existence proof.
\end{proof}

We now obtain the rotational analogue of
\cref{lem:total-minicontainer-area}.

\begin{lemma}
\label{lem:total-minicontainer-area-rot}
For a set of stars $\mathcal{P}$, in polynomial time, we can compute a
rotation of each star and a packing of the rotated stars into a set of
convex mini-containers of total area
$O(\log n)\cdot\opt_{\mathrm{rot}}(\mathcal P)$.
Every mini-container $B$ satisfies
\[
   \Span_z(B)=O(\Span_z(P))
\]
for every direction $z$ and every star $P$ that $B$ contains, where
$P$ is taken in its rotated position.
\end{lemma}
\begin{proof}
Rotate each star so that its distinguished radius points horizontally to
 the right, and let $\widehat{\mathcal P}$ denote the resulting family.
Rotations preserve $r_P$ and $w_P$, so the size groups do not change.

As in \cref{lem:total-minicontainer-area}, pack each star satisfying
$r_P<r_{\max}/n$ or $w_P<w_{\max}/n$ individually into the convex
hull of its rotated copy. These hulls have total area
$O(r_{\max}w_{\max})=O(\opt_{\mathrm{rot}}(\mathcal P))$ by
\cref{lem:rmax-wmax}; taking a convex hull does not change any span.

For every remaining size group $\mathcal G$, run
\cref{cor:star-size-class-algorithm} on $\widehat{\mathcal G}$.
The resulting mini-containers have total area
$O(\opt_{\mathrm{trans}}(\widehat{\mathcal G}))$;
indeed, the bound of that corollary is no larger than a constant
multiple of the minimum area of a single convex container.
Their span guarantee is precisely \cref{lem:star-mini-container-spans}.

It remains to bound the sum over the size groups.
Consider an optimal rotational packing of $\mathcal P$, and let
$\mathcal P^*$ denote the stars with the orientations used in this
packing.
Write $\mathcal G^*$ for the corresponding stars of $\mathcal G$.
By \cref{cor:rotation-algorithm-interface},
\[
  \opt_{\mathrm{trans}}(\widehat{\mathcal G})
  =O(\opt_{\mathrm{rot}}(\mathcal G))
  =O(\opt(\mathcal G^*)).
\]
Here the second bound holds because a translational packing of
$\mathcal G^*$ is also a rotational packing of $\mathcal G$.

The proof of \cref{lem:similar-size} in fact gives the stronger
single-container estimate
\[
  \sum_{\mathcal G}\opt(\mathcal G^*)
  =O(\log n)\cdot\opt(\mathcal P^*).
\]
To see this directly, for each class $\mathcal C_{i,j}$ considered in
that proof, \cref{eqn:IG-IC,eqn:IC,eqn:IG} yield
$\sum_{\mathcal G\in\mathcal C_{i,j}}\opt(\mathcal G^*)
=O(\opt(\mathcal C_{i,j}^*))$.
There are $O(\log n)$ such classes for the remaining stars, and
$\opt(\mathcal C_{i,j}^*)\le\opt(\mathcal P^*)$ for each one.
Consequently the total area of the computed mini-containers is
\[
  O\!\left(\sum_{\mathcal G}
     \opt_{\mathrm{trans}}(\widehat{\mathcal G})\right)
  =O(\log n)\cdot\opt(\mathcal P^*)
  =O(\log n)\cdot\opt_{\mathrm{rot}}(\mathcal P).
\]
Adding the singleton mini-containers proves the claim.
\end{proof}

The rotational part of \cref{thm:main} now follows by applying
\cref{thm:convex-packing} to these mini-containers. All stars, including
the singleton ones, have horizontal distinguished radii. Thus each
rotated star has horizontal span $O(r_{\max})$ and vertical span
$O(w_{\max})$, and \cref{lem:total-minicontainer-area-rot} gives the
same bounds for every mini-container. In particular,
$\ell_{\max}h_{\max}=O(r_{\max}w_{\max})
=O(\opt_{\mathrm{rot}}(\mathcal P))$ by
\cref{lem:rmax-wmax}, whose proof is rotation invariant. The convex
packing algorithm therefore returns a rectangle of area
$O(\log n)\cdot\opt_{\mathrm{rot}}(\mathcal P)$.

\section{APX-hardness}\label{sec:apx-hardness}

In this section, we show that star packing with constant-complexity pieces is APX hard, with or without rotations.

\paragraph{The problem MAX-VERTEX-DISJOINT-PATH-COVER. }
We will reduce from a problem called MAX-VERTEX-DISJOINT-PATH-COVER. In the problem MAX-VERTEX-DISJOINT-PATH-COVER$(d)$, we are given a directed graph $G=(V, E)$ where the in- and out-degree of each vertex is at most $d$. We want to determine the largest number of edges in a \emph{path cover}. A path cover is a set of disjoint, directed paths in $G$ (possibly including single-vertex paths) such that every vertex of $G$ is contained in a path.

When $G$ has a Hamiltonian path, then there is a path cover with $|V|-1$ edges, which is the most possible. So we can view this problem as an approximation version of HAMILTONIAN-PATH. If $G$ has no edges, then the largest path cover is of size $0$.

Bosboom, Demaine, Demaine, Hesterberg, Manurangsi and Yodpinyanee \cite{DBLP:journals/jip/BosboomDDHMY17} show that, for some small $\epsilon>0$, MAX-VERTEX-DISJOINT-PATH-COVER is NP-hard to approximate to within a $1-\epsilon$ factor:

\begin{lemma}(\cite{DBLP:journals/jip/BosboomDDHMY17}, Theorem 2)\label{lem:apx:disjpathcover}
For any $\epsilon<\frac{1}{1396640}$, the following is true: for any language $\lambda$ in NP, there is a polynomial-time function that takes as input an instance $\varphi$ of $\lambda$ and outputs an instance $\psi$ of MAX-VERTEX-DISJOINT-PATH-COVER$(2)$. If $\varphi$ is a ``yes'' instance, then $\psi$ has a path cover with $|V|-1$ edges. If $\varphi$ is a ``no'' instance, then the largest path cover in $\psi$ has at most $(1-\epsilon)(|V|-1)$.
\end{lemma}

\paragraph{Reduction from MAX-DISJOINT-PATH-COVER to star packing. }
Our reduction encodes a directed graph using star-shaped polygons.
Each vertex is represented by a polygon with a narrow spike and
up to two slits. The spike of one polygon fits into a slit of
another precisely when the corresponding directed edge exists.
Thus, a directed path can be represented by a tightly nested
stack of polygons.

The main difficulty is to ensure that an arbitrary packing
cannot avoid this encoding. We introduce an additional
container polygon and show that every sufficiently good
packing must place the vertex polygons inside it, in a
common nesting order and with nearly fixed orientations.
Consecutive polygons corresponding to a missing edge
must then be separated by an additional vertical distance.
Consequently, a graph with no nearly complete path cover
requires a container of strictly larger area. 

\begin{figure}
\centering
\includegraphics[page=1]{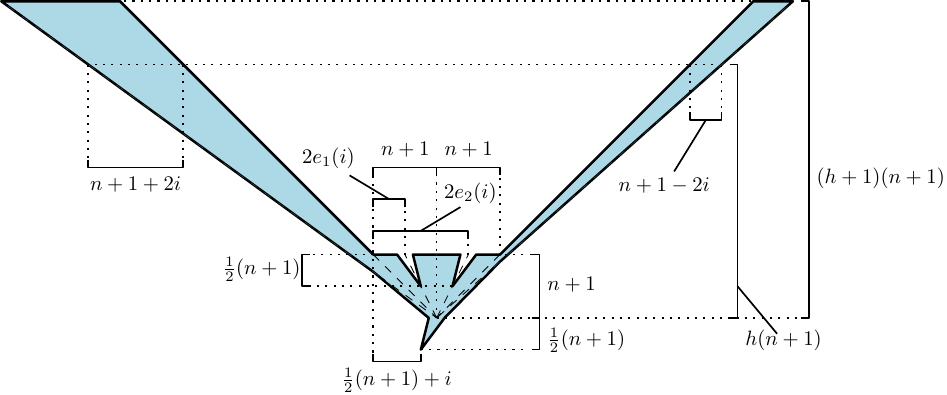}
\caption{ The piece $S_i$ has a spike, with angle determined by $i$, and up to $2$ slits corresponding the edges outgoing from $i$. If $i$ has fewer outgoing edges, then $S_i$ has fewer slits. }
\label{fig:apx:sipiece1}
\end{figure}

\begin{figure}
\centering
\includegraphics[page=2]{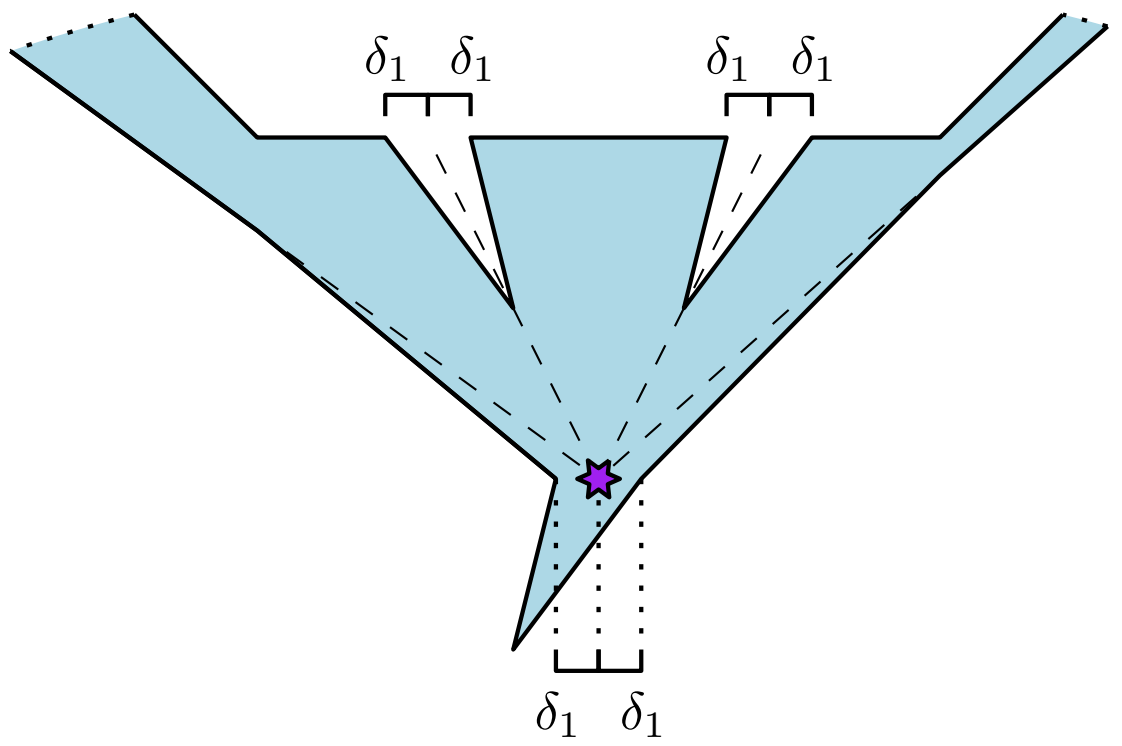}
\caption{ Close up of the core of $S_i$. The width of the spike and the slits is determined by $\delta_1$. The star point for $S_i$ is illustrated. }
\label{fig:apx:sipiece2}
\end{figure}

Now we give the formal reduction. 
We are given a graph $G=(V, E)$, where each vertex has in- and out-degree at most $2$. We want to construct a set of star shaped pieces of constant complexity. Label the vertices in $V$ by $1, \dots, n$. Let $e_1(i)$ and $e_2(i)$ be the outgoing edges from vertex $i$ (if they exist), with $e_1(i)<e_2(i)$. For each vertex $i$, we create one star $S_i$ with at most $17$ vertices. This is illustrated in \Cref{fig:apx:sipiece1,fig:apx:sipiece2}. In addition to $G$, the stars depend on two parameters $\delta_1<1$ and $h>1$, which will be determined later.

\begin{figure}
\centering
\includegraphics[page=3]{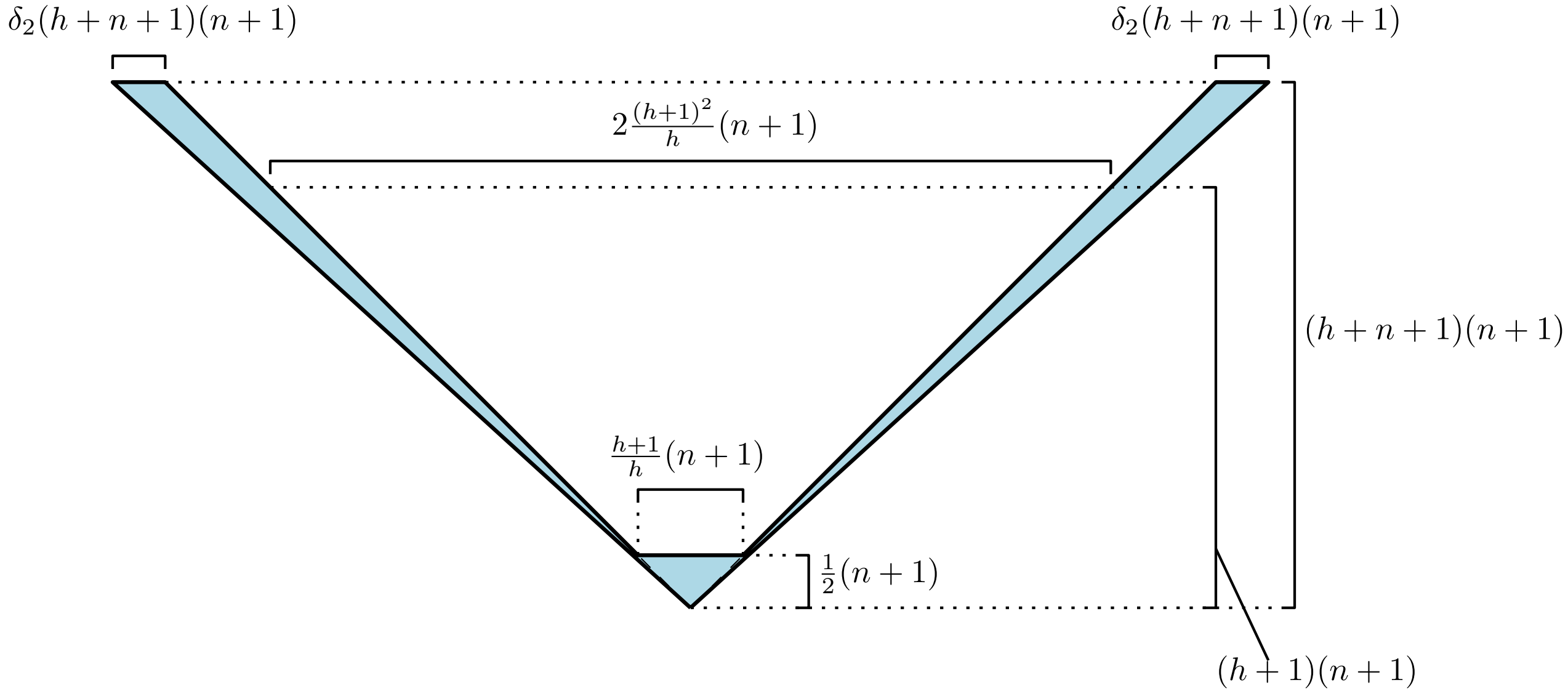}
\caption{ The container piece $C$. The container is symmetric under reflection across a vertical line. }
\label{fig:apx:containerpiece}
\end{figure}

We also make one outer container called $C$, which is the $7$-gon illustrated in \Cref{fig:apx:containerpiece}. This depends on a parameter $\delta_2$, which we will set $\delta_2=\frac{1}{10}$. 

The construction has two different scales. The long arms
control the overall nesting of the polygons, while the
much smaller spikes and slits encode the edges of the
graph. The parameter $\delta_1$ controls the angular
separation between different spike and slit positions.
The parameter $h$ determines the length of the arms
and makes rotations or incorrect nesting expensive.
Finally, the outer polygon $C$ forces the vertex
polygons to occupy a common stack in any packing
whose area is close to $|\operatorname{conv}(C)|$.

\begin{lemma}\label{lem:apx:completeness}
If $G$ has a Hamiltonian path (equivalently, a path cover with $|V|-1$ edges), then all the $S_i$ can be packed inside the convex hull of $C$, even if rotations are not allowed.
\end{lemma}

\begin{proof}
An example is shown in \Cref{fig:apx:goodstack}. If there is an edge from $i$ to $j$, then the spike on $S_j$ fits into one of the slits in $S_i$. So a Hamiltonian path in $G$ can be used to construct a stack of the $S_i$, where each piece has height $n+1$ above the one below it. The total height of the stack, from the bottom of the spike on the lowest $S_i$ to the top of the highest, is $h(n+1)+n(n+1)+\frac12(n+1)$. The slopes of the inner walls of $C$ are such that one of the pieces $S_i$ fits exactly inside $C$, with the base of the spike of $S_i$ exactly $\frac12(n+1)$ units above the base of $C$.

\begin{figure}
\centering
\includegraphics[page=4]{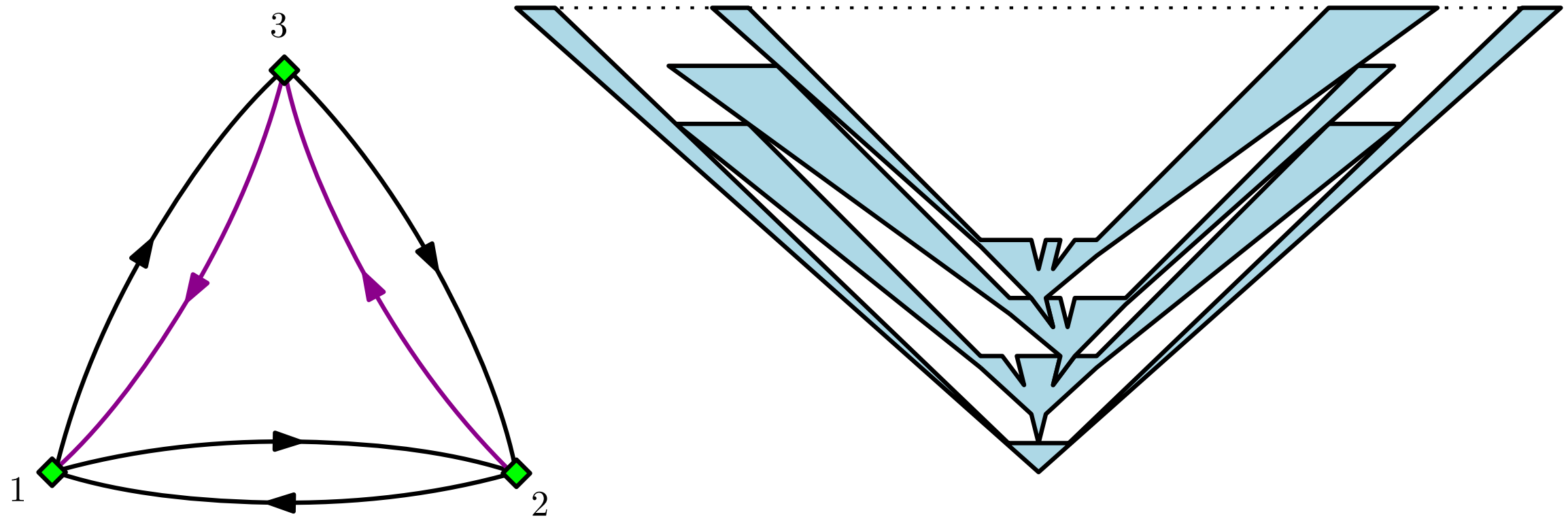}
\caption{ Left: a graph $G$ with max in/out-degree $2$, and a marked Hamiltonian path. Right: the corresponding $S_i$ stacking inside the container piece. }
\label{fig:apx:goodstack}
\end{figure}
\end{proof}

\begin{figure}
\centering
\includegraphics[page=5]{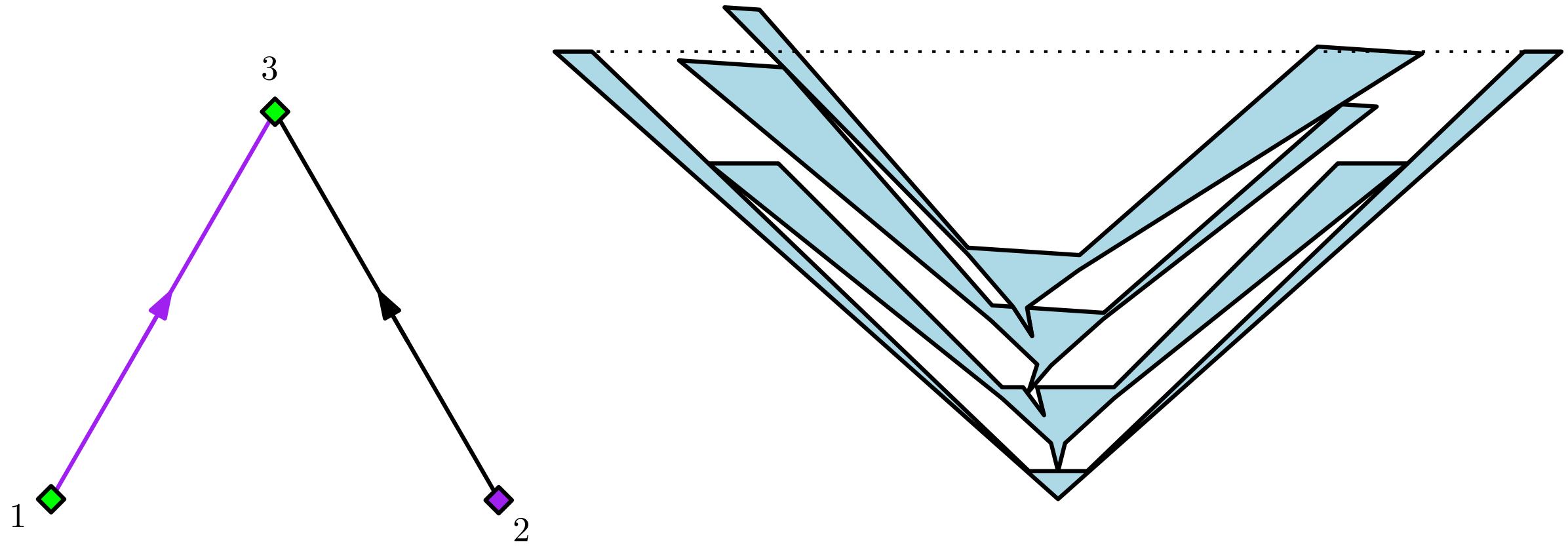}
\caption{ Left: a graph $G$ where the optimal disjoint path cover has $|V|-2$ edges. The path cover consists of the edge from $1$ to $3$ and a $0$-length path containing vertex $2$. Right: the pieces $C$ and $S_i$ that we construct packed according to this path cover. Piece $S_1$ is stacked inside $S_2$, but there is no edge from $2$ to $1$, so the spike of $S_1$ does not fit into one of the slits in $S_2$. For sufficiently chosen $\delta$ and $h$, we will show that that stacking a pair of pieces like this requires us to use extra height proportional to the length of a spike. }
\label{fig:apx:badstack}
\end{figure}

We now want to show that, if the best disjoint path cover in $G$ has $(1-\epsilon)(|V|-1)$ edges, then any packing of the $C$ and $S_i$ has cost at least $1+\Omega(\epsilon)$ times the cost of the packing described above. The idea is illustrated in \Cref{fig:apx:badstack}. 

We now fix some packing, and orient it so that $C$ is oriented as in \Cref{fig:apx:containerpiece}. The area of the convex hull of $C$ is:

\[A_0=\left(\frac{h+1}{h}+\delta_2\right)(h+n+1)^2(n+1)^2\]

We write $h=qn$, where $q\ge 4000$ is some constant to be determined later. We suppose $n\ge 100$, so we have $q^2n^4\le A_0\le \left(\frac{q+1}{q}\right)^3\left(\frac{n+1}{n}\right)^4(1+\delta_2)q^2n^4\le \frac{3}{2}q^2n^4$. We want to show that, if the packing fits into a convex container of area larger than $A_0$ by at most small small constant factor, then we can find a good disjoint path cover for $G$. We start by proving a few lemmas.

Note that allowing rotations could, in principle, let a spike enter
a slit that does not encode an edge of the graph.
Our first step rules out this possibility for large
rotations. The long arms of each $S_i$ are almost as
tall as the outer polygon $C$. Rotating $S_i$
substantially forces part of it to extend beyond $C$,
increasing the area of their convex hull. Hence a
packing of sufficiently small area must keep every
$S_i$ close to its intended orientation.

\begin{lemma}\label{lem:apx:similarangles}
If the convex hull of the packing has area less than $A_0+\frac12qn^4$, then no $S_i$ is rotated by more than $\theta\ge \frac{6}{q}$ (radians, in either direction) compared to the orientation in \Cref{fig:apx:sipiece1}.
\end{lemma}

\begin{proof}
Suppose $S_i$ is rotated by $\theta\ge \frac{6}{q}$ compared to the orientation in \Cref{fig:apx:sipiece1}. If $\theta\le \frac{\pi}{4}$ then there are points in $S_i$ that have $y$ coordinates differing by at least $(h+1)(n+1)(\cos(\theta)+\sin(\theta))\ge (h+1)(n+1)(1+\frac{3}{q})$, as illustrated in \Cref{fig:apx:rotations1} (top). 

When $\frac{\pi}{4}\le \theta\le \frac{3\pi}{4}$, then \Cref{fig:apx:rotations1} (bottoms) shows that there are a pair of points with vertical distance $2(h+1)(n+1)\sin(\theta)\ge (h+1)(n+1)(1+\frac{1}{10})$.

If $S_i$ is rotated by $\frac{3\pi}{4}\le \theta\le \pi-\frac{6}{q}$, then we can flip the picture and get the same bounds on the height difference, only depending on $\theta-\pi$ instead of $\theta$. 

\begin{figure}
\centering
\includegraphics[page=6]{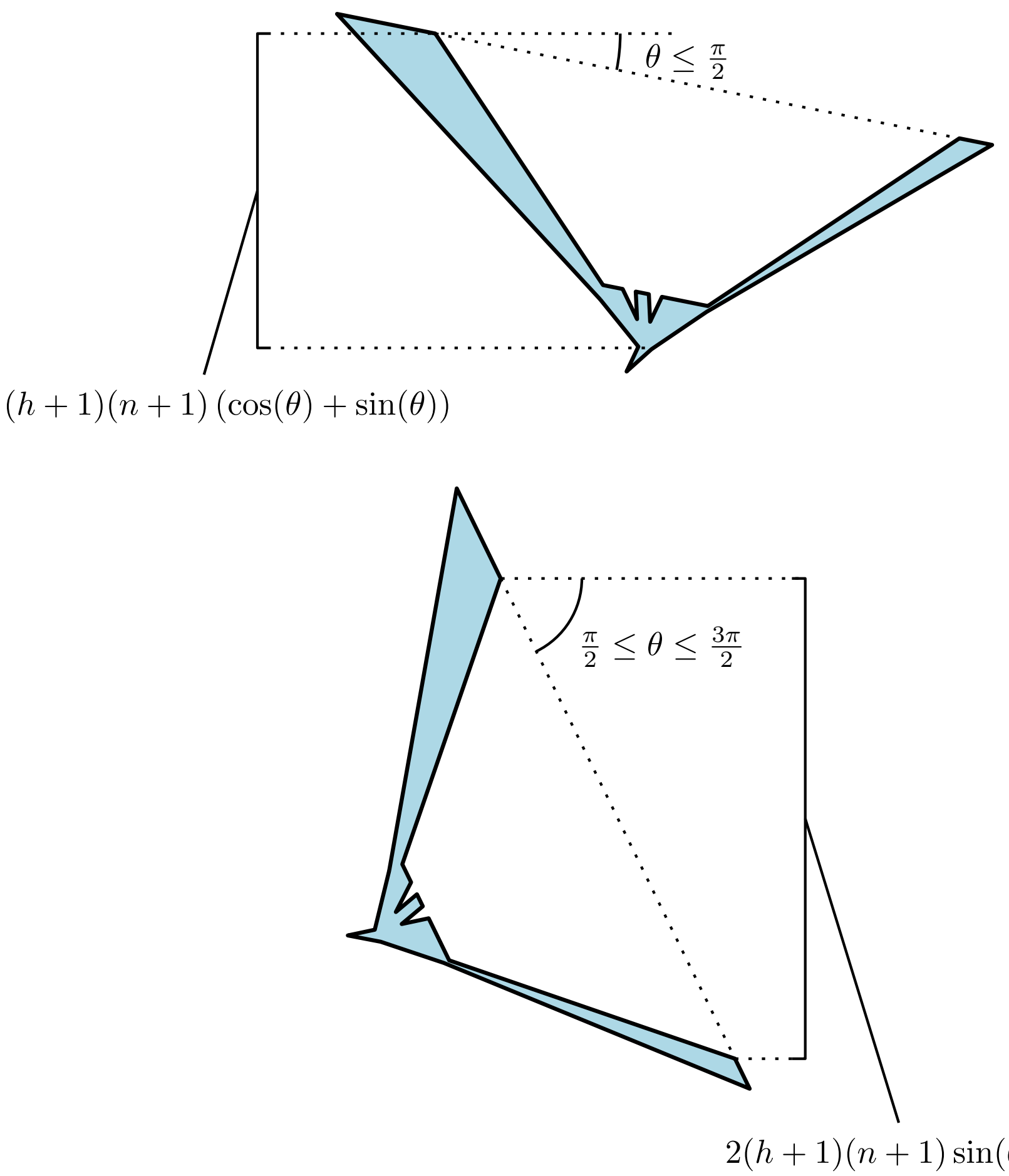}
\caption{ Finding pairs of points with large vertical distance when $S_i$ is rotated. }
\label{fig:apx:rotations1}
\end{figure}

Now $C$ has height $(h+n+1)(n+1)=((q+1)n+1)(n+1)\le (1+\frac{2}{q})(n+1)(h+1)$. So if $S_i$ is rotated by more than $\frac{6}{q}$ but less than $\pi-\frac{6}{q}$, then it sticks out above or below $C$ by at least $\kappa=\frac{1}{q}(h+1)(n+1)$. This means that one of the $3$ types of triangles shown in \Cref{fig:apx:rotations2} is contained in the convex hull of $S_i$ and $C$. Such a triangle has area at least $\frac{1}{2q}(h+1)^2(n+1)^2\ge \frac{1}{2}qn^4$.

\begin{figure}
\centering
\includegraphics[page=7]{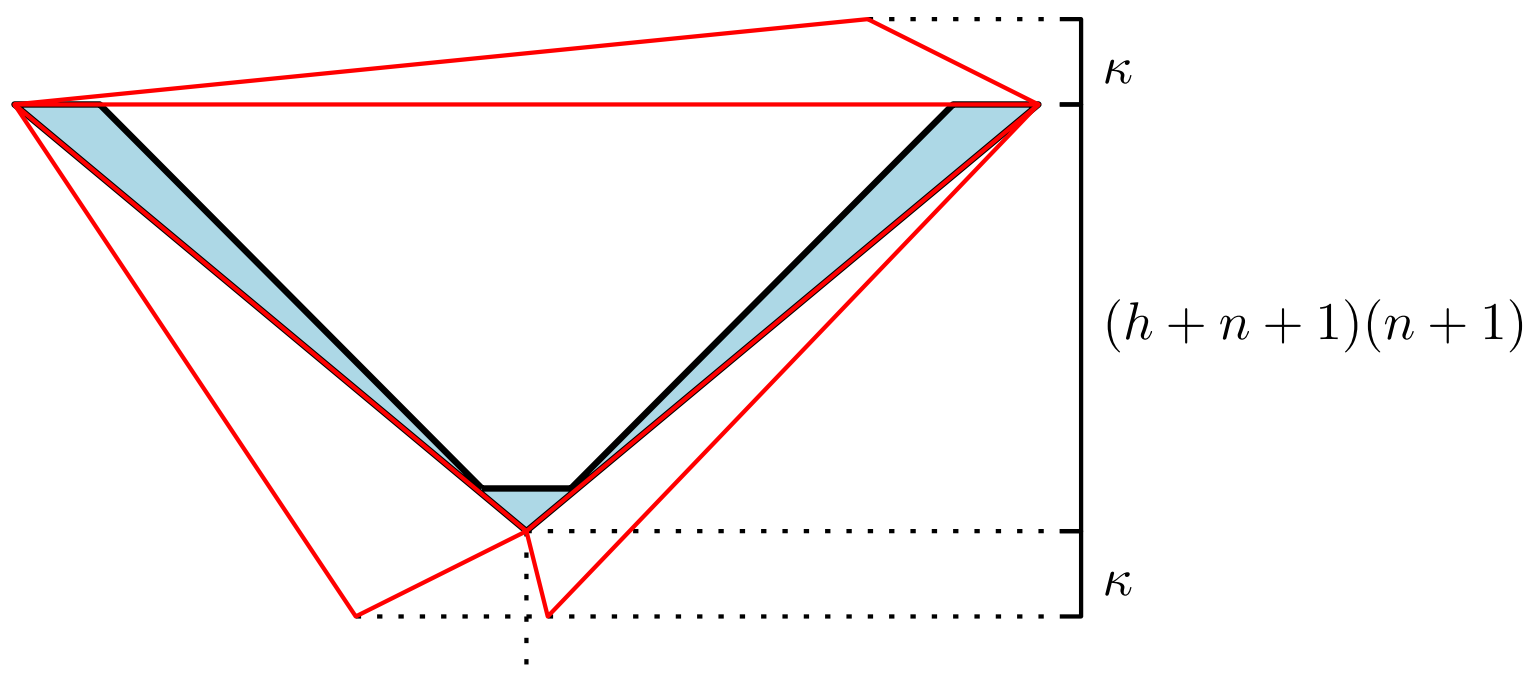}
\caption{ If two points on $S_i$ have vertical difference $\ge (1+\frac{1}{1000})(h+1)(n+1)$, then there is either a point above the top of $C$ by at least $\kappa$, below the bottom of $C$ by at least $\kappa$ and left of the middle of $C$, or below the bottom of $C$ by at least $\kappa$ and right of the middle of $C$. So the convex hull of $S_i$ and $C$ contains one of the triangles shown. }
\label{fig:apx:rotations2}
\end{figure}

The remaining case is when $S_i$ is rotated by close to $\pi$ relative to $C$. The only cases to analyze are the ones shown in \Cref{fig:apx:rotations3}, which clearly exceed the allowed area.

\begin{figure}
\centering
\includegraphics[page=8]{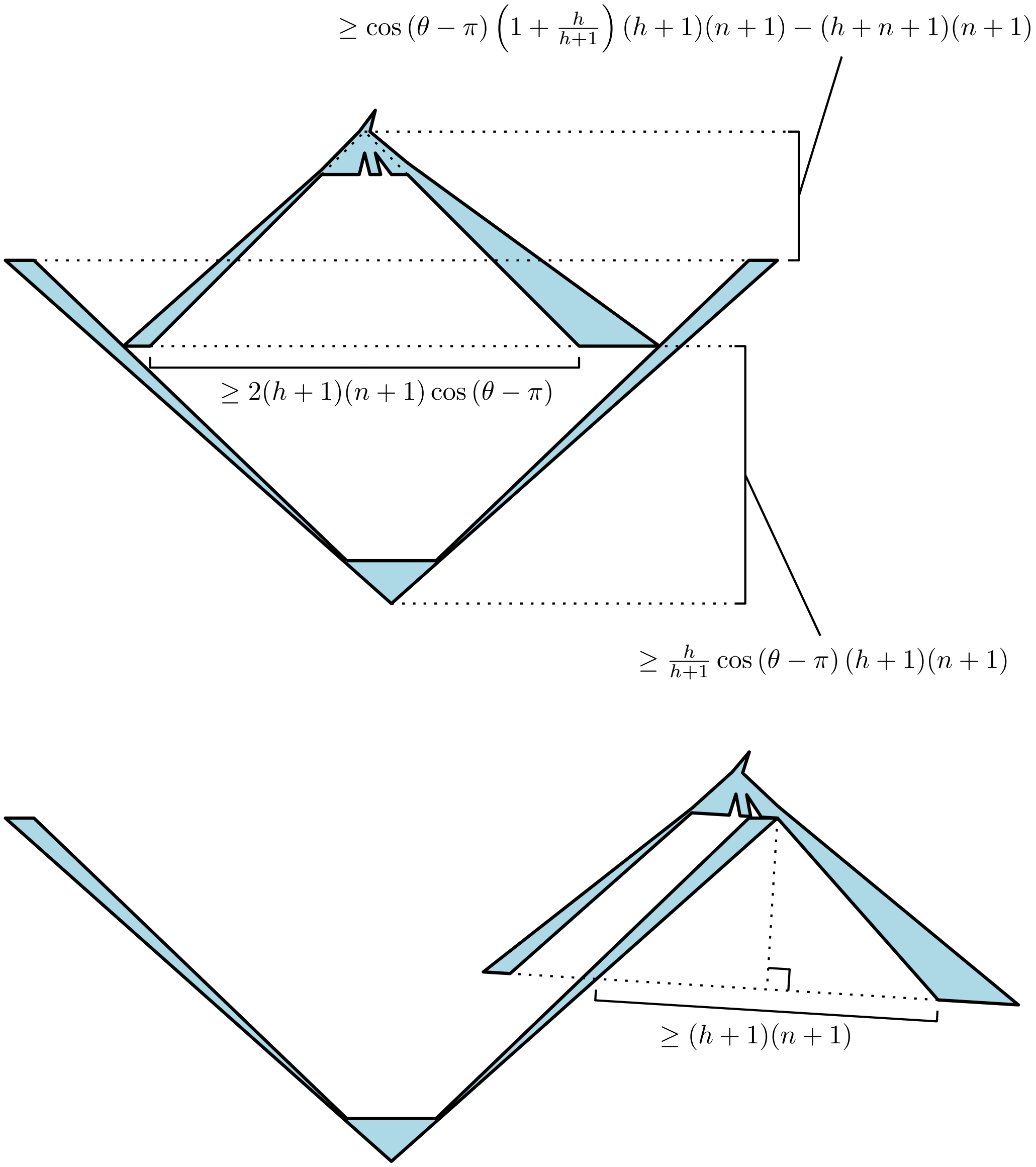}
\caption{ If $S_i$ is close to being completely upside-down, then $S_i$ and $C$ have a convex hull with area much larger than $A_0$. }
\label{fig:apx:rotations3}
\end{figure}
\end{proof}

Each piece has a \emph{room}, which is the large concave region of that piece.
We say that a piece $X$ is stacked inside of $Y$ if the star point of $X$ is inside of the room of $Y$.

\begin{lemma}\label{lem:apx:sinsideC}
If the convex hull of the packing has area less than $A_0+\frac{1}{2}qn^4$, then all of the $S_i$ are stacked inside of $C$. 
\end{lemma}

\begin{proof}
It is sufficient to consider a single piece $S_i$ and $C$. Suppose that $S_i$ is not stacked inside $C$. Since \Cref{lem:apx:similarangles} applies, the only way for the convex hulls of $S_i$ and $C$ to overlap significantly is if $C$ is stacked inside of $S_i$, as in \Cref{fig:apx:sioutsideC1}.

\begin{figure}
\centering
\includegraphics[page=9]{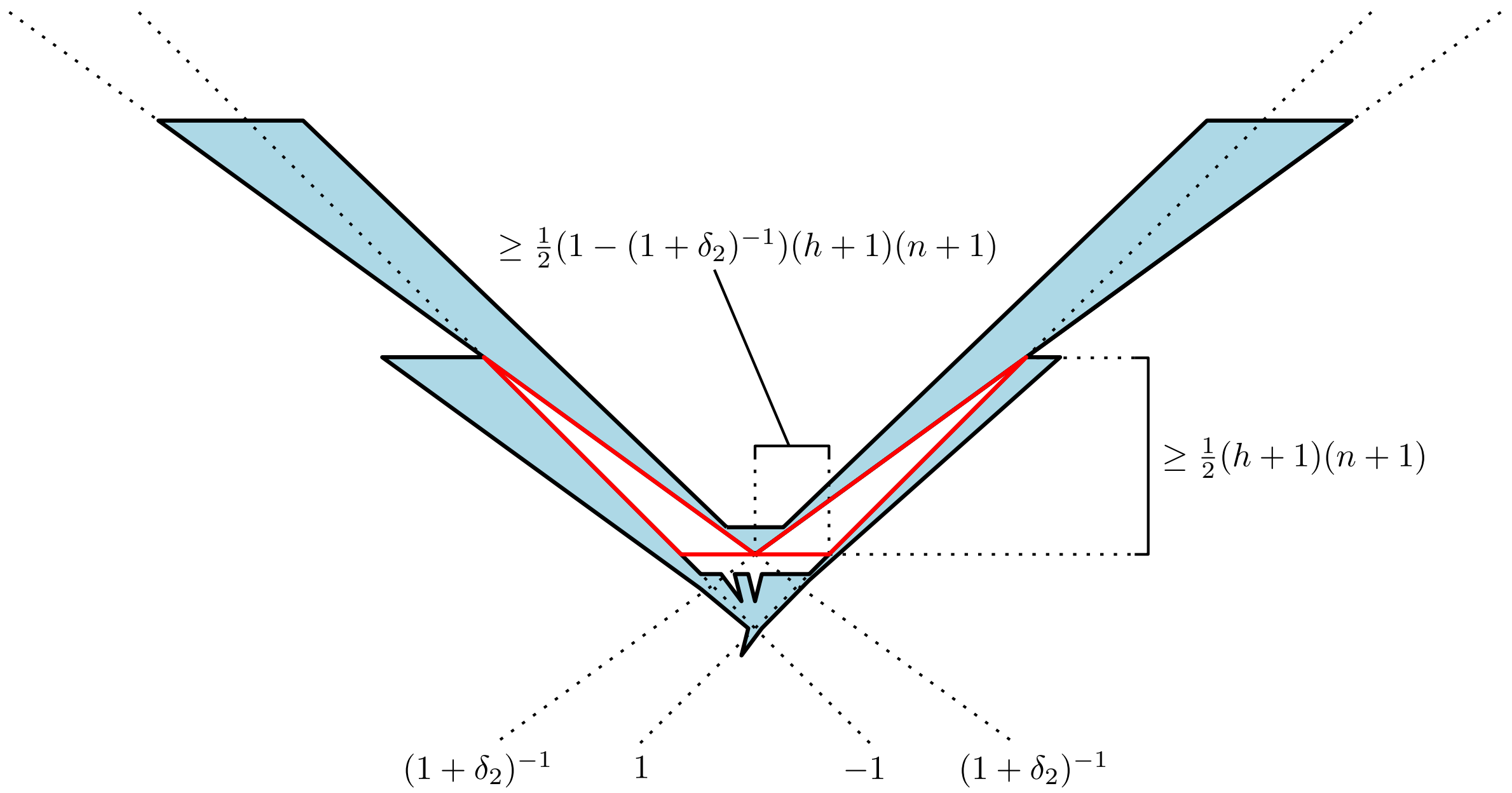}
\caption{ Trying to stack $S_i$ outside of $C$. This incurs an extra cost depending on $\delta_2$. }
\label{fig:apx:sioutsideC1}
\end{figure}

We translate $S_i$ so that the top left and right vertices of the room of $S_i$ touch the outer walls of $C$, as in the figure. The inner walls of $S_i$ have slopes $\frac12$ and $-\frac12$, while the outer walls of $C$ have slopes $\frac12(1+\delta_2)^{-1}$ and $-\frac12(1+\delta_2)^{-1}$. So out of the two red triangles marked in the figure, one of them has a slope difference of at least $(1-(1+\delta_2)^{-1})\ge \frac{1}{2}\delta_2$ between the two long edges. The height of this triangle is close to $(h+1)(n+1)$ (for simplicity, say at least $\frac12(h+1)(n+1)$), so it has area at least:

\[\frac{1}{8}\delta_2(h+1)^2(n+1)^2\ge \frac{1}{8}\delta_2q^2n^4\]

Since $\delta_2\ge \frac{8}{q}$, this is at least $\frac12qn^4$, as required.
\end{proof}

We can now show that the pieces form a stack.

\begin{lemma}\label{lem:apx:formsstack}
If the convex hull of the packing has area at most $A_0+\frac12qn^4$, then there is an ordering $t(1), \dots, t(n)$ of $1, \dots, n$ such that $S_{t(j)}$ is stacked inside $S_{t(i)}$ for every $j>i$.  
\end{lemma}

\begin{proof}
\Cref{lem:apx:similarangles,lem:apx:sinsideC} apply. We create a directed graph $\Lambda$ with vertices $1, \dots, n$ with a directed edge from $i$ to $j$ if $S_j$ is stacked inside $S_i$. The orientations of the $S_i$ are close to the ones shown in \Cref{fig:apx:sipiece1}, so if $S_j$ is stacked inside $S_i$ then the star point of $S_j$ must be above the star point of $S_i$.
It follows that $\Lambda$ is acyclic. To prove the claim, we just need to show that $\Lambda$ is complete.

Consider two pieces $S_i$ and $S_j$. If neither is stacked inside the other, then the two triangles shown in \Cref{fig:apx:stackgraphcomplete} do not overlap. Each of these triangles has area $(h+1)^2(n+1)^2$. Since $A_0\le \frac32q^2n^4$, this would mean that the convex hull of the packing exceeds $A_0+\frac12q^2n^4$.

\begin{figure}
\centering
\includegraphics[page=10]{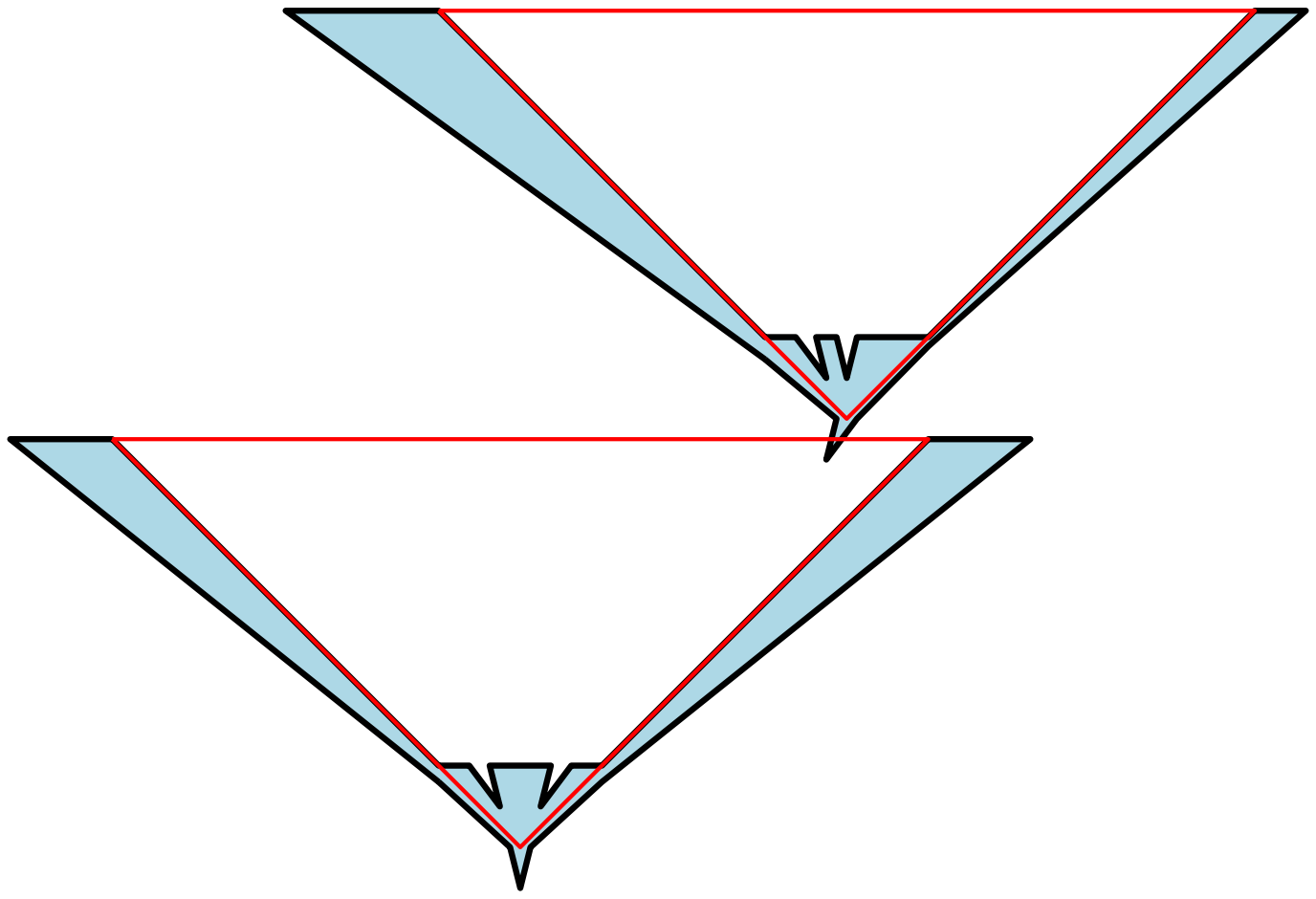}
\caption{ One of $S_i$ or $S_j$ must be stacked in the other, otherwise the triangles shown have area larger than $A_0+\frac12qn^2$. }
\label{fig:apx:stackgraphcomplete}
\end{figure}
\end{proof}

Recall that we orient the packing so that $C$ has the same orientation as in \Cref{fig:apx:containerpiece}. We set $\delta_1=\frac{1}{64n}$. 
Consider two consecutive polygons $S_i,S_j$ in the
nesting order. If $(i,j)\in E$, the spike of $S_j$
can enter a matching slit of $S_i$, allowing the
two polygons to nest closely.
If $(i,j)\notin E$, we show that $S_j$ must be
placed substantially higher. There are two cases.
If the relative rotation is small, the spike cannot
penetrate deeply into a nonmatching slit because
their angles are separated. If the relative rotation
is larger, the long arms themselves force a large
vertical separation. In either case, a missing edge
incurs an additional height of $\Omega(n)$.
We can now perform the key calculation:

\begin{lemma}\label{lem:apx:soundness}
Suppose the conditions of \Cref{lem:apx:formsstack} are satisfied. If $S_j$ is stacked inside $S_i$, then the star point of $S_j$ has height at least $\left(1-\frac{12}{q}\right)(n+1)$ above the star point of $S_i$. If additionally there is no edge from $i$ to $j$ in $G$, then the star point of $S_j$ has height at least $\left(1-\frac{12}{q}\right)(n+1)+\frac12\left(1-\frac{36}{q}\right)n$ above the star point of $S_i$.
\end{lemma}

\begin{proof}
The conditions of \Cref{lem:apx:similarangles} are satisfied, so all the $S_i$ have orientation close to the one shown in \Cref{fig:apx:sipiece1}. If $S_{j}$ is stacked inside $S_i$, the the lowest possible height it can have is when the spike on $S_j$ fits exactly into one of the slits in $S_i$, as in \Cref{fig:apx:goodstack}. If $S_i$ is oriented as in \Cref{fig:apx:sipiece1}, then the star point of $S_j$ has height $n+1$ above that of $S_i$ in this case. Note that the horizontal offset between the star points of $S_i$ and $S_j$ is at most the vertical offset (so at most $n+1$ here). Since $S_i$ can be rotated by $\theta \le \frac{6}{q}$ compared to the orientation in \Cref{fig:apx:sipiece1}, the star point of $S_j$ has height at least $(\cos(\theta)-\sin(\theta))(n+1)\ge \left(1-\frac{12}{q}\right)(n+1)$ above that of $S_i$ (note $q\ge 4000$).

This type of stacking only exists when there is an edge from $i$ to $j$ in $G$. So suppose that there is no such edge. The difference in angle between edges of different possible slits in an $S_i$ is at least:

\[\arctan\left(\frac{n-\frac12\delta_1}{n+1}\right)-\arctan\left(\frac{n-1+\frac12\delta_1}{n+1}\right)\]

\noindent which is at least $\frac{1}{4}\cdot\frac{1-\delta_2}{n+1}\ge \frac{1}{8n}$. First, suppose that the orientations of $S_i$ and $S_j$ differ by an angle less than $\frac{1}{16n}$. Since there is no edge from $i$ to $j$, there is no slit in $S_i$ that corresponds to the spike on $S_j$, and the walls of spike on $S_j$ differ in angle from the walls of any slit in $S_i$ by an angle of at least $\frac{1}{16n}$. If we orient $S_i$ as in \Cref{fig:apx:sipiece1}, then the spike in $S_j$ sticks into a slit in $S_i$ by a vertical distance of at most $2\frac{\delta_1}{\sin\left(\frac{1}{16n}\right)}$, as illustrated in \Cref{fig:apx:soundness1}. Since $n\ge 100$, we have $\sin(\frac{1}{16n})\ge \frac{1}{32n}$, so this is at most$64\delta_1n=1$. 

\begin{figure}
\centering
\includegraphics[page=11]{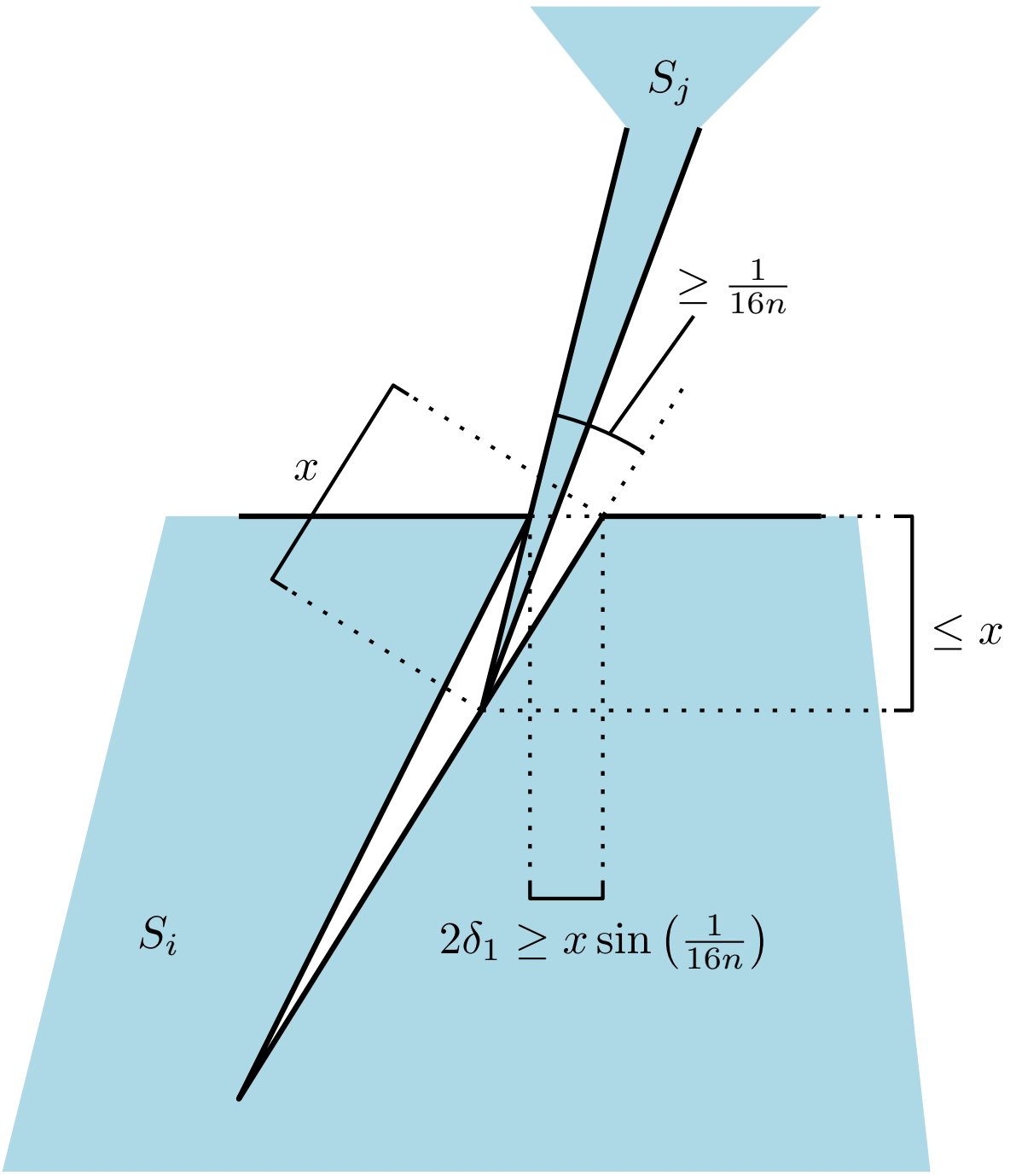}
\caption{ If $S_i$ and $S_j$ differ in angle by $\le \frac{1}{16n}$, then the spike in $S_j$ can't go far into a non-matching slit in $S_i$. }
\label{fig:apx:soundness1}
\end{figure}

So the star point of $S_j$ has a height of $(n+1)+\frac12n\left(\cos(\phi)-\sin(\phi)\right)$ above that of $S_i$, where $\phi\le \frac{12}{q}$ is the angle between the orientations of $S_i$ and $S_j$. So this is at least $(n+1)+\frac12\left(1-\frac{24}{q}\right)n$.

We changed the orientation of $S_i$ by at most $\frac{6}{q}$ here, so we need to change it back, and we get a height difference of at least $\left(1-\frac{12}{q}\right)(n+1)+\frac12\left(1-\frac{36}{q}\right)n$.

Now consider the case that the orientation of $S_j$ differs from that of $S_i$ by more than $\frac{1}{16n}$. Again, we orient $S_i$ and $S_j$ so that it is as in \Cref{fig:apx:sipiece1}. In this case, the star point of $S_j$ is at least $\sin\left(\frac{1}{16n}\right)(h+1)(n+1)$ above the star point of $S_i$, as illustrated in \Cref{fig:apx:soundness2}. Since $(h+1)\ge 4000n$, this is at least $250(n+1)$. Even after orienting $S_i$ back to its original orientation, this is much larger than the bound claim by the lemma.

\begin{figure}
\centering
\includegraphics[page=12]{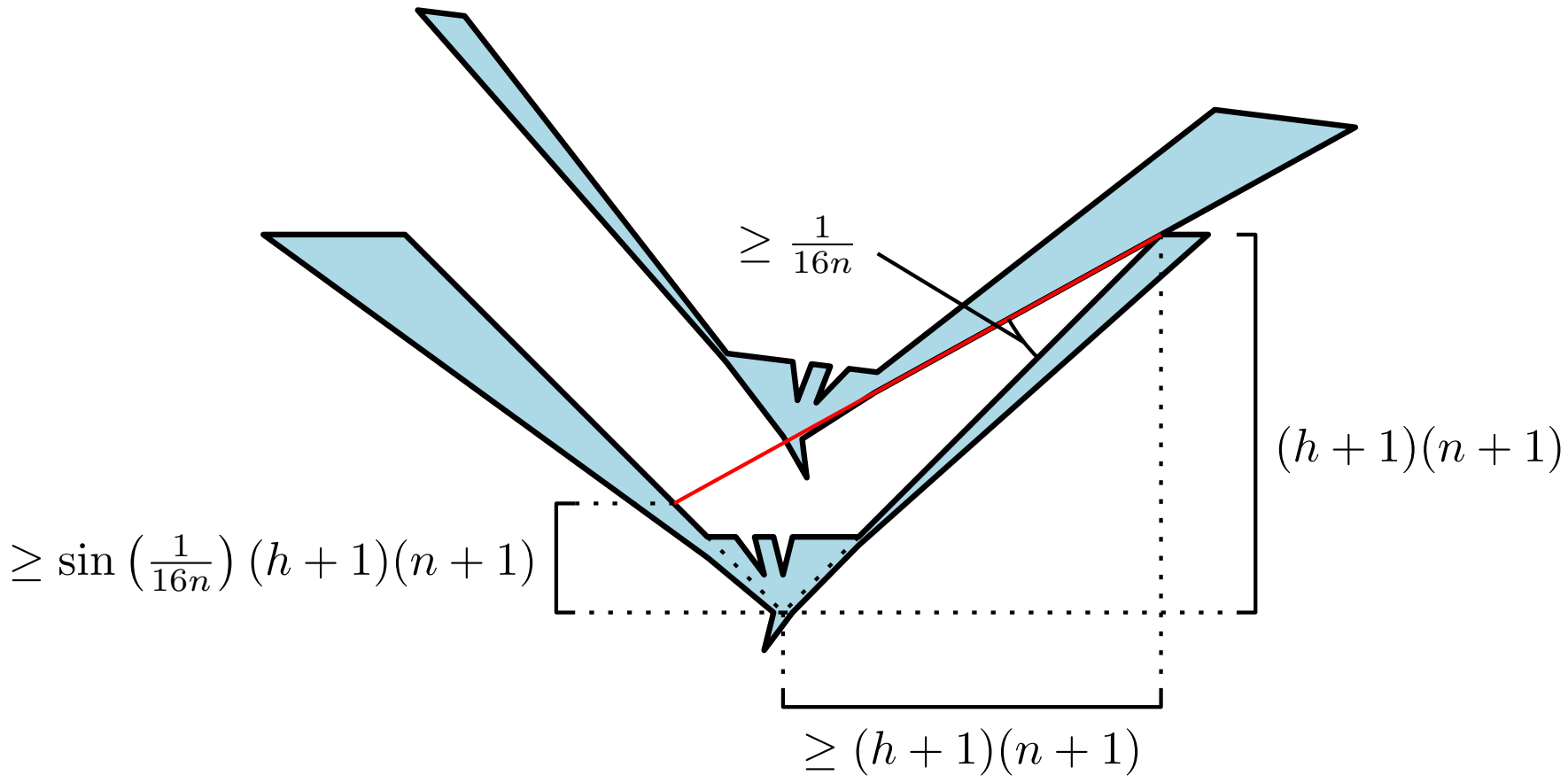}
\caption{ If $S_j$ is rotated by more than $\frac{1}{16n}$ compared to $S_i$, then its star point must be above the red line shown in the figure. }
\label{fig:apx:soundness2}
\end{figure}
\end{proof}

We can now prove that constant complexity star packing is APX-hard:

\begin{theorem}\label{thm:apx:main}
For some $\epsilon>0$, it is NP-hard to approximate the star packing to within a $(1+\epsilon)$ factor, even if the stars are polygons with at most $17$ vertices.
\end{theorem}

\begin{proof}
Let $\epsilon=\frac{1}{1396641}$ be a constant small enough for \Cref{lem:apx:disjpathcover} to apply. We reduce from a MAX-VERTEX-DISJOINT-PATH-COVER instance $G=(V, E)$, with $|V|=n$, produced by \Cref{lem:apx:disjpathcover}. The construction is as above, with $q=\frac{1000}{\epsilon}$. If $G$ has a Hamiltonian path, then \Cref{lem:apx:completeness} shows that we can pack the pieces in a convex container of area $A_0$, where $A_0$ is the area of $C$ and satisfies $q^2n^4\le A_0\le \frac32q^2n^4$. 

Write $\alpha=\frac{1}{q}=\frac{1}{1000}\epsilon$. Write $\beta=\frac{\alpha}{q}$. We suppose that $n$ is large enough that $\frac{1}{n}\le \frac{1}{100}\epsilon$.

Suppose now that there exists a packing with convex hull of area at most $(1+\frac{2}{3}\beta)A_0\le A_0+\alpha qn^4$. Since $\alpha\le \frac12$, \Cref{lem:apx:formsstack} applies, and there is an ordering $t(1), \dots, t(n)$ of $1, \dots, n$ such that $S_{t(i+1)}$ is stacked in $S_{t(i)}$ for each $i=1, \dots, n-1$. We construct a path cover of $G$ as follows: for each $i=1, \dots, n-1$, we add the edge from $t(i)$ to $t(i+1)$ to the cover if it exists. This gives a set of disjoint paths in $G$. For any vertex not covered, we add the zero-length path containing only that vertex. Let $m$ be the number of edges in this disjoint path cover.

By applying \Cref{lem:apx:soundness} to each pair $S_{t(i)}$ and $S_{t(i+1)}$ and summing, the difference in height between the star point of $S_{t(1)}$ and the star point of $S_{t(n)}$ is at least $(1-\frac{36}{q})\left((n-1)(n+1)+\frac12(n-1-m)n\right)$. So the difference in height between the star point of $S_{t(1)}$ and a point on the top opening of $S_{t(n)}$ is at least $(1-\frac{36}{q})((n-1)(n+1)+\frac12(n-m)n)+(h+1)(n+1)$. The height difference between a point on the top of $S_{t(n)}$ and the star point in $S_{t(n)}$ is at least $(h+1)(n+1)$, even if $S_{t(n)}$ is rotated slightly (see \Cref{lem:apx:similarangles}).

The height of $C$ is $(h+n+1)(n+1)$. So some $S_i$ sticks out above the top of $C$ by at least:

\[(1-\tfrac{36}{q})((n-1)(n+1)+\tfrac12(n-1-m)n)-n(n+1)\]

If there is a point in some $S_i$ a distance of $\kappa$ above the top of $C$, then the convex hull of the packing has area at least $A_0+(h+1)(n+1)\kappa$. Since we know the area is at most $A_0+\alpha qn^4$, we have $\kappa\le \alpha n^2$. So:

\[\tfrac12(1-\tfrac{36}{q})(n-1-m)n-\tfrac{36}{q}(n-1)(n+1)-2(n+1)\le \alpha n^2\]
\[\tfrac12(n-1-m)n-\tfrac{54}{q}(n-1)(n+1)-2(n+1)\le \alpha n^2\]
\[\frac12\frac{n-1-m}{n-1}-\frac{54}{q}-\frac{2}{n}\le 2\alpha\]

For the last step, note that we are assuming $n$ is large.

If $G$ does not have a Hamiltonian path, then \Cref{lem:apx:disjpathcover} says that the best path of $G$ cover has at most $(1-\epsilon)(n-1)$ edges. In this case, $n-1-m\ge \epsilon(n-1)$, so:

\[\frac12\cdot\frac{n-1-m}{n-1}-\frac{54}{q}-\frac{2}{n-1}\ge \frac12\epsilon-\frac{1}{10}\epsilon-\frac{1}{10}\epsilon>2\alpha\]

By the above, this is incompatible with the area of the convex hull being smaller than $(1+\frac{2}{3}\beta)A_0$. So if the area of the convex container is at most $(1+\frac{2}{3}\beta)A_0$, then we can conclude that there exists a Hamiltonian path. So it is NP hard to approximate star packing to within a factor of $1+\frac23\beta$. Since $\beta=\frac{\alpha}{q}=\frac{1}{q^2}=\frac{1}{10^6}\epsilon^2$ does not depend on $n$, this establishes the claim.
\end{proof}

\section*{AI Methodology}
\addcontentsline{toc}{section}{AI Methodology}

ChatGPT was used to obtain some of the crucial ideas.
The model suggested the argument with two test lines in the proof of \Cref{thm:star-local-decomposition-strong}, but in the context of packing wedges, i.e., degenerate weakly simple polygons that are the union of two line segments sharing an endpoint.
The authors managed to generalize the argument to arbitrary star-shaped polygons and use it to design an $O(\log^2 n)$-approximation algorithm for stars.
The model also suggested a version of the argument that our algorithm gives an $O(\log n)$-approximation rather than an $O(\log^2 n)$-approximation, even though we have $\Theta(\log^2 n)$ dyadic size scale combinations of $r,w$ (\cref{lem:logn}).
Finally, the model suggested the argument for the rotational algorithm and suggested to obtain APX-hardness by reducing from MAX-DISJOINT-PATH-COVER, but the actual reduction was found by the authors.
All ideas from the model have been carefully studied, adjusted and rewritten by the authors.
ChatGPT and Claude have also been used as writing aids.

\section*{Acknowledgements}
\addcontentsline{toc}{section}{Acknowledgments}

M.~Abrahamsen, J.~Conradi, J.~Stade, and H.~Zhang are supported by the Independent Research Fund Denmark, grant 1054-00032B, and by the Carlsberg Foundation, grant CF24-1929.
A.~Khan is partly supported by Google India Research Award, Ittiam Systems CSR grant, the Walmart Center for Tech Excellence at IISc (CSR Grant WMGT-23-0001), and SERB Core Research Grant (CRG/2022/001176) on ``Optimization under Intractability and Uncertainty''.
S.~Wong is supported by the European Union's Marie Sk{\l}odowska-Curie Actions Postdoctoral Fellowship, grant number 101146276.
S.~Yan is supported by VILLUM Foundation Grant 54451, Basic Algorithms Research Copenhagen (BARC).

\bibliographystyle{alphaurl-doi}
\bibliography{bib}

\newpage

\appendix

\section{Proof of \cref{cor:bin-strip}}\label{app:cor:bin-strip}

\begin{proof}[Proof of \cref{cor:bin-strip}]
Let $\mathcal M$ be the set of mini-containers computed by \cref{lem:total-minicontainer-area} (or \Cref{lem:total-minicontainer-area-rot} in the rotational case), and let $\opt_{\mathrm{bin}}$ and $\opt_{\mathrm{strip}}$ denote the optimum of the bin and strip packing instance, respectively.
Recall that plain $\opt$ is the area of the smallest convex container.
We assume that the strip is vertical, i.e., the strip is $[0,1]\times[0,\infty)$ and its length is measured in the $y$-direction.
The bins of an optimal bin packing, placed side by side, form a rectangle of area $\opt_{\mathrm{bin}}$, and the used part of the strip in an optimal strip packing is a rectangle of area $\opt_{\mathrm{strip}}$.
Hence $\opt\le\min\{\opt_{\mathrm{bin}},\opt_{\mathrm{strip}}\}$.
By \cref{lem:total-minicontainer-area} (or \Cref{lem:total-minicontainer-area-rot} in the rotational case), the total area of $\mathcal M$ is $O(\log n)\cdot\opt$, and thus at most $O(\log n)\cdot\opt_{\mathrm{bin}}$ and $O(\log n)\cdot\opt_{\mathrm{strip}}$.

In the rotational setting, we first rotate each mini-container so that its minimum width is attained in the vertical direction.
Let $\ell_{\max}$ and $h_{\max}$ be the maximum horizontal and vertical spans of the mini-containers.
Since $\opt_{\mathrm{strip}}$ is at least the vertical span of any star (in the rotational setting, its minimum width), \eqref{eqn:span-mini-container} gives $h_{\max}=O(\opt_{\mathrm{strip}})$.
Moreover, every mini-container has diameter $O(\delta)$, and we choose $\delta$ such that $\ell_{\max},h_{\max}\le\frac14$.

We apply \cref{thm:convex-packing} to $\mathcal M$ and obtain an axis-aligned rectangle $R$ of area $O(\log n)\cdot\opt+O(\ell_{\max}h_{\max})$.
Since $\ell_{\max}\le1$ and $h_{\max}=O(\opt_{\mathrm{strip}})$, we have $|R|=O(\log n)\cdot\opt_{\mathrm{strip}}$, and since $\ell_{\max}h_{\max}\le1\le\opt_{\mathrm{bin}}$, we have $|R|=O(\log n)\cdot\opt_{\mathrm{bin}}$.
It remains to cut $R$ into pieces that fit into the strip or the bins.

Let us first consider strip packing.
Let $R$ be an $a\times b$ rectangle, and note that $\ell_{\max}\le a$ and $h_{\max}\le b$.
If $a\ge\frac12$, we cut $R$ by vertical lines into $\lceil 2a\rceil\le 4a$ pieces of width at most $\frac12$ and height $b$.
If $a<\frac12$, we cut $R$ by horizontal lines into $\lceil b/h_{\max}\rceil$ pieces of width $a$ and height at most $h_{\max}$.
We assign each mini-container to the piece that contains the lower left corner of its axis-aligned bounding box.
Then the mini-containers assigned to a piece lie in that piece extended by $\ell_{\max}$ to the right and by $h_{\max}$ upwards.

In the first case ($a\ge\frac12$), each extended piece has width at most $\frac12+\ell_{\max}<1$ and height at most $b+h_{\max}\le 2b$.
We place the extended pieces on top of each other, which gives length at most $4a\cdot 2b=8ab=O(|R|)$.

In the second case ($a<\frac12$), each extended piece has width at most $a+\ell_{\max}\le 2a<1$ and height at most $2h_{\max}$.
We place the extended pieces in rows, with $\lfloor 1/(2a)\rfloor\ge 1/(4a)$ pieces side by side in each row.
The number of rows is at most $4a\lceil b/h_{\max}\rceil+1\le 4ab/h_{\max}+3$, using $4a<2$.
Since each row has height at most $2h_{\max}$, this gives length at most $8ab+6h_{\max}=O(|R|+h_{\max})$.

In both cases, the length is $O(\log n)\cdot\opt_{\mathrm{strip}}$, since $|R|=O(\log n)\cdot\opt_{\mathrm{strip}}$ and $h_{\max}=O(\opt_{\mathrm{strip}})$.

For bin packing, we first construct a strip packing of length $L=O(|R|+h_{\max})=O(|R|+1)$ as above.
We cut the strip by horizontal lines at distance $\frac12$ and assign each mini-container to the part containing the lower left corner of its bounding box.
The mini-containers assigned to a part lie in a rectangle of width $1$ and height at most $\frac12+h_{\max}<1$, which fits into a bin.
This gives at most $2L+1=O(|R|+1)=O(\log n)\cdot\opt_{\mathrm{bin}}$ bins, since $\opt_{\mathrm{bin}}\ge1$.
\end{proof}

\section{Proof of \cref{thm:general-macc-to-bin}}\label{app:thm4}

Recall the statement of this theorem. 

\binreduction*

\begin{proof}
The proof mostly follows the structure of \cref{sec:star-packing}. 
Most of the arguments follow directly from the case of stars, and we will omit repeated details and focus only on necessary modifications for general polygons. 

As we did for stars, we divide the input polygons into size classes and then fine types, show that packing each fine type separately will only incur an $O(\log n)$ extra factor in the approximation ratio. 
The mini-packing within each fine type can be simply reduced to a bin-packing problem by an affine transformation. 
Let $\mathcal{P}$ be the set of polygons to be packed. 

~ \\ \noindent \textbf{Parameterize general polygons.}
Let $P$ be an arbitrary polygon. We define $a_P, b_P$ to be the farthest pair of points in $P$, measured in Euclidean distance. 
Let $c_P$ be the farthest point from line $\overline{a_P b_P}$ in $P$, measured in Euclidean distance as well. 
If there are multiple candidates for $a_P, b_P$ and $c_P$, we choose the triple that maximizes the lexicographical order of vector $\langle x(a_P), y(a_P), x(b_P), y(b_P), x(c_P), y(c_P)\rangle$. 
Note that segment $a_P b_P$ may not be inside $P$, which is different from the case of radii $q_Q R_Q$ for star $Q$. 
We define
\[
\Delta_P = \|a_P b_P\|, \qquad w_P = \text{dist}(c_P, \overline{a_P b_P}).
\]

\newcommand{\conv}{\text{Conv}}
Similar to the use of triangle $q_Q L_Q R_Q$ for star $Q$, triangle $a_P b_P c_P$ is also an ``approximation'' of $\conv(P)$, where $\conv(P)$ is the convex hull of set $P$. 

\begin{fact}
Any polygon $P$ lies in a rectangle of size $\Delta_P \times (2w_P)$. 
\end{fact}

To obtain the $O(\alpha\log n)$-approximation algorithm, we use the same strategy as packing stars. 
Let $\Delta_{\max} = \max_{P \in \mathcal{P}} \Delta_P$ and $w_{\max} = \max_{P \in \mathcal{P}}w_P$. 
Similar to stars, we partition $\mathcal{P}$ into dyadic size groups $\mathcal{G}_{a,b}$ of scales $\Delta, w$ that contains all polygons $P$ with 
\[
\Delta_P \in (r/2, r], \quad w_P \in (w/2, w], \quad \Delta = \frac{\Delta_{\max}}{2^a}, \quad w = \frac{w_{\max}}{2^b}. 
\]
In each size group $\mathcal{G}_{a,b}$, we partition polygons into fine types based on the orientation of the diameter $\overrightarrow{a_P b_P}$, where the angular difference in each fine type is at most $\frac{\pi w}{50\Delta}$. 
This definition is the same as stars when we replace the role of radii by diameter. 
For each fine type $\tau$, we define $\Vec{d}_{\tau}$ and $\Vec{u}_{\tau}$ similarly as we did for stars. 

Next, we show a similar structural packing lemma for general polygons. 

\begin{lemma}[Similar to \cref{thm:star-local-decomposition-strong}]
\label{lem:polygon-local-decomposition-strong}
Let $\mathcal P$ be a set of polygons in one size group with scales $\Delta,w$, and let $\Pi$ be a packing of $\mathcal{P}$ in a convex container $C$.
For any $0 < c < 0.1$ and sufficiently small $\varepsilon>0$, there exists a partition
\[
\mathcal{P} = \mathcal{P}_1\mathbin{\dot\cup} \cdots \mathbin{\dot\cup}\mathcal{P}_t, 
\]
such that each $\mathcal{P}_j$ only contains stars of one fine type $\tau_j$ and $\mathcal{P}_j$ fits into a rectangle of dimensions $O(\Delta) \times O(w)$ aligned with $\Vec{u}_{\tau}$ and $\Vec{d}_{\tau}$. 
\end{lemma}

\begin{proof}
The proof follows the test line argument in the proof of \cref{thm:star-local-decomposition-strong}. 
Let $T$ be the enclosing vertical parallelogram of $C$. 
Treat $a_P b_P$ to be the ``radii'' of $P$, we can use the same way to partition $T$ into vertical strips and take the same set of test lines for each of the vertical strip. 
See \cref{fig:stabbing-segment}. 
The only part that does not hold is the counting bound on the number of occupied pairs for each vertical strip. 

Let $(a, b)$ be an occupied pair that all polygons in it intersects interval $s_{1,a}$ on $\ell_1: x = x_0 + c\Delta$ and interval $s_{2,b}$ on $\ell_2: x = x_0 + 2c\Delta$. 
For each fine type $\tau$ and all polygons $P$ in this occupied pair $(a, b)$ that falls into this fine type, there is a bounding box $R$ of dimensions $(K'\Delta) \times (K'w)$ for some universal constant. 
We also know that and its short side is along $\Vec{d}_{\tau}$, which is at most $2$ degree away from vertical.
Let $s'_{1, a} = R \cap \ell_1$ and $s'_{2, b} = R \cap \ell_2$. 
We know both $s'_{1, a}$ and $s'_{2, a}$ has length at most $\frac{K'w}{\cos 2^{\circ}} \le 1.01K'w$. 
So $s'_{1,a}$ and $s'_{2,b}$ touches at most $A = \Theta_{K', \eps}(1)$ intervals on $\ell_1$ and $\ell_2$ separately. 

Let $P$ be some polygon packed in $R$. 
Since $a_P b_P$ crosses both $\ell_1$ and $\ell_2$, without loss of generality, assume $a_P$ lies to the left of $\ell_1$ and $b_P$ lies to the right of $\ell_2$. 
Then there must be a path $\pi$ in $P$ that connects $a_P$ and $b_P$. 
Therefore, there must be a subpath $\pi'_{a,b} \subseteq \pi$ in $P$ connecting $s'_{1, a}$ and $s'_{2, b}$ between $\ell_1$ and $\ell_2$. 
We will call this subpath $\pi'_{a,b}$ the induced subpath of occupied group $(a, b)$. 
These subpaths have to be crossing free due to the feasibility of this packing. 
Therefore, for any $a_1 \le a_2 - 2A$ and $b_1 \ge b_2 + 2A$, $(a_1, b_1)$ and $(a_2, b_2)$ cannot be both occupied, because otherwise their induced subpath must cross. 
This suggest that the number of occupied pairs is $O(m)$. 

Combined with all other parts, this lemma holds. 
\end{proof}

This part of packing into mini-containers can be rephrased into a bin packing problem. 
For each fine type $\tau$, let $f_{\tau}$ be a linear transformation defined by $f(K\Delta\Vec{u}_{\tau}) = (0, 1)$ and $f(Kw\Vec{d}_{\tau}) = (1, 0)$. 
Let $\mathcal{P}_{\tau}$ be the set of polygons in fine type $\tau$. 
The algorithmic problem for packing $\mathcal{P}_{\tau}$ into rectangles of dimensions $(K\Delta) \times (Kw)$, is equivalent to packing $f(\mathcal{P}_{\tau})$ into minimum number of unit axis-aligned squares. 
Each polygon after this linear transformation has diameter at most $\frac{4}{K}$ and convex hull area at least $\frac{1}{8K^2}$. 
Therefore, we can apply the assumed algorithm to obtain an $\alpha$-approximation in polynomial time. 

Then we need to generalize the $O(\log n)$ analysis of the total optimum size across all size groups. 

\begin{lemma}[Similar to \cref{lem:logn}]
For any set of polygons $\mathcal{P}$, 
\[
\sum_{\mathcal{G}:\text{ size group}}\opt(\mathcal{G}) = O(\log n) \cdot \opt(\mathcal{P}). 
\]
\end{lemma}

\begin{proof}
This proof follows the structure of \cref{sec:star-local-grouping}. 
For each polygon $P$, let $T_P$ be the triangle $a_Pb_Pc_P$. 
We define the core of $P$ in the same way as we did for stars, where replacing the role of $q, L, R$ by $a, c, b$ respectively. 
Let $\Vec{e}_P$ be the unit vector from $a_P$ to $b_P$, $\Vec{v}_P$ be the unit vector perpendicular to $\Vec{e}_P$ and pointing towards $c_P$. 
Then the core $K_P$ of $P$ is the triangle $a'_P b'_P c'_P$ that
\[
a'_P = (a_P + b_P)/2 + w_P \Vec{e}_P / 10, \qquad b'_P = a'_P + (b_P - a_P)/10, \qquad c'_P = a'_P + (c_P - a_P)/10. 
\]
Similarly, $K_P \subset T_P$ and any point $p \in K_P$ is at least $w_P/20$ away from $\partial T_P$. 

First, we show that the cores are disjoint for polygons from some different size groups. 

\begin{lemma}[Similar to \cref{lem:core-disjoint}]
\label{lem:core-disjoint-polygon}
For any two polygons $P_1$ and $P_2$, if $\Delta_{P_1} > 10\Delta_{P_2}$ and $w_{P_2} > 40w_{P_2}$, then $K_{P_1} \cap K_{P_2} = \emptyset$. 
\end{lemma}

\begin{proof}
By contradiction, assume $p \in K_{P_1} \cap K_{P_2}$, $r_{P_1} > 10r_{P_2}$, and $w_{P_2} > ?w_{P_2}$. 
Without loss of generality, we choose the coordinates such that $\Vec{e}_{P_1}$ is horizontal and $\Vec{v}_{P_1}$ is vertical, and $a_P$ to be the origin. 
Then $P_1$ is contained in rectangle $R = [0, \Delta_{P_1}] \times [-w_{P_1}, w_{P_1}]$. 
Let $\pi$ be a path in $P_1$ from $a_P$ to $b_P$. 
See \cref{fig:core-disjoint-polygon}. 

\begin{figure}
\centering
\includegraphics[page=14]{figs.pdf}
\caption{Proof of \cref{lem:core-disjoint-polygon}. By contradiction, assume $p \in K_{P_1} \cap K_{P_2}$. Since $P_1 \subseteq {\color{orange}R}$, and ${\color{red}B_2} \subseteq \conv(P_2) \subseteq {\color{blue}B_1}$, $P_2$ must cross $\pi$. }
\label{fig:core-disjoint-polygon}
\end{figure}

Since $p \in P_2$, $P_2$ lies in $B_1 = \text{Ball}(p, 2\Delta_{P_2})$, which lies in the vertical strip $x \in [0.2\Delta_{P_1}, 0.8\Delta_{P_2}]$. 
So $B_1$ must be cut by $\pi$ into multiple connected components, with at least the upper part and lower part beyond $R$ separated. 
By connectivity of $P_2$, $P_2$ must be in one of the connected components, so $P_2$ cannot reach both the upper and lower parts beyond $R$. 
Consider $B_2 = \text{Ball}(p, w_{P_2}/20)$. 
Since $w_{P_2} > 40w_{P_1}$, $B_2$ must cross both the upper and lower boundary of $R$. 
Recall that $B_2 \subset K_{P_2} \subseteq \conv(P_2)$, so $P_2$ must simultaneously reach both the upper and lower parts beyond $R$, which leads to a contradiction. 
\end{proof}

Then we show that all polygons in the same size group touching a core can be bounded into a matching size rectangle. 

\begin{lemma}[Similar to \cref{lem:core-rectangle}]
\label{lem:core-rectangle-polygon}
Let $P$ be a polygon in the size group with scales $\Delta$ and $w$. There is an $O(\Delta) \times O(w)$ rectangle that contains all polygons $P'$ in the same group such that $K_{P'} \cap K_P \neq \emptyset$.
\end{lemma}

\begin{proof}
Similar to the case of stars, it's sufficient to show that all $P'$ such that $K_P \cap K_{P'} \neq \emptyset$ have similar orientation as $P$. 
Let $\theta$ be the smaller angle formed by $\overline{a_P b_P}$ and $\overline{a_{P'} b_{P'}}$. 
By contradiction, assume $\theta > 1000w/\Delta$. 
Then consider the two bounding rectangles of size $\Delta \times (2w)$ of $P$ and $P'$. 
Without loss of generality, assume $a_P$ is the origin and $\overline{a_P b_P}$ is horizontal. 
See \cref{fig:core-disjoint-polygon}. 
Since $p \in K_{P'}$, $|y(a_{P'})| > 0.4r_{P'}\sin\theta - 2w > w$ and $|y(b_{P'})| > 0.4r_{P'}\sin\theta - 2w > w$, so one of $a_{P'}$ and $b_{P'}$ is above $y = w$, and the other is below $y = -w$. 
Therefore, for any path $\pi$ in $P$ connecting $a_P$ and $b_P$ and any path $\pi'$ in $P'$ connecting $a_{P'}$ and $b_{P'}$, $\pi$ and $\pi'$ must cross. 
Therefore, $\theta = O(w/\Delta)$, and one can create an axis-aligned bounding $O(\Delta) \times O(w)$ rectangle for all polygons with their core touching $K_P$. 

\begin{figure}[!htb]
\centering
\includegraphics[page=15]{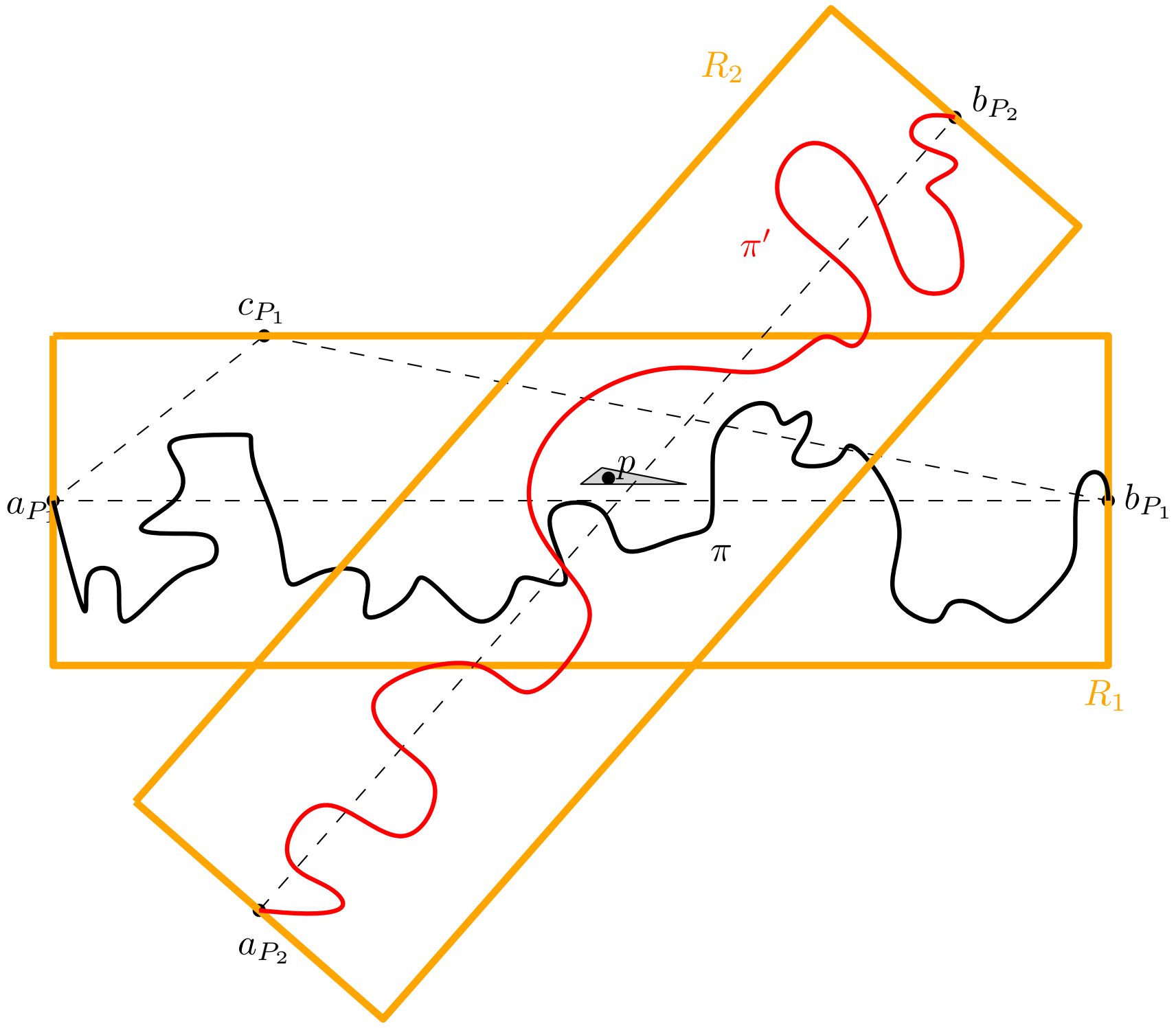}
\caption{Proof of \cref{lem:core-rectangle-polygon}. $p \in K_P \cap K'_P$ and the angle between $\overline{a_P b_P}$ and $\overline{a_{P'}b_{P'}}$ are larger than $1000w/\Delta$, then the two paths $\pi$ and $\pi'$ in $P$ and $P'$ connecting the two ends of their diameter must cross. }
\label{fig:core-rectangle-polygon}
\end{figure}
\end{proof}

Following the same independent set argument, we get a full proof. Repeated details are omitted. 
\end{proof}

Therefore, we obtain a polynomial time $O(\alpha \log n)$-approximation algorithm for packing $n$ polygons into MACC. 
\end{proof}

\end{document}